\documentclass[aps,onecolumn,pre,notitlepage,twocolumn,footinbib,longbibliography,10pt]{revtex4-1}

\usepackage{amsmath}
\usepackage{amsfonts}
\usepackage{amssymb}

\usepackage[shortlabels]{enumitem}

\usepackage[toc,page]{appendix}

\usepackage{bm}

\usepackage{graphics}
\usepackage{epsfig}
\usepackage{colortbl}
\usepackage{color}
\usepackage{nicefrac}
\usepackage{mathrsfs}
\usepackage{bbm}
\usepackage{wrapfig}

\usepackage{xcolor}
\usepackage{hyperref}

\makeatletter
\def\@seccntformat#1{\@ifundefined{#1@cntformat}%
	{\csname the#1\endcsname\quad}
	{\csname #1@cntformat\endcsname}
}
\makeatother

\catcode`\@=11
\def\numberbysection{\@addtoreset{equation}{section}
	\def\theequation{\thesection.\arabic{equation}}}

\definecolor{gruen}{rgb}{0,0.625,0}
\definecolor{rot}{rgb}{0.75,0,0}

\begin{document}
	
	\title{Equilibrium properties of two-species reactive lattice gases on random catalytic chains}
	
	\author{Dmytro Shapoval$^{1,2}$,  Maxym Dudka$^{1,2,3}$, Olivier B\'enichou$^{4}$, and Gleb Oshanin$^{4}$}

	\affiliation{
		\mbox{\vbox{$^1$Institute for Condensed Matter Physics, National Academy of Sciences of Ukraine, 1 Svientsitskii Street, UA-79011 Lviv, Ukraine}} \\
		\mbox{\vbox{$^2$ ${\mathbb L}^4$ Collaboration \& Doctoral College for the Statistical Physics of Complex Systems, Leipzig-Lorraine-Lviv-Coventry, Europe}} \\
		\mbox{\vbox{$^3$Institute of Theoretical Physics, Faculty of Physics, University of Warsaw, Pasteura 5, 02-093 Warsaw, Poland}}\\
		\mbox{\vbox{$^4$Sorbonne Universit\'e, CNRS, Laboratoire de Physique Th\'eorique de la Mati\`ere Condens\'ee (UMR CNRS 7600), 4 Place Jussieu, 75252 Paris Cedex 05, France}}
	}


	\begin{abstract}
		We focus here on the thermodynamic properties of  adsorbates formed by two-species $A+B \to \oslash$ reactions 
		on a one-dimensional infinite lattice with heterogeneous "catalytic" properties. 
		In our model 
		hard-core $A$ and $B$ particles undergo continuous exchanges with their reservoirs and react 
		when dissimilar species appear at neighboring lattice sites in presence of a "catalyst." 
		The latter is modeled by supposing either that
		randomly chosen bonds in the lattice promote reactions (Model I) or that reactions are activated by randomly chosen lattice sites (Model II). 
		In the case of annealed disorder in spatial distribution of a catalyst 
		we calculate the pressure of the adsorbate by solving three-site (Model  I) or four-site (Model II) recursions obeyed by the corresponding 
		averaged grand-canonical partition functions. 
		In the case of  quenched disorder, we use two complementary approaches 
		to find \textit{exact} expressions for the pressure. 
		The first approach
		is based on direct 
		combinatorial arguments. In the second approach, we frame the model in terms of random matrices; 
		the pressure is then  represented as 
		an averaged logarithm of the trace of 
		a product
		of random $3 \times 3$ matrices -- either
		uncorrelated (Model I) or sequentially correlated (Model II). 
		\vspace{0.3cm}
		
		
	\end{abstract}

	\maketitle
	
	\section{Introduction}\label{I1}
	
	Many processes in nature depend on 
	reactions which take place only upon an encounter of two dissimilar species in 
	presence of a third body, a "catalyst,"
	and are chemically inactive otherwise. For diverse systems, a considerable knowledge 
	of equilibrium and out-of-equilibrium properties of such reactions is accumulated (see, e.g., Refs.~\onlinecite{bond,Dav03,evans}). 
	
	This kind of reaction, 
	which we will call here catalytically activated reactions (CARs),
	has attracted a great deal of attention from the statistical physics community 
	following a pioneering paper by Ziff, Gulari, and Barshad \cite{Ziff86}. 
	The authors studied a catalytically activated two-species $A+B \to \oslash$ reaction, and  
	revealed a surprising cooperative behavior 
	with ensuing phase transitions. A review of advancements in
	this direction can be found in Refs. ~\onlinecite{evans,marro} 
	and in the recent Ref.~\onlinecite{dud}. 
	
	Most of available 
	analysis, which used 
	a statistical physics approach to modeling CARs along the lines proposed in Ref. \onlinecite{Ziff86},
	focused on situations in which a catalytic substrate 
	has homogeneous catalytic properties. Indeed, the latter was typically considered as an ideal surface bounding a three-dimensional bath, and it was stipulated that any encounter of reactive particles  
	at any point on this surface leads to an instantaneous reaction event.  In this approach, only a few   
	works \cite{blum,tox,Osh,Osh2,OshB,OshB2,pop,pop2,dud} 
	addressed the question how a spatial heterogeneity 
	of a catalyst affects the behavior of CARs. 
	These 
	works, however, covered only a limited number of particular cases 
	such that a general understanding is lacking at present.

	In this paper we study the equilibrium properties of adsorbates formed in the course
	of catalytically activated two-species $A+B \to \oslash$ reactions, which take place
	on a one-dimensional lattice possessing 
	\textit{heterogeneous} catalytic properties. 
	We model the latter by supposing that either
	some fraction of \textit{bonds}  in the lattice prompts the reaction (see Fig. \ref{fig1}), while the rest of bonds are inert (Model I), or a catalyst is represented as an array of randomly chosen lattice \textit{sites} (see Fig. \ref{figg1}), which possess such a catalytic property (Model II). In both models, 
	particles of two species, $A$  and $B$,  are in thermal contact with their vapor phases acting as reservoirs maintained, respectively, at constant chemical potentials. The particles thus undergo continuous exchanges with their reservoirs -- they steadily 
	adsorb onto empty lattice sites, and spontaneously desorb from the lattice. 
	In Model I, the $A$ and $B$ particles appearing simultaneously on neighboring 
	sites connected by a catalytic bond, immediately react and the product desorbs. In Model II, neighboring $A$ and $B$ particles react and the product desorbs, if one of them (or both) resides on a catalytic site. 
	The $A$ and $B$ pairs appearing 
	on neighboring 
	sites, which either are connected by a noncatalytic bond (Model I),  or both are 
	noncatalytic  (Model II), 
	do not enter into a reaction. 
	
	Viewed from a statistical physics perspective, our analysis here concerns thermodynamic properties of a ternary mixture of $A$ and $B$ particles, and voids, on a
	one-dimensional lattice in contact with reservoirs of particles. In this mixture, 
	in addition to on-site hard-core interactions preventing a multiple occupancy of each site, 
	particles of dissimilar species experience (temperature-independent) infinitely large repulsive interactions once they appear on neighboring sites connected by a catalytic bond (Model I), or reside on neighboring sites, at least one of which is catalytic (Model II). 
	
	Whenever all the bonds or sites are catalytic, and only one type of particles is present in the system, i.e., for single-species $A+A \to \oslash$ reactions,  
	the reactive constraint evidently implies that  particles cannot occupy the neighboring sites.  Such models are well known (see, e.g., two-dimensional hard-squares or hard-hexagons models in Ref. \onlinecite{Bax82}) and exhibit a phase transition from a disordered phase into an ordered one at a certain value of the chemical potential.  When only some fraction of bonds or sites is catalytic, in the annealed disorder case the reactive constraint 
	becomes less restrictive and an infinite repulsion between the neighboring particles is replaced by a soft one. In principle, here
	the particles can reside on the neighboring sites, but there is a penalty to pay. 
	As evidenced by a recent Bethe-lattice analysis \cite{dud}, 
	critical behavior in this situation becomes richer. In particular, in the case of catalytic bonds  one observes a direct phase transition and a reentrant transition into a disordered phase, which both are continuous. In the case of catalytic sites, a continuous phase transition into an ordered phase is followed by a reentrant transition into a disordered one, which
	can be continuous or of the first order, depending on the concentration of a catalyst. 
	In one-dimensional systems,  
	the model of single-species CARs has been solved exactly for an arbitrary mean concentration of the catalytic sites or bonds, for the cases of both annealed and quenched disorder \cite{Osh,Osh2,OshB,OshB2}.

	For $A+B \to \oslash$ reactions only the case of annealed disorder in spatial distribution of the catalytic bonds was studied  \cite{pop,pop2}.  It was shown that
	the Hamiltonian of the system with such a CAR can be mapped 
	onto a general spin-$1$ model \cite{Bax82}.  On a honeycomb lattice, for equal chemical potentials of both species, and also under some 
	additional restrictions on the amplitude of repulsive interactions, the Hamiltonian associated with the two-species CAR reduces to an exactly solvable version of  a general spin-$1$ model \cite{hor,wu}. It was then demonstrated in Refs. \onlinecite{pop,pop2} that for equal chemical potentials of both species this CAR 
	exhibits
	a continuous symmetry-breaking transition with large
	fluctuations and progressive coverage of the entire lattice
	by either $A$ or $B$ species only.

	Here, in our analytical approach to two-species CARs on a one-dimensional lattice with heterogeneous catalytic properties, we proceed in the following way. For the case of annealed disorder in spatial distribution of a catalyst, we derive recursion schemes obeyed by
	the corresponding averaged grand-canonical partition functions, and solve them by standard means. 
	In the case of catalytic \textit{bonds}, the recursions extend over three sites, while in the case of catalytic \textit{sites} these are effectively the four-site recursions.  In a more complicated case of quenched disorder, we use two complementary approaches. In the first one, 
	we invoke rather involved but straightforward combinatorial arguments to split the lattice with a given distribution of 
	a catalyst into an array of disjoint fully connected completely catalytic clusters. Then, taking advantage of the 
	expression for the grand-canonical partition function of the model on a finite completely catalytic chain, obtained in
	Ref. \onlinecite{popes}, and calculating the weights of fully connected completely catalytic clusters of a given length, we write an exact expression for the disorder-averaged pressure. In the second approach, we use a matricial representation of the pressure,  
	by writing it as a logarithm of the trace -- the Lyapunov exponent -- of an infinite product of random three-by-three matrices. In the case of catalytic bonds these matrices are mutually uncorrelated, while in the case of catalytic sites they have sequential, pairwise correlations. We show that in such a representation the disorder-averaged pressure can be calculated exactly. We note parenthetically that exact expressions for the Lyapunov exponents are known 
	for some particular classes of random matrices (see, e.g., Refs. \onlinecite{angelo,texier}). We thus provide here nontrivial examples of random correlated matrices for which such an analysis can be carried out exactly.
	
	The paper is outlined as follows: In Sec. \ref{model} we formulate our model of catalytically activated $A+B \to \oslash$ reactions 
	and introduce basic notations. We distinguish between the case of randomly placed catalytic bonds and a more complicated case of randomly placed catalytic sites. 
	In Sec. \ref{partition}, we write the grand-canonical partition functions of Model I and Model II, discuss our analytical approaches, and present exact results for the disorder-averaged values of the partition functions (appropriate for the annealed disorder in placement of the catalytic bonds or sites), and for the disorder-averaged values of a logarithm of the partition functions (appropriate for the case of quenched disorder in placement of the catalytic bonds or sites). Next, in Sec. \ref{pressure} we analyze the behavior of the disorder-averaged pressure, densities, and compressibilities of the two-species adsorbates.  
	In Sec. \ref{conc} we conclude with a brief recapitulation of our results. The details of intermediate calculations and some of the results and figures are presented in the Appendixes.

	\section{Model}
	\label{model}
	
	Consider a one-dimensional 
	lattice containing $N$ adsorption sites, (in what follows we will turn to the limit $N \to \infty$),
	which is in thermal equilibrium with a mixed vapor phase of $A$ and $B$ particles. 
	Particles of both species undergo continuous exchanges with their respective vapor phases and
	adsorb onto \textit{empty} lattice sites, i.e. there may be at most a single particle (either $A$ or $B$) at each lattice site,
	and desorb spontaneously from the lattice. The vapor phases are maintained at constant chemical potentials
	$\mu_A$ and $\mu_B$,  
	and the corresponding activities are defined as $z_{A}=\exp{(\beta \mu_{A})}$ and $z_{B}=\exp{(\beta \mu_{B})}$, 
	where $\beta$ is the reciprocal temperature 
	measured in units of the Boltzmann constant $k_{B}$. 
	
	Further on, we introduce reactions between the adsorbed $A$ and $B$ particles. We distinguish between the cases of catalytic bonds and of catalytic sites.
	
	\subsection*{Model I. Catalytic bonds}
	\label{modelI}
	
\begin{figure}[htb!]
		\begin{center}
			\includegraphics[width=0.75\hsize]{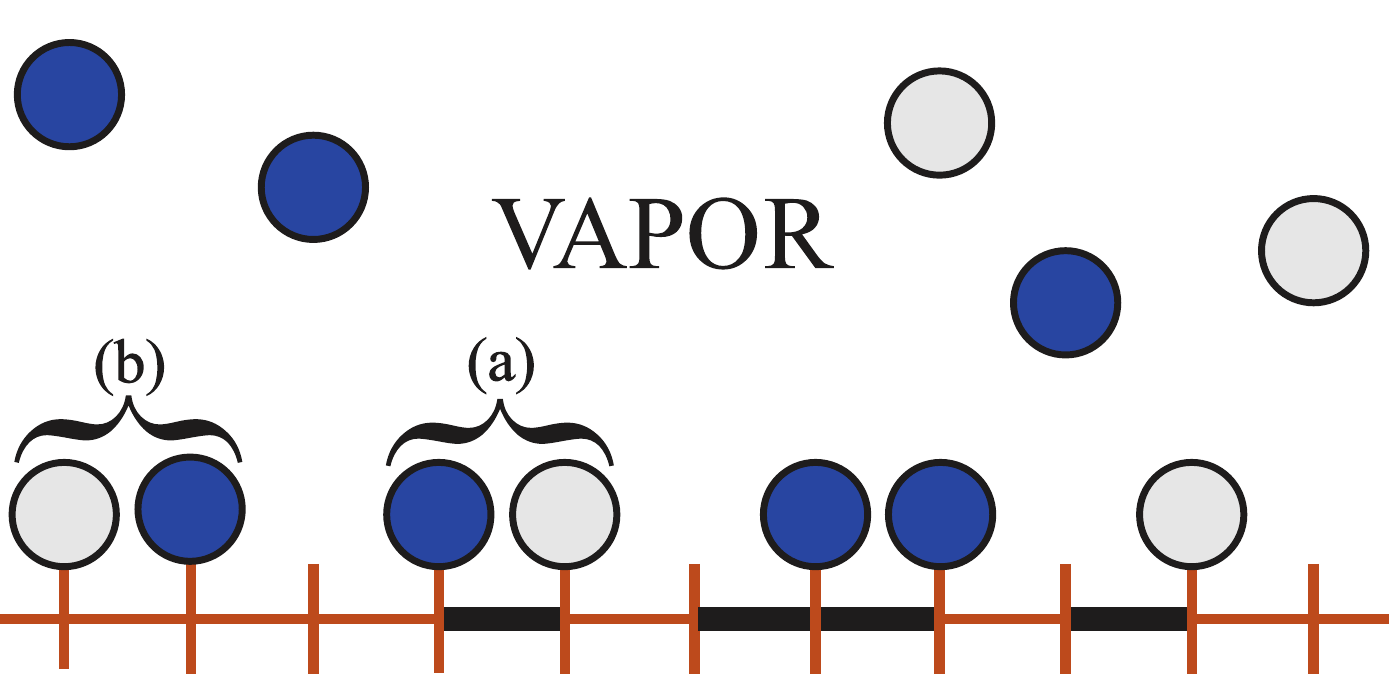}
		\end{center}
		\caption{One-dimensional lattice containing $N$ adsorption sites  in contact with vapor phases of $A$ and $B$ particles (blue and gray circles, respectively). Some fraction of bonds between the neighboring sites possesses special catalytic properties (thick black lines). $A$ and $B$ particles undergo continuous exchanges with their vapor phases, maintained at constant chemical potentials $\mu_A$ and $\mu_B$, respectively, adsorb onto empty lattice sites and desorb from the lattice. $A$ and $B$ particles appearing simultaneously at neighboring sites connected by a catalytic bond [case (a)] 
			react instantaneously, and the reaction product leaves the system. $A$ and $B$ particles adsorbed on the neighboring sites connected by a noncatalytic bond coexist [case (b)].}
		\label{fig1}
	\end{figure}

	In Model I, we choose completely at random some fraction of  \textit{bonds} of the lattice, (i.e., the intersite segments), and stipulate 
	that these selected bonds possess catalytic properties. We depict such catalytic bonds in Fig. \ref{fig1} by 
	thick black lines. Further on, we suppose that 
	$A$ and $B$ particles, which appear simultaneously 
	on the neighboring sites connected by a catalytic bond, instantaneously react, $A+B \to \oslash$, and the reaction product $\oslash$ leaves the system. 
	$A$ and $B$ particles 
	occupying simultaneously the neighboring sites connected by a \textit{noncatalytic} bond, harmlessly coexist. 
	In what follows, we focus on equilibrium properties of the two-species adsorbate, formed on a one-dimensional lattice in the course of the $A+B \to \oslash$ reaction in the presence of such catalytic bonds, 
	considering the case of a random \textit{annealed} and of a random \textit{quenched} disorder in placement
	of the catalytic bonds. The partition function of Model I is written below in Sec. \ref{partitionI}, where 
	we also present \textit{exact} results for its disorder-averaged value (appropriate for the annealed disorder case) and  for the disorder-averaged value of a logarithm of the partition function (appropriate for the quenched disorder case).

	\subsection*{Model II. Catalytic sites}
	\label{modelII}
	
\begin{figure}[htbp]
		\begin{center}
			\includegraphics[width=0.75\hsize]{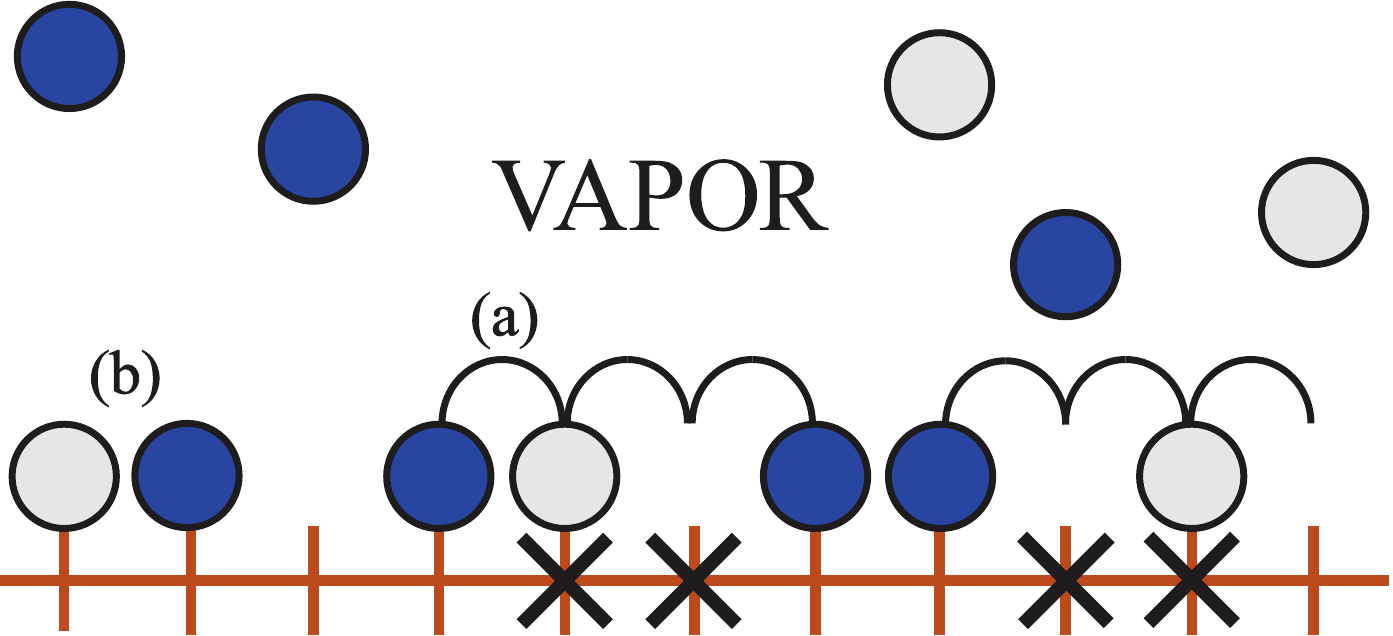}
		\end{center}
		\caption{One-dimensional lattice containing $N$ adsorption sites in contact with vapor phases. Sites with catalytic properties are marked by thick crosses. $A$ and $B$ particles are depicted bt blue and gray circles, respectively. The particles' configuration, which corresponds to an immediate reaction, is realized in case (a), while in case (b) the neighboring $A$ and $B$  particles do not react.}
		\label{figg1}
	\end{figure}

	In Model II, we choose, again completely at random, some fraction of the lattice \textit{sites} and stipulate that these selected sites possess catalytic properties. In this case, which we depict in Fig. \ref{figg1}, $A$ and $B$ particles appearing simultaneously 
	at neighboring lattice sites enter into an irreversible $A+B \to \oslash$ reaction instantaneously,
	if at least one of them resides on a catalytic site. As in  Model I, the reaction product leaves the system. 
	A pair of neighboring $A$ and $B$ particles harmlessly coexist, if they reside on noncatalytic sites. 
	As in Model I, we focus on equilibrium properties of the two-species adsorbate, formed on a one-dimensional lattice with
	a disordered catalytic substrate represented as an array of catalytic sites.
	We again consider the cases of \textit{annealed} and of \textit{quenched} disorder in placement
	of the catalytic sites. The partition function of Model II is presented in Sec. \ref{partitionII} below, as well as its disorder-averaged value and  the disorder-averaged value of its logarithm.

	\section{Partition functions of a two-species adsorbate}
	\label{partition}
	
	\subsection*{Model I}
	\label{partitionI}
	
	To specify positions of the catalytic \textit{bonds}, we introduce a random Boolean variable $\zeta_{i}$, such that it equals $1$ if the
	bond connecting the site $i$ and the adjacent site $i+1$ is catalytic, and equals $0$, otherwise. If the number of catalytic bonds in a chain with $N$ sites is $N_b$, then the fraction $p$ of such bonds is $p = N_b/N$. We assume that $p$ 
	is finite in the thermodynamic limit $N \to \infty$, and thus represents the mean concentration of the catalytic bonds. Random variables $\zeta_{i}$ are uncorrelated for different $i$, and 
	the probability $P(\zeta_i)$ that a given bond is catalytic is 
	\begin{align}
	\label{probI}
	P(\zeta_i) = p \, \delta_{\zeta_i, 1} + (1 - p) \, \delta_{\zeta_i,0} \,,
	\end{align}
	where $\delta_{a,b}$ is the Kronecker $\delta$, such that $\delta_{a,a} = 1$ and zero, otherwise.
	Next, 
	let $n_i$ and $m_i$ be two Boolean occupation variables. We use a convention that $n_i = 1$ ($m_i = 1$) if the site $i$ is occupied by an $A$ (a $B$) particle and is zero otherwise. Then, in thermal equilibrium and for a given realization of an array of random variables $\zeta_{i}$, the grand-canonical partition function of Model I defined on a finite lattice with $N$ adsorbing sites reads
	{\setlength{\belowdisplayskip}{5pt} \setlength{\belowdisplayshortskip}{5pt}
		\setlength{\abovedisplayskip}{5pt} \setlength{\abovedisplayshortskip}{5pt}
		\begin{align}
		\label{partI}
		Z^{(I)}_{N}[\zeta_{i}] &= \sum_{\{n_i,m_i\}}  \prod^{N-1}_i z_A^{n_i} z_B^{m_i} \big(1- n_i m_i \big) \nonumber\\
		& \times \big(1 - \zeta_i n_i m_{i+1}\big) \big(1 - \zeta_i m_i n_{i+1}\big) \,,
		\end{align} }
	where the sum with the subscript $\{n_i,m_i\}$ runs over all possible values of occupation variables. Note that the factor $\big(1- n_i m_i \big)$ in Eq. \eqref{partI} excludes the configurations in which $A$ and $B$ particles reside on the same site.

	\subsubsection*{Model I. Annealed disorder.}
	\label{annealedI}
	
	In the case of annealed disorder in placement of catalytic bonds,  
	the calculations are rather lengthy but very straightforward. Relegating the details to Appendix~\ref{SMIa}, we find that the disorder-averaged value of the grand-canonical partition function 
	{\setlength{\belowdisplayskip}{5pt} \setlength{\belowdisplayshortskip}{5pt}
		\setlength{\abovedisplayskip}{5pt} \setlength{\abovedisplayshortskip}{5pt}
		\begin{align}
		\label{I}
		Z^{(I)}_N = \left\langle Z^{(I)}_N[\zeta_{i}]\right\rangle_{\zeta} 
		\end{align}}
	is given, in the leading in the limit $N \to \infty$ order, by
	\begin{align}
	\label{partannealedI}
	\!Z^{(I)}_N {=} \exp\left({-} N \left[2 \sqrt{r_1} \sin\left(\frac{1}{3} \arcsin\left(\frac{q_1}{r_1^{3/2}}\right)\right) { - }\frac{2 {-} p}{3 p}\right]\right) \,,
	\end{align}
	where the parameters $r_1$ and $q_1$ are functions of the mean concentration $p$ of catalytic bonds, and of the activities $z_A$ and $z_B$. These parameters obey
	{\setlength{\belowdisplayskip}{5pt} \setlength{\belowdisplayshortskip}{5pt}
		\setlength{\abovedisplayskip}{5pt} \setlength{\abovedisplayshortskip}{5pt}
		\begin{align}
		\label{r}
		q_1 & = \frac{2 (2 - p)^3 z_A z_B + 27 p + 9 (2- p) (1 + z_A + z_B)}{54 p^3 z_A z_B} \,, \nonumber\\
		r_1 & = \frac{3 (1 + z_A + z_B) +(2 - p)^2 z_A z_B}{9 p^2 z_A z_B} \,.
		\end{align}}
	Even in this simplest case $Z^{(I)}_N$ is  rather nontrivial.

	\subsubsection*{Model I. Quenched disorder.}
	\label{quenchedI}

	In the case of quenched disorder in spatial distribution of catalytic bonds, we use two complementary approached in order to calculate exactly the disorder-averaged logarithm of the partition function in Eq. \eqref{partI}. In the first approach, we  decompose the substrate into an array of disjoint completely catalytic clusters, as was done in Ref.~\onlinecite{OshB} for a more simple single-species $A+A \to \oslash$ reaction. In this case, a single completely catalytic cluster consists of a sequence of consecutively placed catalytic bonds of a prescribed length, not interrupted by any noncatalytic bond, and having two noncatalytic bonds at its extremities. 
	We use combinatorial arguments to
	calculate the statistical weights of such clusters.
	
	In our second approach, we map the Hamiltonian of Model I onto the Hamiltonian of the Blume-Emery-Griffiths spin-$1$ model \cite{Blum71,Bax82}, and then represent, by introducing an appropriate transfer-matrix $V_{i,i+1}$, the averaged logarithm of the partition function in Eq. \eqref{partI} as
	\begin{align}
	\label{a}
	\left \langle \ln Z^{(I)}_N[\zeta_{i}] \right \rangle_{\zeta} = \left \langle \ln \left({\rm Tr} \prod_{i}^N V_{i,i+1}\right)  \right \rangle_{\zeta} \,,
	\end{align}
	i.e., as the averaged logarithm of the trace 
	of a product of mutually independent, symmetric $3 \times 3$ random matrices 
	\begin{align}
	\label{transI}
	V_{i,i+1} = \begin{pmatrix}
	z_{A}    & \sqrt{z_{A}} & (1-\zeta_{i})\sqrt{z_{A}z_{B}} \\
	\sqrt{z_{A}} & 1            & \sqrt{z_{B}} \\
	(1-\zeta_{i})\sqrt{z_{A}z_{B}} & \sqrt{z_{B}} & z_{B}
	\end{pmatrix}.
	\end{align}
	As demonstrated in Appendix~\ref{SMIb}, the  expression \eqref{a} can be calculated analytically due to the fact
	that for $\zeta_i = 0$ the corresponding transfer matrix has rank $1$ \cite{luck}.
	
	Relegating 
	the details of intermediate calculations to Appendix~\ref{SMIb}, we find that in the leading in the limit $N \to \infty$ 
	order,  the disorder-averaged 
	value of the logarithm of the grand-canonical partition function  is given by
	\begin{align}
	\label{partquenchedI}
	\!\left \langle \ln Z^{(I)}_N[\zeta_{i}] \right \rangle_{\zeta} {=} \frac{1{-}p}{p} \sum_{K=1}^{N} p^{K} \Big((1{-}p)(N{-}K) {+} p {+}1\Big) \ln Z_{K},
	\end{align}
	where $Z_{K}$ is the grand-canonical partition function of a completely catalytic finite chain containing $K$ bonds. It is given explicitly by \cite{popes}
	\begin{align}
	\label{partcompletelycatalytic}
	Z_{K} &= \frac{t_2 t_3 + t_1}{(t_1-t_2)(t_1-t_3)} \, \frac{1}{t_1^K} + \frac{t_1 t_3 + t_2}{(t_2 - t_1)(t_2-t_3)} \, \frac{1}{t_2^K}  \nonumber\\
	& + \frac{t_1 t_2 + t_3}{(t_3-t_1)(t_3-t_2)} \, \frac{1}{t_3^K} 
	\,,
	\end{align}
	where
	\begin{align}
	\label{t-s}
	t_1 &= 2 \sqrt{r_1} \cos\left(\frac{\pi}{6} + \frac{1}{3} \arcsin\left(\frac{q_1}{r_1^{3/2}}\right) \right) - \frac{1}{3} \,, \nonumber\\
	t_2 &= 2 \sqrt{r_1} \sin\left(\frac{1}{3} \arcsin\left(\frac{q_1}{r_1^{3/2}}\right) \right) - \frac{1}{3} \,, \nonumber\\
	t_3 &= - 2 \sqrt{r_1} \cos\left(-\frac{\pi}{6} + \frac{1}{3} \arcsin\left(\frac{q_1}{r_1^{3/2}}\right) \right) - \frac{1}{3} \,,
	\end{align}
	with $r_1$ and $q_1$ defined in Eqs. \ref{r} with $p$ set equal to $1$.
	
	\subsection*{Model II}
	\label{partitionII}
	
	To specify the catalytic properties of lattice \textit{sites} in Model II,  
	we assign to each site a random variable $\eta_{i}$, such that $\eta_{i}=1$ if the $i$-th site is catalytic,  and $\eta_{i}=0$, otherwise. 
	For computational convenience, we add two additional noncatalytic sites at the extremities of the $N$-site chain, 
	i.e., $\eta_{0}=0$ and $\eta_{N + 1} = 0$. We suppose next that the number 
	of such catalytic sites in the $N$-site chain is $N_s$, such that the parameter  $p = N_s/N$  can be thought of as their mean concentration. We assume that this latter property stays finite in the thermodynamic limit $N \to \infty$ meaning that $N_s$ is extensive. Random variables $\eta_i$ are uncorrelated at different sites, and the probability $P(\eta_i)$ that a given site is catalytic is given by
	\begin{align}
	\label{probII}
	P(\eta_i) = p \, \delta_{\eta_i, 1} + (1 - p) \, \delta_{\eta_i,0} \,,
	\end{align}
	where $\delta_{a,b}$ is the Kronecker $\delta$ (see Eq. \eqref{probI}).
	Next, let Boolean variables $n_i$ and $m_i$ denote the occupation variables for $A$ and $B$ particles;  $n_{i} (m_{i}) =1$, 
	if the site $i$ is occupied by an $A$ (a $B$) particle,  $n_{i} (m_{i}) = 0$ if there is no $A$ ($B$) particle at the site $i$.
	In the case when both $n_{i}=0$ and $m_{i}=0$, the site is vacant.
	Then, for a given realization of random variables $\{\eta_i\}$,  
	the grand-canonical partition function $Z^{(II)}_{N}[\eta_i] $ of Model II reads, 
	\begin{align}
	\label{partII}
	Z^{(II)}_{N}[\eta_i] & =  \sum_{\{n_{i},m_{i}\}} \prod_{i}^{N} z_A^{n_i} z_B^{m_i} \Big(\big(1 - n_i m_i\big) \nonumber\\
	& \times \big(1{-}\eta_{i} n_{i}m_{i{-}1}\big) \big(1{-}\eta_{i{+}1} n_{i}m_{i{+}1}\big) \nonumber \\ & \times \big(1{-}\eta_{i} m_{i}n_{i{-}1}\big)\big(1{-}\eta_{i{+}1} m_{i}n_{i{+}1}\big)\Big) \,.
	\end{align}
	As in Model I, the factor $\big(1 - n_i m_i\big) $ ensures
	that configurations when both $n_{i}=1$ and $m_{i}=1$, are excluded.

	\subsubsection*{Model II. Annealed disorder.}
	\label{annealedII}
	
	The disorder-averaged grand-canonical partition function $Z^{(II)}_{N}[\eta_i]$ can be evaluated directly, by deriving appropriate four-site recursion relations obeyed by the grand-canonical partition function. The procedure is described in detail in Appendix~\ref{SMIIa} and gives, for any $N$, 
	\begin{align}
	Z^{(II)}_{N} = \left\langle Z^{(II)}_{N}[\eta_i] \right\rangle_{\eta} = \frac{\gamma_1}{l_1^N} + \frac{\gamma_2}{l_2^N} + \frac{\gamma_3}{l_3^N} + \frac{\gamma_4}{l_4^N} + \frac{\gamma_5}{l_5^N} \,,
	\end{align}
	where $\gamma_j$ are the amplitudes (see Appendix~\ref{SMIIa}), 
	while $l_j$ are the roots of the fifth-order algebraic equation:
	\begin{align}
	\label{roots}
	&\frac{1}{z_A z_B} - \frac{1 + z_A + z_B}{z_A z_B} \,  l + (2 - p)  p \, l^2 + \nonumber\\
	&+  p \left(2 - p + (1 - p)^2 \left(z_A + z_B\right)  \right) \, l^3 -\nonumber\\
	& - (1 - p)^2 p^2  z_A z_B \, l^4 - (1-p)^2 p^2 z_A z_B \, l^5 = 0 \,.
	\end{align}
	This equation cannot be solved explicitly in the general case, and one has to resort to a numerical analysis. On the other hand, the
	asymptotic behavior of the roots can be established analytically in some limiting cases (see Appendix~\ref{SMIIa}). We note, however, that Eq. \eqref{roots} simplifies considerably in the symmetric case $z = z_A = z_B$; here, the fifth-order equation \eqref{roots} factorizes into
	a product of a quadratic and cubic polynomials of $l$ (see Appendix~\ref{SMIIa}). Then, in the leading in the limit $N \to \infty$ order, one has
	{\setlength{\belowdisplayskip}{5pt} \setlength{\belowdisplayshortskip}{5pt}
		\setlength{\abovedisplayskip}{5pt} \setlength{\abovedisplayshortskip}{5pt}
		\begin{align}
		\label{partannealedII}
		Z^{(II)}_{N} &= \exp\Bigg(- N \Bigg[2 \sqrt{r_2} \sin\left(\frac{1}{3} \arcsin\left(\frac{q_2}{r_2^{3/2}}\right)\right) \nonumber\\
		& - \frac{1}{3} \left(1 - \frac{1}{(1 - p) z}\right)\Bigg]\Bigg) \,,
		\end{align}}
	where $r_2$ and $q_2$ are rational 
	functions of the mean concentration $p$ of the catalytic sites and of the activity $z$. Explicitly, these parameters are given by 
	{\setlength{\belowdisplayskip}{5pt} \setlength{\belowdisplayshortskip}{5pt}
		\setlength{\abovedisplayskip}{5pt} \setlength{\abovedisplayshortskip}{5pt}
		\begin{align}
		r_2 &= \frac{3 \big(1+ 2 z\big) - p \big(2 + \big(11 - 5 p - (1 - p)^2 z \big) \, z\big)}{27 p (1 - p)^2 z^2} \,, \nonumber\\
		q_2 &= \frac{1}{54 (1 - p)^3 p z^3} \Big(7 p - 9 + 3 (1 - p) (6 - 7 p) z + \nonumber\\ &+ 3 (1 - p)^2 (6 - 5 p) z^2 +
		+ 2 p (1 - p)^3  z^3\Big) \,.
		\end{align}}
	Asymptotic behavior of $Z^{(II)}_{N}$ is discussed in Appendix~\ref{SMIIa}.
	
	\subsubsection*{Model II. Quenched disorder.}
	\label{quenchedII}
	
	In the quenched disorder case we concentrate on the disorder-averaged logarithm of the grand-canonical partition function. To perform the averaging exactly, we follow two complementary approaches, which are discussed in detail in Appendix~\ref{SMIIb}. In the first approach, we use rather sophisticated combinatorial arguments, decomposing a disjoint array of catalytic sites into effectively completely catalytic clusters and calculating the corresponding statistical weights of such clusters. In this case, a completely catalytic cluster has a more complicated geometry, than in the case of random catalytic bonds, because here the reactive interactions involve effectively three sites (see below). 
	
	In the second approach, we exploit a formal relation between our Model II (similarly as was done for Model I) and the Blume-Emery-Griffiths spin-$1$ model \cite{Blum71,Bax82} with a particular choice of the interaction parameters. This permits us to represent the desired property as
	{\setlength{\belowdisplayskip}{5pt} \setlength{\belowdisplayshortskip}{5pt}
		\setlength{\abovedisplayskip}{5pt}
		\setlength{\abovedisplayshortskip}{5pt}
		\begin{align}
		\left\langle \ln Z^{(II)}_{N}[\eta_i] \right\rangle_{\eta} =  \left\langle \ln {\rm Tr} \prod_{i=1}^{N} V_{i-1,i} V_{i,i+1} \right\rangle_{\eta} \,,
		\end{align}}
	where the transfer matrices $V_{i, j}$  are defined as
	{\setlength{\belowdisplayskip}{5pt} \setlength{\belowdisplayshortskip}{5pt}
		\setlength{\abovedisplayskip}{5pt} \setlength{\abovedisplayshortskip}{5pt}
		\begin{align}
		\label{transII}
		V_{i,j} = \begin{pmatrix}
		z_{A}^{1/2}    & z_{A}^{1/4}  & \varepsilon_{i,j}(z_{A}z_{B})^{1/4} \\
		z_{A}^{1/4} & 1            & z_{B}^{1/4} \\
		\varepsilon_{i,j}(z_{A}z_{B})^{1/4} & z_{B}^{1/4} & z_{A}^{1/2}
		\end{pmatrix} \,,
		\end{align}}
	with $\varepsilon_{i,j} = (1-\eta_{i})(1-\eta_{j})$ and 
	the subscript $i, j$ denoting 
	pairs of the nearest-neighboring sites. In such a representation, a disorder-averaged logarithm of the grand-canonical partition functions
	can be thought of as the Lyapunov index of a product of random $3 \times 3$ matrices, which are
	consecutively correlated; for any $i$ the products $V_{i-1,i} V_{i,i+1}$
	involve the same random variable $\eta_i$, and, hence, they do not decouple (in contrast to Model I).
	
	We show in Appendix~\ref{SMIIb} that the desired thermodynamic 
	property admits the following exact, 
	(in the leading in the limit $N \to \infty$ order), form:
	{\setlength{\belowdisplayskip}{5pt} \setlength{\belowdisplayshortskip}{5pt}
		\setlength{\abovedisplayskip}{5pt} \setlength{\abovedisplayshortskip}{5pt}
		\begin{align}
		\label{partquenchedII}
		\left\langle \ln Z^{(II)}_{N}[\eta_i] \right\rangle_{\eta} = \sum_{K=1}^{N} {\omega}_{K,N}(p) \ln Z_{K} \,,
		\end{align}}
	where $Z_K$ is a grand-canonical partition function of a completely catalytic chain containing $K$ sites, which is defined in Eq. \eqref{partcompletelycatalytic}, while 
	$\omega_{K,N}(p)$ is the statistical weight of a completely catalytic cluster, a $K$-cluster, formed by $K$ catalytic sites appearing in an $N$-site chain (see Appendix~\ref{SMIIb} for more details). Formally, such a  $K$-cluster
	is denoted as a subset of $n$ ($0 < n \leq \lfloor (K-1)/2 \rfloor$, with $\lfloor \ldots \rfloor$ being the floor function), consecutive
	intervals $l_{r+1}, l_{r+2}, l_{r+3},..., l_{r+n}$ from an entire set $\{l_n\}$ of the intersite intervals, where all the intervals $l_{r+i}$, $i=1, \ldots, n$, 
	are greater than unity, obey the "conservation" law of the form $\sum_{i=1}^n l_{r+i} = K - 1$
	and are  bounded by two intervals
	$l_r$ and $l_{r+n+1}$ of unit length.
	For all $N$ and $K$ except for $K=1$ and $K=N$, $\omega_{K,N}(p)$ is given by
	\begin{align}
	\label{weightkn}
	&\omega_{K,N}(p) = p^{(K-1)/2} (1-p)^{(K+3)/2}  \Bigg(2F_{K}\left(\sqrt{\frac{p}{1-p}}\right) \nonumber\\
	&+ (1-p)(N-K-1)F_{K-2}\left(\sqrt{\frac{p}{1-p}}\right) \Bigg),
	\end{align}
	while for $K=N$ and $K = 1$  it obeys
	\begin{align}
	\label{weightnn}
	&\omega_{N,N}(p) = p^{N/2} (1-p)^{N/2} \Bigg(\sqrt{\frac{p}{1-p}}F_{N}\left(\sqrt{\frac{p}{1-p}}\right) + \nonumber\\&+2F_{N-1}\left(\sqrt{\frac{p}{1-p}}\right) + \sqrt{\frac{p}{1-p}}F_{N-2}\left(\sqrt{\frac{p}{1-p}}\right)\Bigg) \,, \nonumber \\
	&\omega_{1,N}(p)  = N (1 - p)^3 + 2 p (1 - p)^2 \,, 
	\end{align}
	respectively, 
	where $F_{n}(x)$ are the Fibonacci polynomials 
	\begin{align}
	F_{n}(x) = \sum_{l=0}^{\lfloor (n-1)/2 \rfloor} \begin{pmatrix} n-l-1 \\ l \end{pmatrix} x^{n-2l-1}\,.
	\end{align}
	The expression \eqref{partquenchedII} attains the following explicit, albeit complicated,
	form in the symmetric case $z {=} z_A {=} z_B$:
	{\setlength{\belowdisplayskip}{5pt} \setlength{\belowdisplayshortskip}{5pt}
		\setlength{\abovedisplayskip}{5pt} \setlength{\abovedisplayshortskip}{5pt}
		\begin{align}
		\label{Psitesquen0}
		&\frac{1}{N}  \left\langle \ln Z^{(II)}_{N}[\eta_i] \right\rangle_{\eta} = (1-p)^{3}\ln{(1 + 2z)} + \nonumber\\
		&+  p (1 - p)^{2} \ln{\left(\frac{1 + 3 z + \sqrt{1 + z(6 + z)}}{2\sqrt{1 + z(6 + z)}}\right)} - \nonumber \\
		&- p (p^{2} - 3 p + 3) \ln{\left(\frac{\sqrt{1 + z(6 + z)} - (1 + z)}{2z}\right)} - \nonumber\\
		&-\frac{p (1-p)^4}{\sqrt{p (4-3p)}}  \sum_{m=0}^{N} \left(\frac{1}{X_{+}^m} - \frac{1}{X_{-}^m} \right) \nonumber\\
		&\times \ln{\left(1-  \frac{1 + 3 z - \sqrt{1 + z(6 + z)}}{1 + 3 z + \sqrt{1 + z(6 + z)}}\left( - \frac{t_{2}}{t_{1}} \right)^{m+3}\right)} \,,
		\end{align}}
	where
	\begin{align}
	X_{\pm} = - \frac{1}{2 (1-p)} \left(1 \mp \sqrt{\frac{(4- 3p)}{p}}\right) \,,
	\end{align}
	and $t_{1}$ and $t_2$ are defined in Eqs. \eqref{t-s}.
	
	Expressions \eqref{partannealedI}, \eqref{partquenchedI}, \eqref{partannealedII}, and \eqref{partquenchedII} [as well as Eq. \eqref{Psitesquen0}] constitute our main exact analytical results. They will serve us as the basis for the analysis of characteristic thermodynamic properties of the two-species adsorbates.

	\section{Disorder-averaged pressure, densities and the compressibilities of a two-species adsorbate}
	\label{pressure}

	For Model I and Model II, the disorder-averaged pressure in the case of {\sl{annealed}} disorder is given by 
	\begin{align}
	\label{Pannealed}
	P^{(ann)}_k&\equiv P^{(ann)}_k  (T, z_A, z_B) = \nonumber\\
	&=\frac{1}{\beta} \lim_{N \rightarrow \infty} \frac{1}{N}\ln \langle Z_{N}^{(k)} [\alpha_i]\rangle_{\alpha} \,,
	\end{align}
	where  the subscripts and superscripts $k=\{I,II\}$ as well as  arguments
	$\alpha_i=\{\zeta_i,\eta_i\}$  correspond to Model I or II.

	In the \textit{quenched} disorder case  the disorder-averaged pressure formally obeys
	\begin{align}
	\label{Pquenched}
	P^{(quen)}_k &\equiv P^{(quen)}_k  (T, z_A, z_B) = \nonumber\\
	&=\frac{1}{\beta} \lim_{N \rightarrow \infty} \frac{1}{N}\langle\ln Z_{N}^{(k)}[\alpha_i]\rangle_{\alpha} .
	\end{align}
	As for the mean particles' densities and the compressibilities of the $A$ and $B$ phases 
	in a two-species adsorbate, we note that our results indicate that the pressure is a symmetric function of $z_A$ by $z_B$ (see the Appendixes). Hence, it suffice to consider the thermodynamic properties 
	of one of the species only. In what follows, we focus on the $A$ phase.  For the latter,
	the density $n^{(I)}_{A}$ (or $n^{(II)}_{A}$) of the $A$ phase in a two-species adsorbate is defined by
	\begin{align}
	\label{density}
	n^{(k)}_{A} = \frac{\partial P_{k}}{\partial \mu_{A}} \,,
	\end{align}
	where $\mu_{A}$ is the chemical potential corresponding to the activity $z_{A}$.
	Here, in order to determine 
	the mean density in the annealed disorder case, one has to use the expressions \eqref{partannealedI} and \eqref{partannealedII} for the grand-canonical partition function, while in the quenched disorder case the disorder-averaged 
	pressure obtains from Eqs. \eqref{partquenchedI} and \eqref{partquenchedII}.
	In turn, the compressibility of the $A$ phase obeys
	\begin{align}
	\label{compress}
	\varkappa^{(k)}_{A} = \frac{1}{\left(n^{(k)}_{A}\right)^{2}}\frac{\partial n_{A}}{\partial \mu_{A}}\,.
	\end{align}
	Below we discuss the behavior of the disorder-averaged mean densities and of the compressibilities of the $A$ phase
	in the two-species adsorbate. 
	To ease the readability, we plot these characteristic properties as functions of
	system's parameters and emphasize some essential features,  
	avoiding complicated analytical formulas. The latter are often too cumbersome, and are listed in full in the Appendixes.
	
	\begin{figure*}
		\includegraphics[width=0.4\textwidth,height=0.3\textwidth]{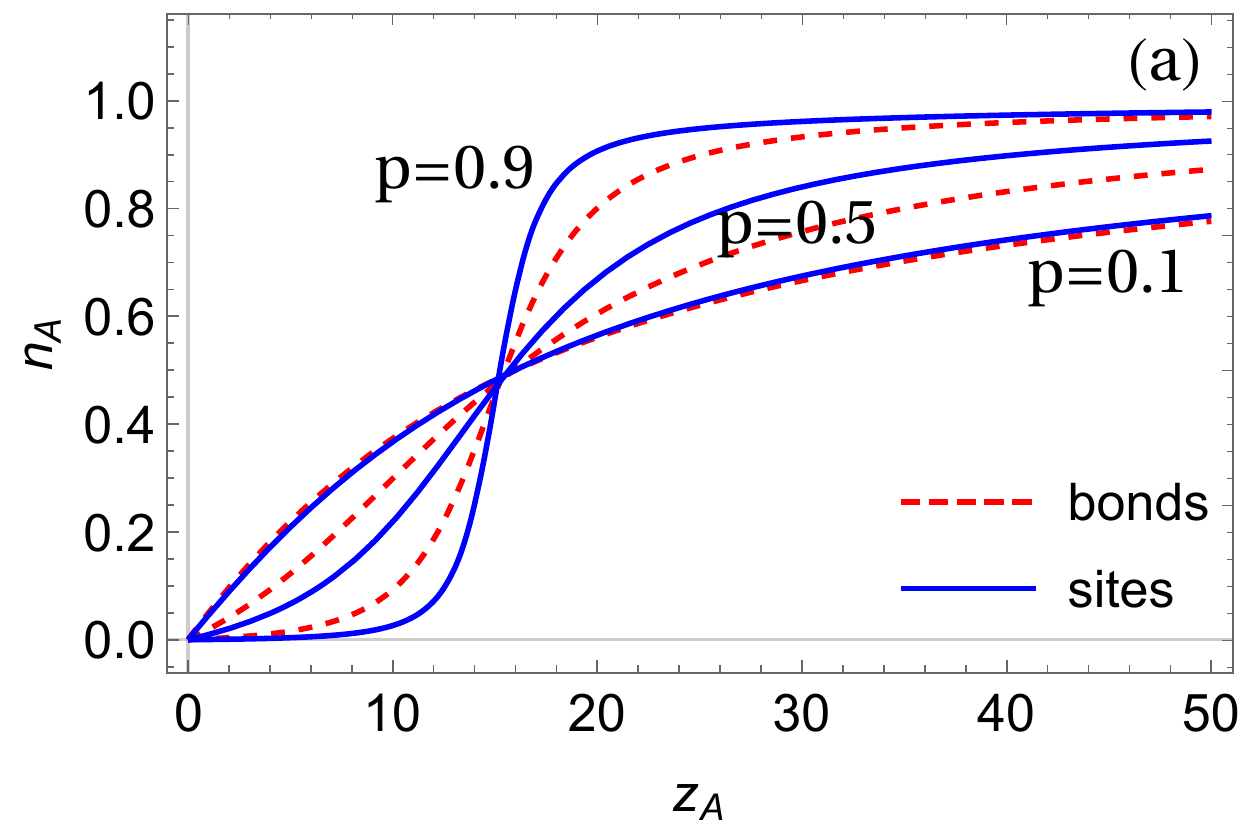}
		\includegraphics[width=0.4\textwidth,height=0.3\textwidth]{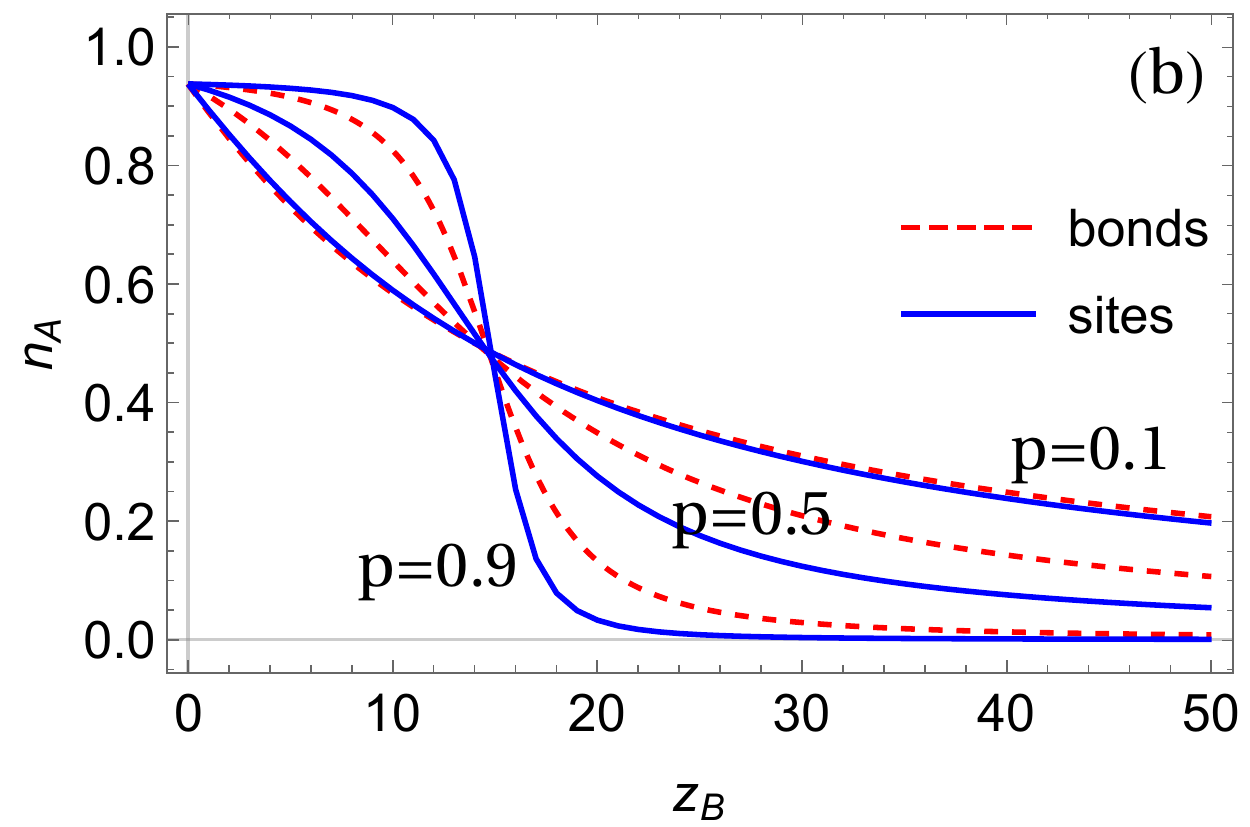}
		\includegraphics[width=0.4\textwidth,height=0.3\textwidth]{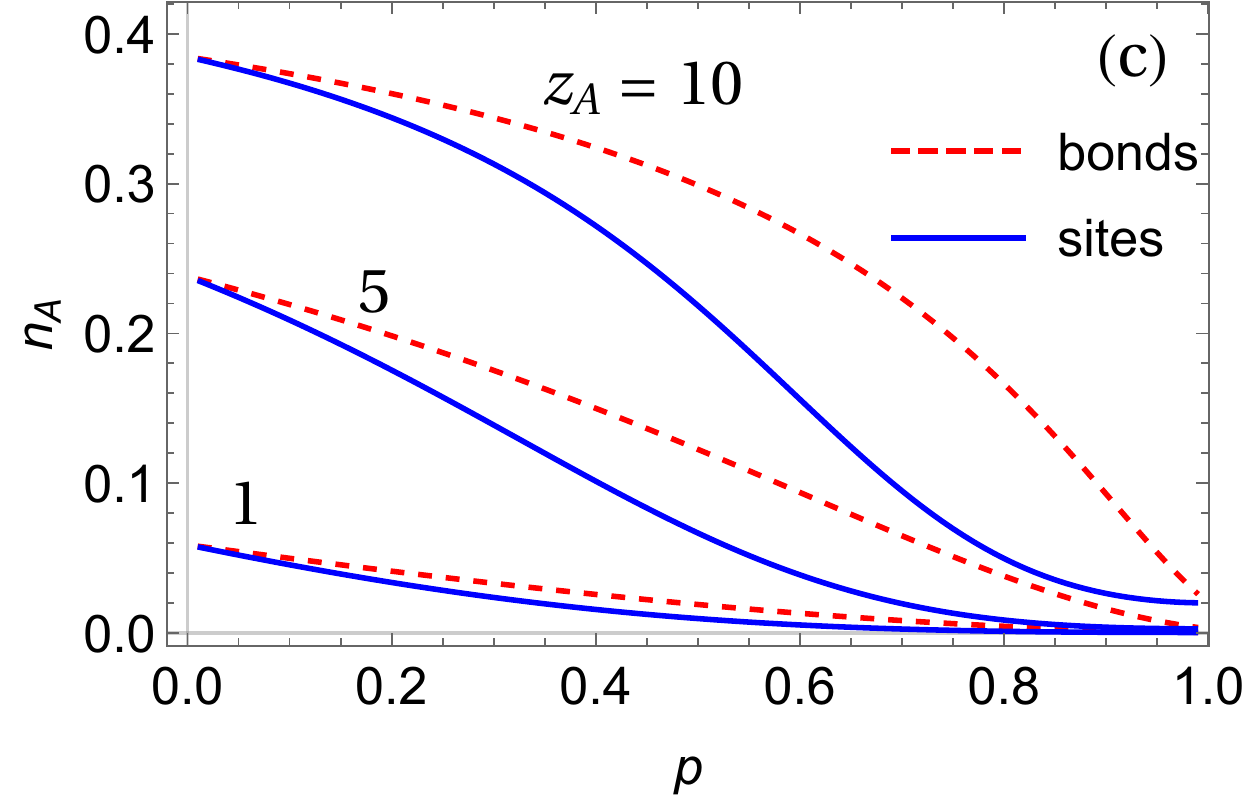}
		\includegraphics[width=0.4\textwidth,height=0.3\textwidth]{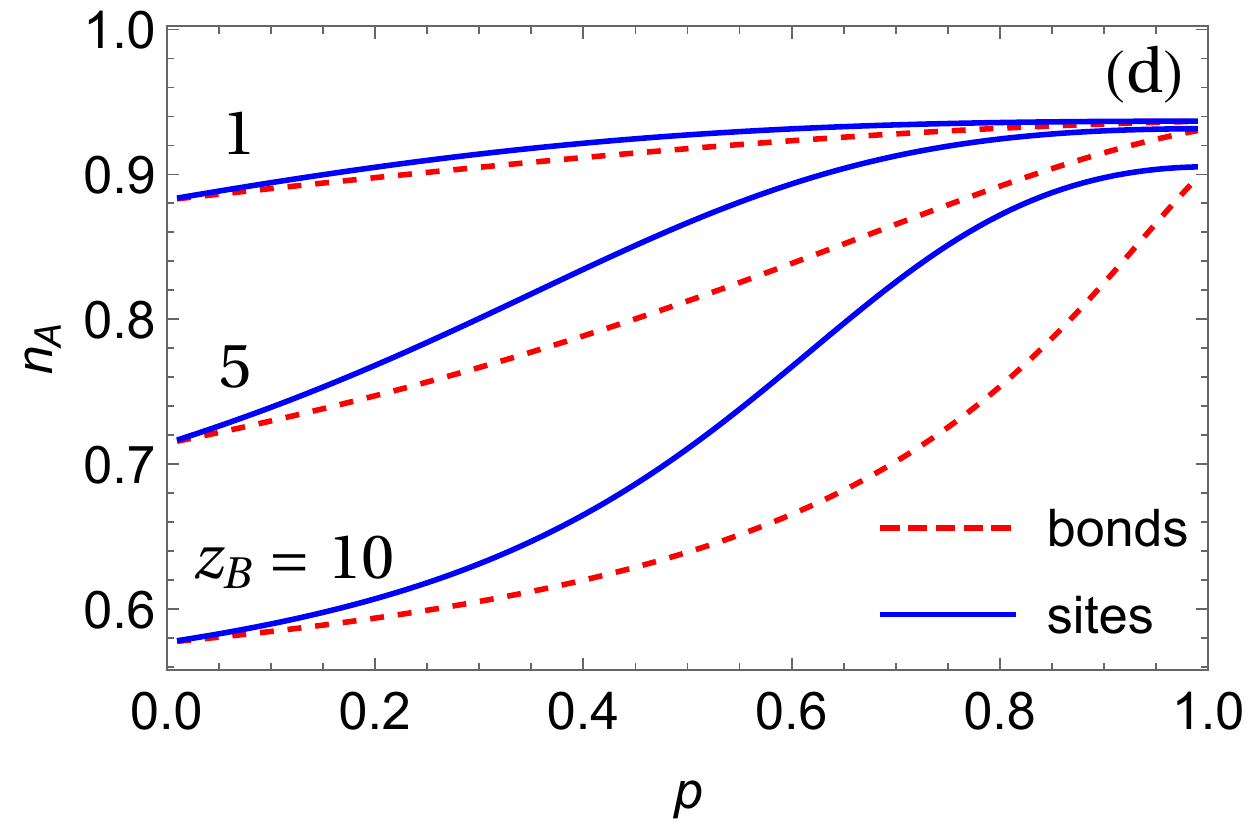}
		\includegraphics[width=0.4\textwidth,height=0.3\textwidth]{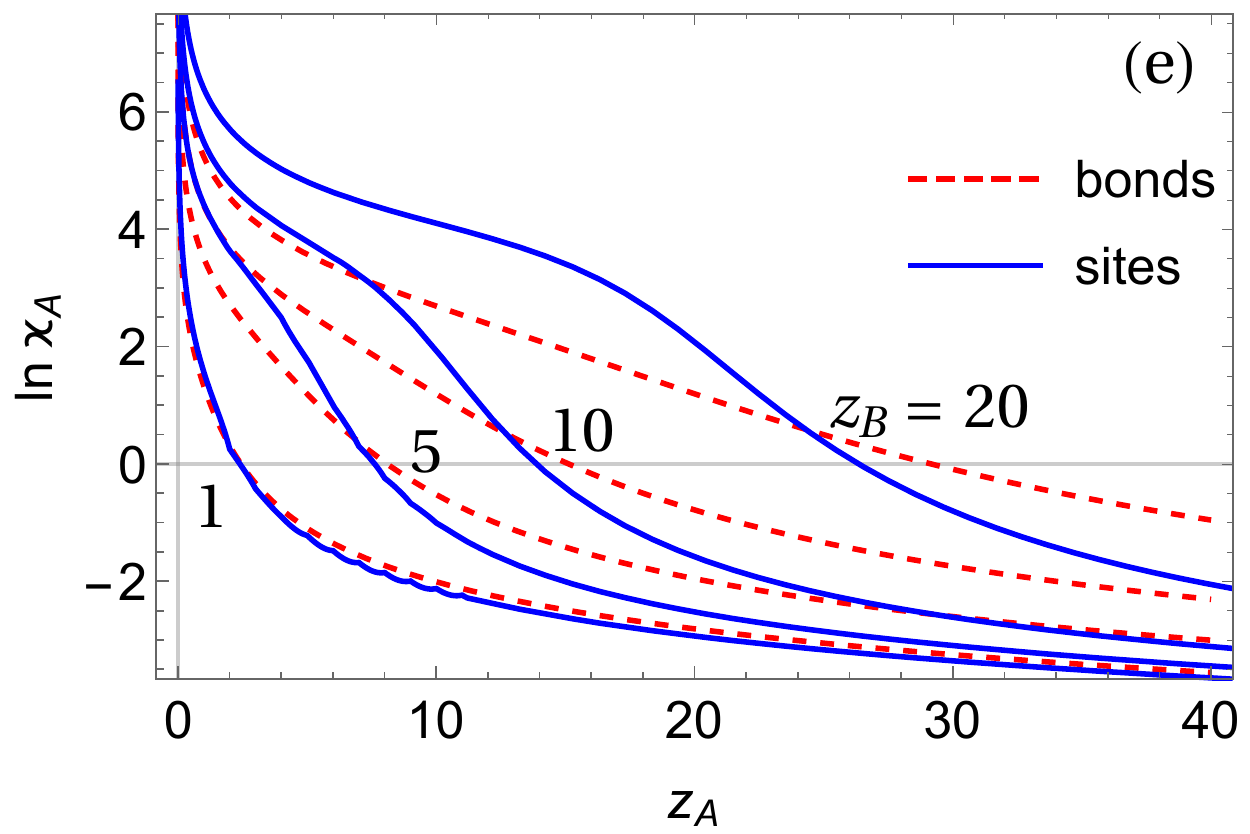}
		\includegraphics[width=0.4\textwidth,height=0.3\textwidth]{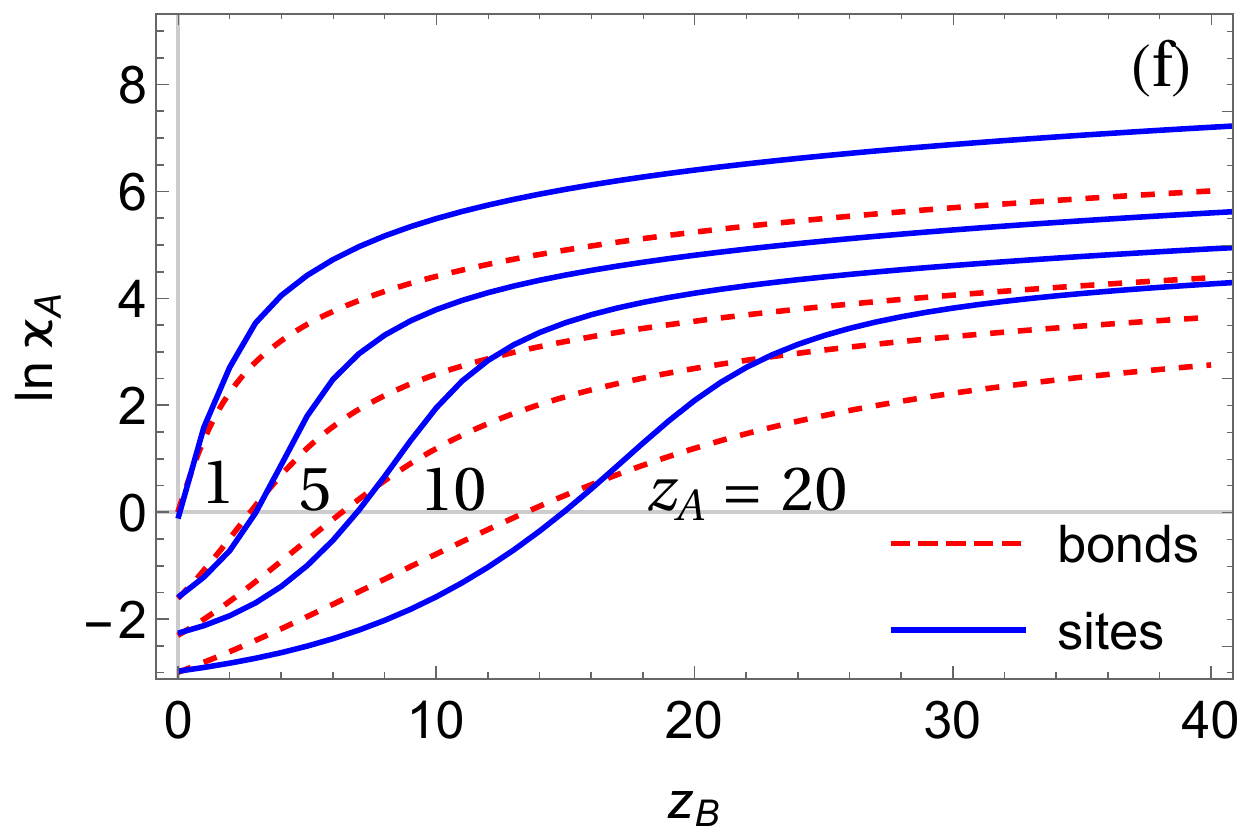}
		\caption{{\em Annealed disorder case}.
			(a), (b) Disorder-averaged density $n_A$ as a function of activity $z_A$ for fixed $z_B = 15$ (a) 
			and as a function of activity $z_B$ for fixed $z_A = 15$ (b), for three values of the mean concentration $p = 0.1, 0.5, 0.9$ of the catalytic bonds (Model I, red dashed curves) and catalytic sites (Model II, blue solid curves).
			(c), (d) Disorder-averaged density $n_A$ as a function of the mean concentration $p$ of catalytic bonds or catalytic sites [the same color code as in panels (a) and (b)] for three values of $z_A$ and $z_B = 15$ (c)
			and for three values of $z_B$ and $z_A = 15$ (d).
			(e), (f) Logarithm of the compressibility $\varkappa_{A}$ for mean concentration $p=0.7$ 
			of the catalytic bonds or sites as a function of the activity $z_A$ (e) or $z_B$ (f). From left to right, the curves correspond to $z_B =1,5,10, 20$ (e), and to $z_A =1,5,10, 20$ (f).
			\label{fig3}
		}
	\end{figure*}

	In Fig. \ref{fig3} we depict the disorder-averaged density $n_A$ and 
	the compressibility $\varkappa_{A}$ of the $A$ phase in the case of \textit{annealed} disorder in placement of the catalytic bonds or sites. In Fig. \ref{fig3} (a) the disorder-averaged density is plotted as a function of the activity $z_A$, at fixed $z_B = 15$, for three values of the mean concentration $p$ of the catalytic bonds (red dashed curves) or catalytic sites (blue solid curves). 
	We observe that $n_A$ is a monotonically increasing function of $z_A$, as it should, 
	being equal to zero at $z_A = 0$ and approaching $1$ as $z_A \to \infty$, 
	which means  that the second phase is squeezed out completely in this limit. 
	In the case of catalytic bonds, the exact large-$z_A$ asymptotic behavior of $n_A$ is rather simple,  
	\begin{align}
	n_A = 1 - \frac{1 + (1 - p)^2 z_B}{z_A}  + O\left(\frac{1}{z_A}\right)\,,
	\end{align}  
	while in the case of catalytic sites $n_A$ has a much more complicated form; in fact, the blue solid curves in Fig. \ref{fig3} are the numerical plots of cumbersome analytical expressions, which we do not manage to simplify into compact forms even in the asymptotic limits.  We see next that at a lowest concentration $p$ (here, $p = 0.1$) the mean density 
	is a rather smooth function, which form resembles the density dependence of binary Langmuir adsorbates of hard-core particles.  
	Here, only a very minor difference between the cases of catalytic bonds or catalytic sites is seen. 
	This difference becomes apparent for an intermediate concentration of catalytic 
	bonds or sites, i.e., for $p = 0.5$, when $n_A$,  as a function of $z_A$, starts to acquire a characteristic $S$-shape form. For largest $p$, (here, $p= 0.9$), this difference is also quite pronounced. Overall, it implies that  
	the precise modeling of a catalyst -- either in the form of catalytic bonds or in the form of catalytic sites -- is physically a relevant issue. We also remark that the larger $p$ is, the more abrupt is the variation of $n_A$ with $z_A$.  We observe that for $p=0.9$, upon an increase of $z_A$, the mean density $n_A$ does not exhibit any significant change in its value up to a certain threshold $z^*_A$, when it starts to increase steeply, within a narrow interval of values of $z_A$, up to almost $1$ and then again does not exhibit any significant change in its value.  This abrupt change in the behavior is more pronounced, for the same value of $p$, in the  case of catalytic sites than in the case of catalytic bonds. 
	Surprisingly enough, curves for cases of both catalytic bonds and catalytic sites, for different values of $p$, cross each other nearly at the same point in a vicinity of  $z_A\approx z_B$ for the present scale of the picture.
	
	Further on, in  Fig. \ref{fig3} (b) we plot $n_A$ as a function of the activity of the other component, for a fixed value of its own activity, $z_A = 15$. We observe here an inverse scenario showing now how the $A$ component gets squeezed out by the other component when the activity of the latter increases. For smallest concentration of catalytic bonds or sites, $n_A$ decreases very smoothly, and no apparent difference between two models is observed. This difference is much more noticeable for higher values of $p$, as well as the abrupt variation of $n_A$ with $z_B$. In particular, for $p = 0.9$ we again observe that $n_A$ stays almost constant (close to $1$) upon a gradual increase of $z_B$ up to a certain threshold value  $z^*_B$, and then, when the activity $z_B$ overpasses this value, $n_A$ abruptly drops down to almost zero value meaning that the $A$ phase fades out almost completely for finite $z_B$.  
	
	In Figs. \ref{fig3} (c) and \ref{fig3} (d), we present the dependence of the disorder-averaged density $n_A$ on the concentration of the catalytic bonds or catalytic sites, for several values of the activity. In Fig. \ref{fig3} (c) we fix $z_B = 15$  and plot $n_A$ as a function of $p$ for $z_A = 1, 5$ and $10$. In Fig. \ref{fig3} (d), conversely, we fix $z_A = 15$  and plot $n_A$ as a function of $p$ for $z_B = 1, 5$ and $10$. We observe that $n_A$ is a monotonically \textit{decreasing} function of $p$ at fixed $z_B$, and is a monotonically \textit{increasing} function of $p$ at a fixed $z_A$. 
	Further on, we realize that the behavior of $n_A$ in the case of catalytic sites becomes markedly different from the one in case of catalytic bonds at intermediate concentrations, and is more pronounced the larger is the value of the activity,  regardless if it concerns $z_A$ or $z_B$. 
	
	\begin{figure*}
		\includegraphics[width=0.4\textwidth,height=0.3\textwidth]{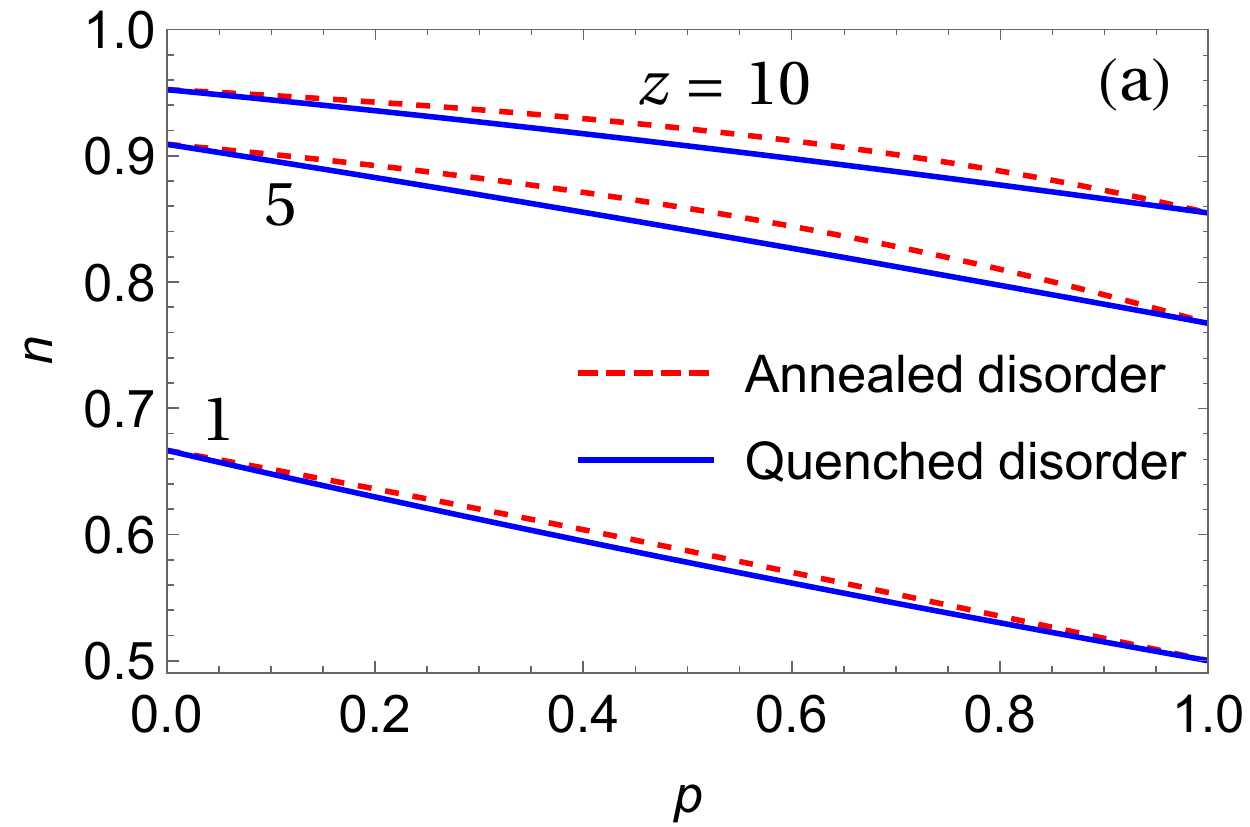}
		\includegraphics[width=0.4\textwidth,height=0.3\textwidth]{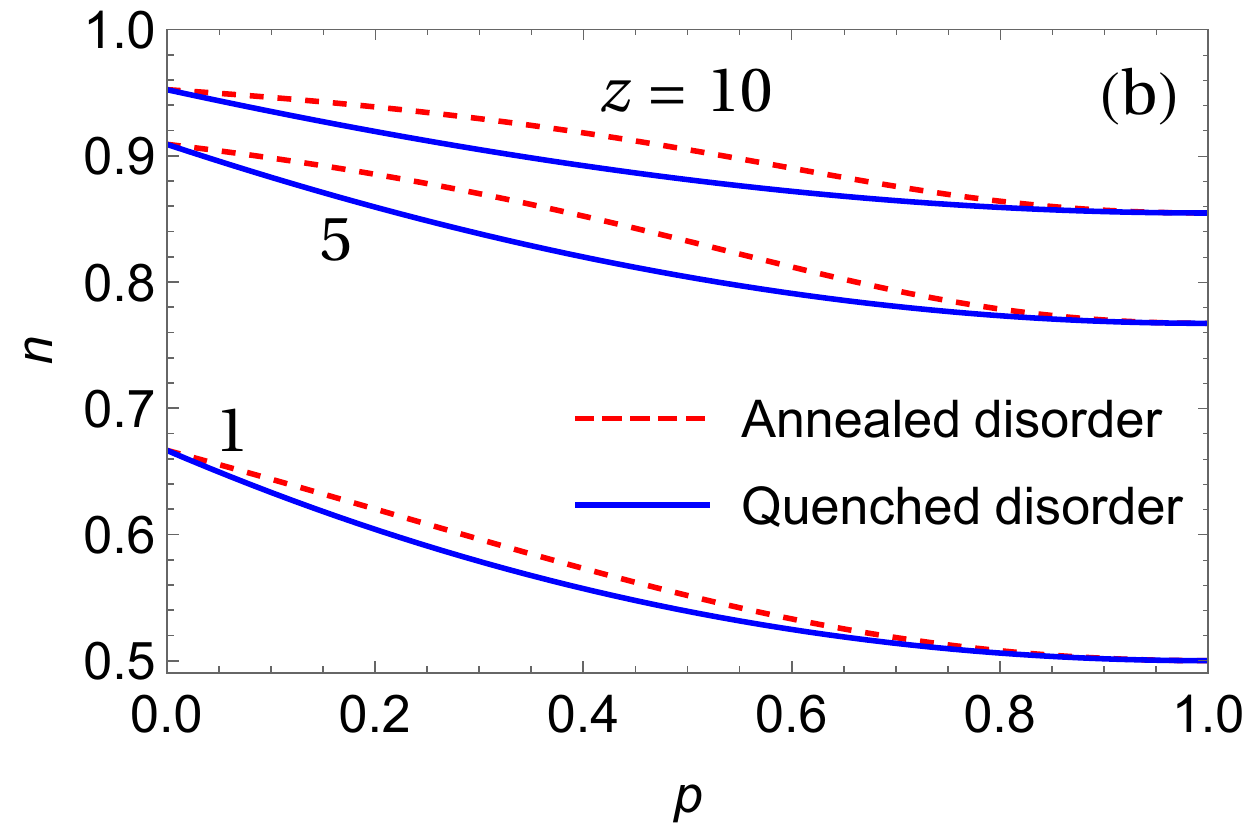}
		\caption{{\em Annealed versus quenched disorder}.
			(a) Disorder-averaged density $n$  for Model I 
			as a function of the mean concentration $p$ of the catalytic bonds
			for three values of activity $z = z_A = z_B$.  
			(b) Disorder-averaged density $n$  for Model II 
			as a function of the mean concentration $p$ of the catalytic sites
			for three values of the activity $z = z_A = z_B$.
			\label{fig4}
		}
	\end{figure*}
	
	In Figs. \ref{fig3} (e) and \ref{fig3} (f), we plot  a logarithm of the compressibility of the $A$ phase as a function of the activity $z_A$ for several values of $z_B$ [Fig. \ref{fig3} (e)] and as a function of the activity $z_B$ for several stray values of $z_A$ [Fig. \ref{fig3} (f)]. We find that, in general, $\varkappa_{A}$  is a monotonically decreasing function of $z_A$ and a monotonically increasing function of $z_B$. The difference between two models is small for low activities and becomes progressively more apparent for larger $z$. Interestingly enough, in the case of catalytic sites 
	$\varkappa_{A}$  as a function of $z_A$ exhibits a shoulder, which is absent in the case of catalytic bonds.
	
	We finally realize that in the case of quenched disorder in placement of catalytic bonds or catalytic 
	sites the behavior is visually 
	very similar to the annealed disorder case (see Fig. \ref{fig3}), which renders a comparison
	between these two cases of disorder
	rather awkward.
	We 
	thus relegate a corresponding figure to Appendixes ~\ref{SMIc} and ~\ref{SMIIc}.
	Instead, here we compare separately in Fig. \ref{fig4} the behavior in the annealed and quenched disorder cases for Model I [Fig. \ref{fig4} (a)] and for Model II [Fig. \ref{fig4} (b)], for simplicity considering only the symmetric case of equal activities $z_A = z_B = z$. As a consequence, in this symmetric case 
	the disorder-averaged densities $n_A$ and $n_B$  are equal to each other, such that we drop the subscript $A$. Moreover, considering the $P_I$ (or $P_{II}$) as a function of $z$ and performing the derivative in respect to $\mu=\ln z/\beta$ we immediately get the full density of both species. Therefore this full density is given in Fig. \ref{fig4}. We conclude that 
	while the behavior in the annealed disorder case appears to be very different if we consider a catalyst as an array of catalytic bonds, or as an array of catalytic sites, 
	we do not see much difference between the cases of annealed and quenched disorder for each model. This is rather counter-intuitive because the latter case is more involved from a mathematical point of view and the resulting expressions are much more cumbersome.

	\section{Conclusions}\label{conc}
	
	To recapitulate, we studied thermodynamic 
	equilibrium properties of two-species adsorbates formed in the course of two-species $A+B \to \oslash$ reactions, taking place on a one-dimensional lattice with randomly placed catalytic elements. We considered two types of such catalytic elements: namely,  the model with randomly placed catalytic bonds (Model I), which prompt an instantaneous 
	reaction between dissimilar species appearing on neighboring sites connected by such a bond, and the model with randomly placed catalytic sites (Model II); in this case the reaction between dissimilar species occurs instantaneously as soon as at least one of them resides on a catalytic site. As well, two types of disorder were considered: the case when disorder can be viewed as {\sl{annealed}}, and a more complicated case with {\sl{quenched}}, i.e., frozen disorder in spatial distribution of catalytic elements. 
	
	For both types of catalytic elements and for both types of disorder, we found exact solutions.
	For Model I and Model II with annealed disorder, we obtained exact results for the disorder-averaged grand-canonical partition function, 
	and hence, for the pressure of the adsorbate and its thermodynamic derivatives. We also discussed in detail
	asymptotic behavior of the disorder-averaged particle density for small and large values of activities $z_{A}$ and $z_{B}$, as well as its dependence on the concentration of the catalytic bonds or catalytic sites
	(see the Appendixes). 
	In the case of quenched disorder the problem of averaging a logarithm of the grand-canonical
	partition function was solved by two complementary approaches. In the first approach, we reduced the problem 
	to a combinatorial enumeration of all possible fully connected (completely catalytic) 
	clusters with fixed positions of catalytic bonds or sites, and finding 
	exact expressions for the statistical weights of such clusters. In the second approach, we reformulated the models under study in terms of the general spin-$1$ model \cite{Blum71}, which permitted us to represent the disorder-averaged pressure as an averaged logarithm of the trace of an infinite product of random three-by-three matrices -- mutually uncorellated for Model I and having sequential, pairwise correlations in the case of Model II. In such a representation, exact solutions were also found, providing  nontrivial examples of infinite products of random matrices for which the Lyapunov exponent can be calculated in an explicit form. 
	
	
	\section*{Acknowledgents}

	 We wish to thank J.-M. Luck for valuable comments and interest in this work. M.D. acknowledges partial support  from the National Academy of Sciences of Ukraine through Project K$\Pi$KBK 6541230, as well as from the Polish National
	Agency for Academic Exchange (NAWA) through Grant No.
	PPN/ULM/2019/1/00160.

\newpage
	

	\onecolumngrid
	\appendix
	
	\section{Model I}\label{SMI}
	
	\subsection{Annealed disorder}\label{SMIa}
	
	In this subsection we present the derivation of Eq. (\ref{partannealedI}).
	
	We first write the disorder-averaged
	grand-canonical partition function $\langle Z^{(I)}_{N}[\zeta_{i}]\rangle_{\zeta}$  in the form
	\begin{align}
	\label{A1}
	\langle Z^{(I)}_{N}[\zeta_{i}]\rangle_{\zeta} &= \sum_{\{n_{i},m_{i}\}} \exp\left(\beta \mu_{A}\sum_{i}n_{i}\right) \exp\left(\beta \mu_{B}\sum_{i}m_{i}\right) \nonumber\\&\times \prod_{i}^{N-1} \Bigg\langle (1 - n_{i} m_{i}) (1-\zeta_{i} n_{i}m_{i+1})(1-\zeta_{i} m_{i}n_{i+1})\Bigg\rangle_{\zeta_{i}} \,,
	\end{align}
	where the angle brackets with the subscript ${\zeta_{i}}$ denote averaging with respect to the ensemble of  $\zeta_{i}$.
	Since
	$\zeta_{i}$ are independent random variables, 
	and $m_i$  and $n_i$ are Boolean, i.e., they assume only values $0$ and $1$, the averaging in expression \eqref{A1} can be carried out directly to give 
	\begin{align}
	Z^{(I)}_{N} =\langle Z^{(I)}_{N}[\zeta_{i}]\rangle_{\zeta} &= \sum_{\{n_{i},m_{i}\}} \exp\left(\beta \mu_{A}\sum_{i}n_{i}\right) \exp\left(\beta \mu_{B}\sum_{i}m_{i}\right) \nonumber\\
	& \times \prod_{i}^{N-1}(1 - n_{i} m_{i}) \Big( 1-p(m_{i} n_{i+1} + n_{i} m_{i+1})\Big).
	\end{align}
	The next step consists in the derivation of appropriate  recursion relations obeyed by the grand-canonical partition function. Here we follow closely the line of thought proposed in Ref. \cite{popes}. Let us define two auxiliary partition functions, $Z_{N}^{(A)}$ and $Z_{N}^{(B)}$, which differ from the grand-canonical partition function in that they obey some additional constraints. The function $Z_{N}^{(A)}$ is
	constrained by the condition that the site $i=N$ is occupied by an $A$ particle (i.e., $n_{N} = 1$, and $m_{N} = 0$), 
	while  $Z_{N}^{(B)}$ - by the condition that this site is occupied by 
	a $B$ particle (i.e., $m_{N} = 1$, and $n_{N} = 0$). One evidently has
	\begin{eqnarray}
	Z_{N}^{(A)} = Z_{N}\Big{|}_{m_{N}=0}^{n_{N}=1} = z_{A} \sum_{\{n_{i},m_{i}\}} z_{A}^{\sum_{i=1}^{N-1} n_{i}} z_{B}^{\sum_{i=1}^{N-1} m_{i}} \prod_{i}^{N-2}\left[ 1-p \ (m_{i} n_{i+1} + n_{i} m_{i+1})\right] \left( 1-p \ m_{N-1}\right),
	\end{eqnarray}
	\begin{eqnarray}
	Z_{N}^{(B)} = Z_{N}\Big{|}_{n_{N}=0}^{m_{N}=1} = z_{B} \sum_{\{n_{i},m_{i}\}} z_{A}^{\sum_{i=1}^{N-1} n_{i}} z_{B}^{\sum_{i=1}^{N-1} m_{i}} \prod_{i}^{N-2}\left[ 1-p \ (m_{i} n_{i+1} + n_{i} m_{i+1})\right] \left( 1-p \ n_{N-1}\right).
	\end{eqnarray}
	Then, we have that for $N\geq2$, 
	\begin{equation}
	\label{ZN}
	Z^{(I)}_{N} = Z^{(I)}_{N-1} + Z_{N}^{(A)} + Z_{N}^{(B)}.
	\end{equation}
	Further on, inspecting possible values of the variables $n_{N-1}$ and $m_{N-1}$, we find that for $N \geq 3$ the functions $Z_{N}^{(A)}$ and $Z_{N}^{(B)} $ can be expressed in terms of $ Z_{N-2}$, $Z_{N-1}^{(A)}$ and $Z_{N-1}^{(B)}$  as
	\begin{eqnarray}
	\label{ZNA}
	Z_{N}^{(A)} &=& z_{A}Z_{N-2} + z_{A}(1-p)Z_{N-1}^{(B)} + z_{A}Z_{N-1}^{(A)}.
	\end{eqnarray}
	An analogous expression for $Z_{N}^{(B)} $ is obtained from (\ref{ZNA}) by merely interchanging subscripts and superscripts 'A' $\leftrightarrows$ 'B', which gives
	\begin{eqnarray}
	\label{ZNB}
	Z_{N}^{(B)} &=& z_{B}Z_{N-2} + z_{B}(1-p)Z_{N-1}^{(A)} + z_{B}Z_{N-1}^{(B)}.
	\end{eqnarray}
	Equations (\ref{ZN}), (\ref{ZNA}) and (\ref{ZNB})  satisfy the following initial conditions:
	\begin{eqnarray}
	Z_{1} = 1 + z_{A} + z_{B}, \nonumber
	\end{eqnarray}
	\begin{eqnarray}
	Z_{1}^{(A)} = z_{A}, \qquad Z_{2}^{(A)} = z_{A}\left( 1 + z_{A} + z_{B}(1-p) \right), \nonumber
	\end{eqnarray}
	\begin{eqnarray}
	\label{InCon}
	Z_{1}^{(B)} = z_{B}, \qquad Z_{2}^{(B)} = z_{B}\left( 1 + z_{B} + z_{A}(1-p) \right).
	\end{eqnarray}
	Solution of the recursion in Eqs. (\ref{ZN}), (\ref{ZNA}) and (\ref{ZNB})  with the initial conditions given by (\ref{InCon}) can be found by using the standard generating function technique (see, e. g., Ref. \cite{popes}). One finds then that the generating function  $\mathcal{Z}_{t} = \sum_{N=1}^{\infty}Z_{N} t^{N}$ obeys
	\begin{eqnarray}
	\label{Zt}
	\mathcal{Z}_{t} = \frac{t \mathcal{L}_{1}(t)}{\mathcal{L}_{2}(t)},
	\end{eqnarray}
	where
	\begin{eqnarray}
	\label{L1}
	\mathcal{L}_{1}(t) = \frac{1 + z_{A} + z_{B}}{z_{A} z_{B}} - 2t + 2 (1-p) t - t^{2} + p(1-p)t^{2}, \nonumber
	\end{eqnarray}
	\begin{eqnarray}
	\label{L2}
	\mathcal{L}_{2}(t) = \frac{1}{z_{A}z_{B}} - \frac{1 + z_{A} + z_{B}}{z_{A} z_{B}}t + t^{2} - (1-p)^{2}t^{2} + t^{3} - (1-p^{2})t^{3}.
	\end{eqnarray}
	Denoting next the roots of the cubic polynomial $\mathcal{L}_{2}(t)$
	as
	$t_{1}$, $t_{2}$ and $t_{3}$, such that  $\mathcal{L}_{2}(t)=(t-t_{1})(t-t_{2})(t-t_{3})$, 
	we 
	express Eq. (\ref{Zt}) in terms of elementary fractions and expanding each factor into the Taylor series in powers of $t/t_{j}$, $j=1,2,3$. In doing so, we find that Eq. (\ref{Zt}) can be formally rewritten as
	\begin{eqnarray}
	\label{Ztt}
	\mathcal{Z}_{t} = \sum_{N=1}^{\infty} \left[ \alpha_{1}\left(\frac{t}{t_{1}}\right)^{N} + \alpha_{2}\left(\frac{t}{t_{2}}\right)^{N} + \alpha_{3}\left(\frac{t}{t_{3}}\right)^{N} \right],
	\end{eqnarray}
	where
	\begin{eqnarray}
	\label{alphas}
	\alpha_{1} = \frac{t_{2} t_{3} +p \ [p-(1-p)\ t_{1}] \ t_{1}}{(t_{1}-t_{2})(t_{1}-t_{3})}, \qquad \alpha_{2} = \frac{t_{1} t_{3} +p \ [p-(1-p)\ t_{2}]\ t_{1}}{(t_{2}-t_{1})(t_{2}-t_{3})}, \nonumber
	\end{eqnarray}
	\begin{eqnarray}
	\alpha_{3} = \frac{t_{1} t_{2} +p \ [p-(1-p)\ t_{3}]\ t_{3}}{(t_{3}-t_{1})(t_{3}-t_{2})}.
	\end{eqnarray}
	Comparing Eq. (\ref{Ztt}) with the above presented definition of the generating function, 
	we infer that
	the grand-canonical partition function of a chain with $N$ adsorption sites is given explicitly by
	\begin{equation}
	\label{Zn}
	Z^{(I)}_{N} = \frac{\alpha_{1}}{t_{1}^{N}} + \frac{\alpha_{2}}{t_{2}^{N}} + \frac{\alpha_{3}}{t_{3}^{N}}.
	\end{equation}
	As can be seen from (\ref{Zn}), the behavior of the grand-canonical 
	partition function is entirely determined by the roots $t_{1}$, $t_{2}$, and $t_{3}$. The latter can be conveniently written as \cite{Abr72}
	\begin{eqnarray}
	t_{1,3} = \pm 2\sqrt{r_1}\cos\left(\pm\frac{\pi}{6} + \frac{1}{3}\arcsin(X_1)\right) - \frac{2-p}{3p},
	\end{eqnarray}
	\begin{eqnarray}
	\label{t2}
	t_{2} = 2\sqrt{r_1}\sin\left(\frac{1}{3}\arcsin(X_1)\right) -\frac{2-p}{3p},
	\end{eqnarray}
	where we used shortenings
	\begin{eqnarray}
	r_1&=&\frac{3\ (1+z_{A} + z_{B}) + (2-p)^{2} \ z_{A} z_{B}}{9\ p^{2}\ z_{A}z_{B}},  \nonumber \\  q_1&=&\frac{2\ (2-p)^{3}z_{A}z_{B} + 27p + 9\ (2-p)\ [1+z_{A}+z_{B}]}{54\ p^{3}\ z_{A}z_{B}}, \nonumber
	\end{eqnarray}
	\begin{equation}
	X_1=\frac{q_1}{r_{1}^{3/2}}.
	\end{equation}
	One notices that for all $z_{A, B} > 0$, the difference $q_{1}^{2} - r_{1}^{3} < 0$ and $0 < X_{1} < 1$, which implies that all three roots of the cubic polynomial $\mathcal{L}_{2}(t)$ are real. Moreover, the roots are ordered,
	$t_{1} > t_{2} > t_{3}$ and $|t_{3}| > t_{1}$, and satisfy the following conditions:
	\begin{eqnarray}
	t_{1} t_{2} t_{3} = - \frac{1}{p^{2}\ z_{A} z_{B}} < 0, \qquad t_{1} t_{2} + t_{1} t_{3} + t_{2} t_{3} = -\frac{1 + z_{A} + z_{B}}{p^{2}\ z_{A} z_{B}} < 0.
	\end{eqnarray}
	In the thermodynamic limit $N \to \infty$, the disorder-averaged grand-canonical partition functions  is governed by the smallest positive root (in our case, this is $t_2$) and follows
	\begin{align}
	\label{smpartannealedI}
	\!Z^{(I)}_N {=} \exp\left({-} N \left[2 \sqrt{r_{1}} \sin\left(\frac{1}{3} \arcsin\left(\frac{q_1}{r_{1}^{3/2}}\right)\right) { - }\frac{2 {-} p}{3 p}\right]\right) \,.
	\end{align}
	
	\subsubsection{Pressure, densities and compressibilities }
	
	The disorder-averaged pressure obtains from (\ref{smpartannealedI}), 
	\begin{eqnarray}
	\label{Pann}
	P^{(ann)}=\frac{1}{\beta} \lim_{N \rightarrow \infty} \frac{1}{N}\ln \langle Z_{N}^{(I)} [\zeta_i]\rangle_{\zeta}=-\frac{1}{\beta}\ln\left[2\sqrt{r}\sin\left(\frac{1}{3}\arcsin(X)\right) -\frac{2-p}{3p} \right].
	\end{eqnarray}
	For $p = 1$,  this expression reduces to the result obtained for a completely catalytic chain in Ref. \cite{popes}.
	
	Expressions for the disorder-averaged mean density  $n_{A}^{(ann)}=\frac{\partial P_{I}}{\partial \mu_{A}}$ 
	and for the compressibility
	\begin{equation}
	\varkappa^{(I)}_{A} = \frac{1}{\left(n^{(k)}_{A}\right)^{2}}\frac{\partial n_{A}}{\partial \mu_{A}} \,,
	\end{equation}
	are obtained directly from (\ref{Pann}) by a mere differentiation. They appear to be rather cumbersome. 
	We therefore concentrate on their asymptotic behavior for small values of the activity $z_{A}$ and $z_{ B}$. First, we  consider a situation, when one of two activities is small. In the case when $z_{A} \ll 1$, for a fixed activity $z_{B}$, we obtain
	\begin{eqnarray}
	n_{A}^{(ann)} = \frac{(1 {+} (1{-}p)z_B)^2}{(1{+}z_B)^3} z_A {-} \frac{(1{+}(1{-}p)z_B)^{2} [(3p^{2}{-}6p{+}1)z_B^{2} {-} 2(p^{2} {+}3p {-}1)z_B {+}1]}{(1{+}z_B)^6} z_A^2 +  \mathcal{O}(z_A^{3}),
	\end{eqnarray}
	and thus the compressibility obeys
	\begin{flalign}
	\varkappa_{A}^{(ann)}  &= \frac{(1{+}z_B)^3}{[1{+}(1{-}p)z_B]^{2}z_A} + \frac{p[p^{3}(z_B^{2}{-}3z_B {-} 1)z_B {-} 2p^{2}(2z_B^{2} {-} z_B {-}3)z_B {+} p(6z_B {-} 5) (1{+}z_B)^{2} {-} 4 (1{+}z_B) ]z_B}{(1{+}z_B)^{3}[1{+}(1{-}p)z_B]^{2}} z_A + \mathcal{O}(z_A^{3/2}).&&
	\end{flalign}
	In the case when the activity $z_B \ll 1$, while $z_A$ is fixed we obtain
	\begin{eqnarray}
	\! n_{A}^{(ann)} &{=}& \frac{z_A}{1{+}z_A} {-} \frac{[p^{2}(z_A{-}2)z_A {-} 2p(z_A^{2} {-} 1) {+} (1{+}z_A)]z_A}{(1{+}z_A)^4} z_B \nonumber \\
	&{+}& \frac{[4p^{3}(3z_A^{3} {-} 3z_A^{2} {-} 5z_A {+}1)z_A {-} p^2(16z_A^{2} {-} 19z_A {+}1)(1{+}z_A)^{2} {+} 4p(2z_A{-}1)(1{+}z_A)^{3} {-} 3p^{4}(z_A^{2}{-}3z_{A}{+}1)z_A^{2} {-} (1{+}z_A)^{4}]}{(1{+}z_A)^7}\nonumber \\&{\times}&z_A z_B^{2} 
	{+} \mathcal{O} (z_B^{5/2}),
	\end{eqnarray}
	while the compressibility is given by
	\begin{flalign}
	\varkappa_{A}^{(ann)} = \frac{1}{z_A} + \frac{p^{2}(z_A - 5)z_A^{2} - 2p(z_A^{3} - z_A^{2} - 3z_A - 1) +(1+z_A)^3}{z_A (1+z_A)^3} z_B + \mathcal{O}(z_B^{3/2}).
	\end{flalign}
	
	Next we consider a somewhat 
	more complicated case when either one or both of the activities are large. We start with the analysis of the asymptotic behavior of $t_{2}$ (the smallest positive root)  defined in Eq. (\ref{t2}). Assume that the activity $z_A\gg 1$, while $z_B$ is fixed. Using the identities
	\begin{eqnarray}
	\! \sin\left[\frac{1}{3}\arcsin\left( \frac{\sqrt{z_B}(2{-}p)(9{+}2z_{B}(2{-}p)^{2})}{2(3{+}z_{B}(2{-}p)^{2})^{3/2}} \right) \right] {=} \frac{2{-}p}{2}\sqrt{\frac{z_B}{3{+}z_{B}(2{-}p)^{2}}} \nonumber
	\end{eqnarray}
	and
	\begin{eqnarray}
	\! \cos\left[\frac{1}{3}\arcsin\left( \frac{\sqrt{z_B}(2{-}p)(9{+}2z_{B}(2{-}p)^{2})}{2(3{+}z_{B}(2{-}p)^{2})^{3/2}} \right) \right] {=} \frac{\sqrt{3}}{2}\sqrt{\frac{4{+}z_{B}(2{-}p)^{2}}{3{+}z_{B}(2{-}p)^{2}}},
	\end{eqnarray}
	one finds that $t_2$ has the following asymptotic representation 
	\begin{eqnarray}
	\! t_{2} = \frac{1}{z_A} {-} [1 {+} (1{-}p)^{2}z_{B}] \frac{1}{z_{A}^{2}} {+} [1{+}2\ (1{-}3p{+}2p^2)z_{B} {+} (1{-}p)^{2}(1{-}4p{+}2p^2)z_{B}^{2}]\frac{1}{z_{A}^{3}} + \mathcal{O}\left(\frac{1}{z_{A}^{4}}\right).
	\end{eqnarray}
	Therefore, the pressure in Eq. (\ref{Pann}) obeys
	\begin{eqnarray}
	\beta P^{(ann)} {=} \ln(z_{A}) {+} [1{+}(1{-}p)^{2} z_{B}]\frac{1}{z_{A}} {-} \frac{1}{2}[1 {+} 2\ (1 {-} 4p {+} 3p^{2})z_{B} {+} (1 {-} p)^{2}(1 {-} 6p {+} 3p^{2})z_{B}^{2}] \frac{1}{z_{A}^{2}} {+} \mathcal{O}\left(\frac{1}{z_{A}^{3}}\right).
	\end{eqnarray}
	As a consequence, 
	the disorder-averaged particles density $n_{A}^{(ann)}$ follows
	\begin{eqnarray}
	n_{A}^{(ann)} = 1 - [1+(1-p)^{2} z_{B}]\frac{1}{z_{A}} + [1+2\ (1-4p+3p^{2})z_{B} + (1-p)^{2}(1-6p+3p^{2})z_{B}^{2}] \frac{1}{z_{A}^{2}} + \mathcal{O}\left(\frac{1}{z_{A}^{3}}\right),
	\end{eqnarray}
	while the compressibility in this limit is given by
	\begin{eqnarray}
	\varkappa_{A}^{(ann)} = [1 + (1-p)^{2} z_{B}] \frac{1}{z_{A}}  - 2\ [2\ (1-4p+3p^{2})z_{B} + (1-p)^{2}(1-6p+3p^{2})z_{B}^{2}] \frac{1}{z_{A}^{2}} + \mathcal{O}\left(\frac{1}{z_{A}^{3}}\right).
	\end{eqnarray}
	
	In the limit of large activity $z_{B} \gg 1$ with  $z_{A}$ fixed, we can  rewrite Eq. (\ref{t2}) as follows
	\begin{eqnarray}
	\! t_2 = \frac{1}{z_{B}} {-} [1 {+} (1{-}p)^{2}z_{A}] \frac{1}{z_{B}^{2}} {+} [1{+}2\ (1{-}3p{+}2p^2)z_{A} {+} (1{-}p)^{2}(1{-}4p{+}2p^2)z_{A}^{2}]\frac{1}{z_{B}^{3}} + \mathcal{O}\left(\frac{1}{z_{B}^{4}}\right).
	\end{eqnarray}
	This implies that the disorder-averaged density of the $A$ particles admits the form
	\begin{eqnarray}
	n_{A}^{(ann)} = (1-p)^{2} \frac{z_{A}}{z_{B}} - (1-4p+3p^2)\frac{z_{A}}{z_{B}^{2}} + (1-p)^{2}(1-6p+3p^2)\frac{z_{A}^{2}}{z_{B}^{2}} + \mathcal{O}\left(\frac{1}{z_{B}^{3}}\right),
	\end{eqnarray}
	while the compressibility of the $A$ phase exhibits the following behavior in the leading in $z_B$ order, 
	\begin{eqnarray}
	\varkappa_{A}^{(ann)} = \frac{1}{(1-p)^2} \frac{z_{B}}{z_{A}} + \mathcal{O}\left(\frac{1}{z_{B}}\right).
	\end{eqnarray}
	
	\subsubsection{Expressions for the symmetric case}
	
	In the symmetric case $z_{A} = z_{B} = z$, our expressions simplify considerably. In this case, 
	$\mathcal{L}_{2}(t)$ in Eq. (\ref{L2}) factorizes into a product of a linear and a quadratic equations, 
	\begin{eqnarray}
	\label{L2symm}
	\mathcal{L}_{2}(t) = (1 - p \, z \, t) (1 - (1 + z (2 - p))t - p \, z \, t^{2}) .
	\end{eqnarray}
	One notices that the smallest root, which defines the leading behavior of the grand-canonical partition function in the limit $N \to \infty$,  is the smallest root of the quadratic equation (\ref{L2symm}):
	\begin{eqnarray}
	\label{solsymmquad}
	t_{\pm} = \pm \frac{1}{2 p\, z} \sqrt{(1 + (2 - p)z)^{2} + 4 p \,z} - \frac{1 + (2-p) \, z}{2p \, z},
	\end{eqnarray}
	i.e., $t_{+}$. Therefore, the disorder-averaged pressure in the symmetric case in the thermodynamic limit $N \to \infty$ is simply given by
	\begin{eqnarray}
	\label{symmPann}
	\beta P^{(ann)} = - \ln\left(\frac{1}{2 p\, z} \sqrt{(1 + (2 - p)z)^{2} + 4 p \,z} - \frac{1 + (2-p) \, z}{2p \, z}\right).
	\end{eqnarray}
	
	In the symmetric case, the mean densities 
	of the $A$ and $B$ phases, as well as their compressibilities, are evidently equal to each other.  In the limit  of a small concentration of catalytic bonds, $p\ll1$, the mean density of $A$ and $B$ phases is given by
	\begin{eqnarray}
	\label{nanb0}
	n^{(ann)}(p) = \frac{2z}{1+2z} - \frac{4 z^2}{(1+ 2z)^3} p + \mathcal{O}(p^2),
	\end{eqnarray}
	while in the limit when the system is \textit{almost} completely catalytic, i.e., $p\sim1$, one has
	\begin{eqnarray}
	\label{nanb1}
	n^{(ann)}(p) = \frac{1}{2}\left(1 - \frac{1 - z}{\sqrt{1 + z (6 + z)}}\right) + \frac{4 z^2}{(1 + z (6 + z))^{3/2}} (1-p) + \mathcal{O}((1-p)^2).
	\end{eqnarray}
	Note that in the limit $z \to \infty$, for both small and high $p$, 
	$n^{(ann)}(p) \to 1$, which means that the system becomes completely covered with particles. As shown in Ref. \cite{popes}, which considered only the case  $p \equiv 1$, this happens because the system spontaneously decomposes into clusters containing only one type of particles. We are not in position to unveil an analogous behavior in our case with $p < 1$; this would require a much more sophisticated approach. Note, as well, that the leading term in (\ref{nanb1}) coincides with the result obtained in Ref. \cite{popes}.

	\subsection{Quenched disorder}\label{SMIb}
	
	In this subsection we present the derivation of Eq. (\ref{partquenchedI}). 
	
	First let us consider a combinatorial approach in which
	an array of catalytic bonds is decomposed into a collection of disjoint but completely catalytic clusters. In the case of  quenched disorder, when the positions of the catalytic bonds are fixed, (unlike in the problem with annealed disorder), here we need to perform averaging of  a logarithm of the grand-canonical partition function with a distribution $P(\zeta_{i})$, where the random quenched variable $\zeta_{i}$ is such that
	\begin{equation*}
	\zeta_{i} =
	\begin{cases}
	0, &\text{if $i$ $\in$  $\{X_{n}\}$,}\\
	1, &\text{otherwise,}
	\end{cases}
	\end{equation*}
	where $\{X_{n}\}, \ n=1,2,\ldots, N_{nc}$ are the positions of the noncatalytic bonds. A logarithm of the grand-canonical partition function, averaged over all realizations of the ensemble of $\{\zeta_{i}\}$, can be rewritten as
	\begin{equation}
	\langle\ln Z^{(I)}_{N}[\zeta] \rangle_{\zeta} = \sum_{N_{nc}=0}^{N-1} p^{N-N_{nc}-1}(1-p)^{N_{nc}}\sum_{\{X_{n}\}}\ln Z_{N}(\{X_{n}\}),
	\end{equation}
	where the sum with the subscript $\{X_{n}\}$ signifies that the summation extends over all possible placement of the noncatalytic bonds $N_{nc}$.
	
	Next we introduce a set $N_{nc}+1$ of intervals $\{l_{n}\}$, which define consecutive catalytic bonds such that $l_{n}=X_{n}-X_{n-1} \, ($ with $ \, \, X_{0} = 0)$ and $l_{N_{nc}+1}=N-X_{N_{nc}}$. This means that the first interval includes all sites connected by the catalytic bonds, starting from the boundary site $i=0$ to the nearest noncatalytic bond, the second interval extends from this noncatalytic bond to the next, and so on, and the closing interval $l_{N_{nc}+1}$ goes from the last noncatalytic bond inside the chain to the boundary site $i = N$. Thus, the grand-canonical partition function can be rewritten in this "language" of intervals as follows
	\begin{equation}
	\label{Zquen1}
	\langle\ln Z^{(I)}_{N}[\zeta] \rangle_{\zeta} = \sum_{N_{nc}=0}^{N-1} p^{N-N_{nc}-1}(1-p)^{N_{nc}}\sum_{\{l_{n}\}}\ln Z_{N}(\{l_{n}\}),
	\end{equation}
	where the sum with subscript $\{l_{n}\}$ denotes now the summation over all possible solutions of the Diophantine equation
	\begin{eqnarray}
	l_{1} + l_{2} + l_{3} +\ldots+l_{N_{nc}+1} = N,
	\end{eqnarray}
	in which each $l_{i}\geq1$.
	
	Then, we represent the grand-canonical 
	partition function of the entire chain in form of a sum over partition functions of smaller clusters that contain their own sets of intervals, 
	\begin{equation}
	\langle\ln Z^{(I)}_{N}[\zeta] \rangle_{\zeta} = \sum_{N_{nc}=0}^{N-1} p^{N-N_{nc}-1}(1-p)^{N_nc}\sum_{K=1}^{N} N_{K}(N_{nc}|N) \ln Z_{K},
	\end{equation}
	where $N_{K}(N_{nc}|N)$  defines the total number of fully catalytic clusters containing $K$-sites 
	($K$ clusters)  in all realizations with a fixed number of noncatalytic bonds $N_{nc}$, namely,
	\begin{eqnarray}
	N_{K}(N_{nc}|N)=\sum_{\{l_{n}\}}\mathcal{N}_{K}(\{l_{n}\}|N),
	\end{eqnarray}
	in which the summands $\mathcal{N}_{K}(\{l_{n}\}|N)$ obey the "conservation" law
	\begin{eqnarray}
	\! \mathcal{N}_{1}(\{l_{n}\}|N) {+} 2\mathcal{N}_{2}(\{l_{n}\}|N) {+} 3\mathcal{N}_{3}(\{l_{n}\}|N) {+} {\cdots} {+} N\mathcal{N}_{N}(\{l_{n}\}|N){=}N.
	\end{eqnarray}
	Therefore the disorder-averaged logarithm of a grand-canonical partition function with a  quenched random placement of the catalytic bonds is given by
	\begin{equation}
	\label{Pquen1}
	\langle\ln Z^{(I)}_{N}[\zeta] \rangle_{\zeta} =\sum_{K=1}^{N} \omega_{K,N}(p)\ln Z_{K},
	\end{equation}
	where $\omega_{K,N}(p)$  is the statistical weight of the $K$-clusters, which is defined as
	\begin{eqnarray}
	\omega_{K,N}(p) = \sum_{N_{nc}=0}^{N-1} p^{N-N_{nc}-1}(1-p)^{N_{nc}}N_{K}(N_{nc}|N).
	\end{eqnarray}
	
	Statistical weights $\omega_{K,N}(p)$ can be found in an explicit form as follows. We  first consider the cases of ($K=1$)- and ($K=2$) clusters, and then we will generalize the obtained results for an arbitrary $K$. A ($K=1$) cluster may appear when there is a unit interval $l_{r}=1$. Therefore, the number $\mathcal{N}_{1}(\{l_{n}\}|N)$ of ($K=1$) clusters in the $\{l_{n}\}$-realization is given by
	\begin{eqnarray}
	\mathcal{N}_{1}(\{l_{n}\}|N) = \sum_{r=1}^{N_{nc}+1}\delta(l_{r},1),
	\end{eqnarray}
	where the Kronecker $\delta$ is defined by
	\begin{equation*}
	\delta(k,m) = \frac{1}{2\pi i} \oint_{\mathcal{C}} \frac{d\tau}{\tau^{1+k-m}}=
	\begin{cases}
	1, &\text{if $k=m$,}\\
	0, &\text{otherwise.}
	\end{cases}
	\end{equation*}
	Thus, the total number $N_{1}(N_{nc}|N)$ of ($K=1$) clusters  in all realizations is given by
	\begin{eqnarray}
	N_{1}(N_{nc}|N) &=& \sum_{r=1}^{N_{nc}+1} \sum_{\{l_{n}\}} \delta(l_{r},1) = \frac{1}{2\pi i} \sum_{r=1}^{N_{nc}+1} \sum_{\{l_{n}\}} \oint_{\mathcal{C}} \frac{d\tau}{\tau}\frac{1}{\tau^{l_{r}-1}} = \frac{N_{nc}+1}{2\pi i}\oint_{\mathcal{C}} \frac{d\tau}{\tau} \tau^{(\sum_{r=1}^{N_{nc}}l_{r} - (N-1) )} \nonumber \\
	&=& \frac{N_{nc}+1}{2\pi i}\oint_{\mathcal{C}} \frac{d\tau}{\tau} \left( \frac{\tau}{1-\tau} \right)^{N_{nc}} \tau^{-(N-1)}.
	\end{eqnarray}
	Using the expansion
	\begin{eqnarray}
	\! \left(\frac{1}{1{-}\tau}\right)^{N_{nc}} {=} \sum_{n=0}^{\infty} \begin{pmatrix} n {+} N_{nc} {-}1 \\ N_{nc}{-}1 \end{pmatrix} \tau^{n} {=} \sum_{n=N_{nc}-1}^{\infty} \begin{pmatrix} n \\ N_{nc}{-}1 \end{pmatrix} \tau^{n{-}(N_{nc} {-} 1)},
	\end{eqnarray}
	we obtain the following result:
	\begin{equation*}
	N_{1}(N_{nc}|N) =( N_{nc}+1) \begin{pmatrix} N-2 \\ N_{nc}-1 \end{pmatrix} \times
	\begin{cases}
	1, &\text{if $1\leq N_{nc} \leq N-1$,}\\
	0, &\text{otherwise.}
	\end{cases}
	\end{equation*}
	Hence, the statistical weight $\omega_{1,N}(p)$ of ($K=1$)-clusters is given by the following expression
	\begin{eqnarray}
	\label{weight1}
	\omega_{1,N}(p) = \sum_{N_{nc}=1}^{N-1} p^{N-N_{nc}-1} (1-p)^{N_{nc}} \ (N_{nc}+1) \ \begin{pmatrix} N-2 \\ N_{nc}-1 \end{pmatrix} = (1-p) [(1-p)(N-1) + p + 1].
	\end{eqnarray}
	In the same way,  we find that the statistical weight $\omega_{2,N}(p)$ of ($K=2$)-clusters is given by
	\begin{eqnarray}
	\omega_{2,N}(p) = \sum_{N_{nc}=1}^{N-2} p^{N-N_{nc}-1} (1-p)^{N_{nc}} \ (N_{nc}+1) \ \begin{pmatrix} N-3 \\ N_{nc}-1 \end{pmatrix} = (1-p) \ p\ [(1-p)(N-2) + p +1].
	\end{eqnarray}
	Invoking essentially the same type of combinatorial arguments, 
	we eventually find that the statistical weight $\omega_{K, N}(p)$ of the clusters with $K-1$ bonds obeys
	\begin{eqnarray}
	\label{weight}
	\omega_{K,N}(p) = (1-p) \ p^{K-1}\ [(1-p)(N-K) + p +1].
	\end{eqnarray}
	Therefore, the resulting expression for a disorder-averaged logarithm of the grand-canonical partition 
	function reads 
	\begin{align}
	\label{smpartquenchedI}
	\!\left \langle \ln Z^{(I)}_N[\zeta_{i}] \right \rangle_{\zeta} {=} \frac{1{-}p}{p} \sum_{K=1}^{N} p^{K} \Big((1{-}p)(N{-}K) {+} p {+}1\Big) \ln Z_{K}, 
	\end{align}
	where $Z_{K}$ is the grand-canonical partition function of a completely catalytic finite chain comprising $K$ bonds. An explicit form of $Z_K$ was derived earlier
	in Ref. \cite{popes}.
	The disorder-averaged pressure  in the case of quenched disorder obtains from Eq. (\ref{smpartquenchedI}) by a mere differentiation, 
	\begin{equation}
	\label{Pquen2}
	\beta P^{(quen)}(p)= \frac{1}{N} \frac{1{-}p}{p} \sum_{K=1}^{N} \ p^{K}\ [(1-p)(N-K) + p +1]\ln Z_{K}.
	\end{equation}

	\subsubsection{Symmetric case}
	
	We focus here on the symmetric case $z_{A} = z_{B} = z$. First, we would like to evaluate $Z_K$, a grand-canonical partition function of a completely catalytic chain comprising $K$ bonds. This can be done as follows: 
	To solve the recurrence relations (\ref{ZN}) -- (\ref{ZNB}) one has to find the solutions of the quadratic equation (\ref{L2symm}) for $p=1$. In this case, the generation function $\mathcal{Z}_{t} = \sum_{K=1}^{\infty} Z_{K} t^{K}$ in Eq. (\ref{Zt}) is given by
	\begin{eqnarray}
	\label{Ltsymm}
	\mathcal{Z}_{t} = t \frac{1 + z (2 + t)}{1 - (1 + z) t - z \, t^{2}},
	\end{eqnarray}
	where the roots of a quadratic equation in the denominator are 
	\begin{flalign}
	t_{1} = \frac{1}{2z}((1 + z) + \sqrt{(1 + z)^{2} + 4z}), \qquad t_{2} = \frac{1}{2z}(-(1 + z) + \sqrt{(1 + z)^{2} + 4z}).
	\end{flalign}
	Next, we rewrite Eq. (\ref{Ltsymm}) in terms of elementary fractions, and expand the resulting expression into the Taylor series in powers of $t$. Comparing the obtained expression with the definition of the generation function $\mathcal{Z}_{t}$, we conclude that the grand-canonical partition function of a finite completely catalytic chain with $K$ bonds reads
	\begin{eqnarray}
	\label{ZOBB}
	Z_{K} = \frac{1 + 3 z + \sqrt{1 + z(6 + z)}}{2 \sqrt{1 + z(6 + z)} t_{2}^{K}}\mathcal{L}_{K},
	\end{eqnarray}
	where
	\begin{eqnarray}
	\label{mathcalL}
	\mathcal{L}_{K} = 1 - (-1)^{K} \frac{1 + 3 z -\sqrt{1 + z(6 + z)}}{1 + 3 z + \sqrt{1 + z(6 + z)}} \left(\frac{t_{2}}{t_{1}} \right)^{K}.
	\end{eqnarray}
	Eventually, a logarithm of the grand-canonical partition function (\ref{ZOBB}) can be rewritten as:
	\begin{eqnarray}
	\label{LogZN}
	\ln Z_{K} = \ln \mathcal{L}_{K} + \ln\left(\frac{ 1 + 3z + \sqrt{1 + z(6 + z)}}{2 \sqrt{1 + z(6 + z)}}\right) - K \ln t_{2}.
	\end{eqnarray}
	Now, we rewrite Eq. (\ref{Pquen2}) for a finite $N$ as a sum of three contributions:
	\begin{eqnarray}
	\beta P^{(quen)}(p) = \beta P_{1}^{(quen)}(p) + \beta P_{2}^{(quen)}(p) + \beta P_{3}^{(quen)}(p),
	\end{eqnarray}
	where $P_{1}^{(quen)}(p) $ is the contribution of elementary $(K=1)$ clusters, $P_{2}^{(quen)}(p)$ is the contribution of an    
	$N$ cluster (i.e., a completely catalytic cluster which spans the entire chain with $N$ bonds), and eventually, $P_{3}^{(quen)}(p)$ is  a contribution of remaining, all possible $K$ clusters. 
	In the limit $N \to \infty$, the contribution of $(K=1)$ clusters is given explicitly by
	\begin{eqnarray}
	\beta P_{1}^{(quen)}(p) = \lim_{N\rightarrow \infty}\left(\frac{1}{N} \omega_{1,N}(p)\ln Z_{1}\right) ,
	\end{eqnarray}
	while the contribution of an $N$ cluster obeys 
	\begin{eqnarray}
	\beta P_{2}^{(quen)}(p) = \lim_{N\rightarrow \infty}\left(\frac{1}{N} \omega_{N,N}(p)\ln Z_{N}\right) .
	\end{eqnarray}
	Finally, the contribution of all possible $K$ clusters follows
	\begin{eqnarray}
	\label{startP3}
	\beta P_{3}^{(quen)}(p) = \lim_{N\rightarrow \infty}\left(\frac{1}{N} \sum_{K=2}^{N-1} \omega_{N,K}(p)\ln Z_{K}\right).
	\end{eqnarray}
	Taking into account the result for the statistical weight of $(K=1)$ cluster (\ref{weight1}), we find that
	\begin{eqnarray}
	\label{P1}
	\beta P_{1}^{(quen)}(p) = (1-p)^2 \ln(1 + 2z),
	\end{eqnarray}
	while the contribution of the $N$ cluster for all $0 < p \leq 1$ in the thermodynamic limit is zero:
	\begin{eqnarray}
	\beta P_{2}^{(quen)}(p) = \lim_{N\rightarrow \infty}\left(\frac{1}{N}  (1-p^2) p^{N-1} \ln Z_{N}\right) \equiv 0.
	\end{eqnarray}
	Let us rewrite next Eq. (\ref{startP3}), taking into account that a logarithm of the grand-canonical partition function is given by the expression (\ref{LogZN}). We have
	\begin{eqnarray}
	\label{P3}
	\beta P_{3}^{(quen)}(p) &=& \lim_{N\rightarrow \infty}\Bigg(\frac{1}{N} \ln\left(\frac{1 + 3 z + \sqrt{1 + z(6 + z)}}{2 \sqrt{1 + z(6 + z)}}\right) \sum_{K=2}^{N-1} \omega_{N,K}(p)  - \frac{ \ln t_{2} }{N} \sum_{K=2}^{N-1} K \, \omega_{N,K}(p) \nonumber \\ &-& \frac{1}{N} \sum_{K=2}^{N-1} \omega_{N,K}(p) \sum_{n=1}^{\infty} \frac{(-1)^{nK}}{n} \left(\frac{1 + 3 z - \sqrt{1 + z(6 + z)}}{1 + 3 z + \sqrt{1 + z(6 + z)}} \right)^{n} \left( \frac{t_{2}}{t_{1}}\right)^{nK} \Bigg),
	\end{eqnarray}
	where the function $\mathcal{L}_{K} $ in Eq. (\ref{mathcalL}) is expanded into the Taylor series in powers of $t_2/t_1$, $(t_{2}/t_{1}<1)$. After some tedious but straightforward calculations, we find that the contribution of all possible $K$ clusters (excluding $K = 1$ and $K = N$) reads
	\begin{eqnarray}
	\label{finalP3}
	\beta P_{3}^{(quen)}(p) &=& p (1- p) \ln\left(\frac{1+ 3 z + \sqrt{1 + z(6 + z)}}{2 \sqrt{1 + z(6 + z)}}\right) - p (2-p) \ln\left(\frac{\sqrt{1 + z(6 + z)} - (1 + z)}{2 z}\right) \nonumber \\ &-& p (1-p)^2 \sum_{n=1}^{\infty} \frac{1}{n} \left(\frac{1 + 3 z - \sqrt{1 + z(6 + z)}}{1 + 3 z + \sqrt{1 + z(6 + z)}} \right)^{n} \frac{(t_{2}/t_{1})^{2n}}{1 - p \, (-1)^{n}(t_{2}/t_{1})^{n}}.
	\end{eqnarray}
	Then, taking into account contributions from $\beta P_{1}^{(quen)}(p)$ (\ref{P1}) and $\beta P_{3}^{(quen)}(p)$ (\ref{finalP3}), we find that the disorder-averaged pressure is given by the following expression:
	\begin{eqnarray}
	\label{Pquen}
	\beta P^{(quen)}(p) &=& (1-p)^{2} \ln(1+ 2 z) + p (1- p) \ln\left(\frac{1+ 3 z + \sqrt{1 + z(6 + z)}}{2 \sqrt{1 + z(6 + z)}}\right)- p (2-p) \ln\left(\frac{\sqrt{1 + z(6 + z)} - (1 + z)}{2 z}\right) \nonumber \\ &-& p (1-p)^2 \sum_{n=1}^{\infty} \frac{1}{n} \left(\frac{1 + 3 z - \sqrt{1 + z(6 + z)}}{1 + 3 z + \sqrt{1 + z(6 + z)}} \right)^{n} \frac{(t_{2}/t_{1})^{2n}}{1 - p \, (-1)^{n}(t_{2}/t_{1})^{n}},
	\end{eqnarray}
	which can be rewritten, expanding the denominator in the last term, as
	\begin{eqnarray}
	\label{finalPquen}
	\beta P^{(quen)}(p) &=& (1-p)^{2} \ln(1+ 2 z) + p (1- p) \ln\left(\frac{1+ 3 z + \sqrt{1 + z(6 + z)}}{2 \sqrt{1 + z(6 + z)}}\right) - p (2-p) \ln\left(\frac{\sqrt{1 + z(6 + z)} - (1 + z)}{2 z}\right) \nonumber \\ &+& p (1-p)^2 \sum_{m=0}^{\infty} \, p^{m} \, \ln\left(1 - (-1)^{m} \frac{1 + 3 z - \sqrt{1 + z(6 + z)}}{1 + 3 z + \sqrt{1 + z(6 + z)}} \left(\frac{t_{2}}{t_{1}}\right)^{m + 2}\right).
	\end{eqnarray}
	Last,  in virtue of the expression for $\beta P^{(quen)}(p)$ (\ref{finalPquen}), we have that the disorder-averaged particles density $n^{(quen)}(p)$ in the case of quenched disorder is given exactly by
	\begin{eqnarray}
	\label{nquen}
	n^{(quen)}(p) &{=}& \frac{2 z (1{-}p)^{2}}{1 {+} 2 z} {-} 4 p \, z \frac{(1{+}z)(1{-}p) {+} (2 p {-} 3)\sqrt{1 {+} z(6 {+} z)}}{(1 {+} z (6 {+} z)) (1 {-} z {+} \sqrt{1 {+} z(6 {+} z)})} \nonumber \\ &{+}& p \, (1 {-} p)^{2} z \, \sum_{m {=} 0}^{\infty} p^{m} \left(1 {-} ({-}1)^{m} \frac{1 {+} 3 z {-} \sqrt{1 {+} z(6 {+} z)}}{1 {+} 3 z {+} \sqrt{1 {+} z(6 {+} z)}} \left(\frac{t_{2}}{t_{1}}\right)^{m {+} 2}\right)^{{-}1}  \frac{4(-1)^{m+1} (t_{2}/t_{1})^{m + 1}}{\sqrt{1 + z(6 + z)}}\nonumber \\ &\times&\left(\frac{4 z (t_{2}/t_{1})}{(1 + 3 z + \sqrt{1 + z(6 + z)})^{2}} + \frac{1 + 3 z - \sqrt{1 + z(6 + z)}}{1 + 3 z + \sqrt{1 + z(6 + z)}} \frac{(2 + m)(1 - z)}{(1 + z + \sqrt{1 + z(6 + z)})^{2}}\right).
	\end{eqnarray}
	
	The asymptotic behavior of the disorder-averaged particles density $n^{(quen)}(p)$ (\ref{nquen}) in the limit of a small mean concentration of catalytic bonds, i.e., for $p \ll 1$, for an arbitrary $z$ is given by
	\begin{align}
	\label{nqb0}
	n^{(quen)}(p) &= \frac{2z}{1 {+} 2z} {+} \Bigg(\frac{3}{2} {+} \frac{2}{1 {+} 2 z} {-} 2\frac{1 {+} 2z}{1 {+} 2 z (2 {+} z)} {-} \frac{1 {+} 3 z}{1 {+} z (6 {+} z)}\nonumber\\& {+} \frac{5 {-} 3 z}{2 \sqrt{1 {+} z (6 {+} z)}} {-} \frac{3 {+} 4 z (3 {+} 2 z)}{1 {+} 2 z (1 {+} z)(3 {+} z)} \Bigg) \, p  {+} \mathcal{O}(p^2), 
	\end{align}
	while in the opposite limit of an almost completely catalytic chain, i.e., for $p \sim 1 $,  it follows
	\begin{eqnarray}
	\label{nqb1}
	n^{(quen)}(p) = \frac{1}{2}\left(1 - \frac{1 - z}{\sqrt{1 + z (6 + z)}}\right) - \frac{1}{\sqrt{1 + z (6 + z)}}\left(1 - \frac{1 + 3z}{\sqrt{1 + z (6 + z)}}\right) (1-p) + \mathcal{O}((1-p)^2).
	\end{eqnarray}
	
	\subsubsection{Quenched disorder. Mapping of Model I onto the spin-1 model}
	
	We outline here the essential steps involved in our 
	second approach, which consists in mapping the Hamiltonian associated with the grand-canonical partition function  of Model I onto the Hamiltonian of the classic Blume-Emery-Griffiths spin-1 model (BEG) \cite{Blum71, Bax82}. 
	This mapping onto the BEG model is performed as follows:  Assign to each site $i$,  ($i = 1, \ldots, N$),  of a finite one-dimensional chain 
	a three-state variable $\sigma_{i}$, such that
	\begin{equation}
	\sigma_{i} =
	\begin{cases}
	+1, &\text{if site $i$ is occupied by an $A$ particle,}\\
	-1, &\text{if site $i$ is occupied by a $B$ particle ,} \\
	0, &\text{if site $i$ is vacant.}
	\end{cases}
	\end{equation}
	Standard Boolean occupation numbers $n_{i}$ and $m_{i}$ can be simply formulated in terms of this three-state variable $\sigma_{i}$ as
	\begin{eqnarray}
	\label{occupnumbers}
	n_{i} = \frac{\sigma_{i} + \sigma_{i}^{2}}{2}, \qquad m_{i} = \frac{-\sigma_{i} + \sigma_{i}^{2}}{2},
	\end{eqnarray}
	To somewhat simplify our derivations,  we also impose the boundary conditions $\sigma_{N+1}=\sigma_{1}$.
	
	Define next the couplings between the nearest-neighboring sites 
	\begin{equation}
	J_{i,j} =
	\begin{cases}
	-E_{1} \ (E_{1} \to 0), &\text{for $A-A$ neighbors,}\\
	-E_{2} \ (E_{2} \to 0), &\text{for $B-B$ neighbors,}\\
	+\zeta_{i} E_{3} \ (E_{3}\to \infty), &\text{for $A-B$ or $B-A$ neighbors,}\\
	0,  &\text{otherwise,}
	\end{cases}
	\end{equation}
	where in the parentheses we indicate the limiting value to which the value of the corresponding coupling has to be set equal.
	
	Therefore, the Hamiltonian of Model I can be written as
	\begin{flalign}
	\label{hamilt1}
	{\cal H} = \sum_{<ij>}^{N}[n_{i} n_{j}(-E_{1}) + m_{i} m_{j}(-E_{2}) + \zeta_{i}(n_{i}m_{j} + n_{j} m_{i}) E_{3}] - \sum_{i=1}^{N}(\mu_{A}n_{i} + \mu_{B}m_{i}),
	\end{flalign}
	where summation in the first term extends over all pairs of the nearest-neighboring sites, with each pair taken in account only once. The Hamiltonian (\ref{hamilt1}) can be rewritten using the variable $\sigma_{i}$ to give
	\begin{eqnarray}
	\label{hamilt2}
	{\cal H}&=&- \frac{E_{1}+E_{2}+2\zeta_{i}E_{3}}{4}\sum_{i=1}^{N}\sigma_{i}\sigma_{i+1} -\frac{E_{1}+E_{2}-2\zeta_{i}E_{3}}{4}\sum_{i=1}^{N}\sigma_{i}^{2}\sigma_{i+1}^{2} -\frac{E_{1}-E_{2}}{4}\sum_{i=1}^{N}(\sigma_{i}\sigma_{i+1}^{2}+\sigma_{i+1}\sigma_{i}^{2}) \nonumber \\ &-& \frac{\mu_{A}-\mu_{B}}{2}\sum_{i=1}^{N}\sigma_{i} -\frac{\mu_{A}+\mu_{B}}{2}\sum_{i=1}^{N}\sigma_{i}^{2}.
	\end{eqnarray}
	One recognises next that this is exactly 
	the Hamiltonian of the spin $S=1$ model \cite{Fur77} with the following parameters
	\begin{eqnarray}
	\label{interparam}
	J&=&\frac{E_{1}+E_{2}+2\zeta_{i}E_{3}}{4}, \, K=\frac{E_{1}+E_{2}-2\zeta_{i}E_{3}}{4}, \, C=\frac{E_{1}-E_{2}}{4}, \nonumber \\
	H&=&\frac{\mu_{A}-\mu_{B}}{2}, \, \text{and} \, \Delta=-\frac{\mu_{A}+\mu_{B}}{2}.
	\end{eqnarray}
	Noticing the equivalence of our model at hand with the BEG model, 
	we remark that the values of the parameters appearing in the effective BEG model are a little bit unusual. Our conditions 
	$E_{1}=E_{2}=0$ and $E_{3} \to \infty$, imply that  
	$C=0$, a bilinear exchange constant $J = \zeta_{i} E_{3}/2 \to \infty$ (if $\zeta_{i}=1$), and, finally, a biquadratic exchange constant $K = - \zeta_{i} E_{3}/2 \to -\infty$, with, however, 
	the ratio $J/K$ being constant and equal to $-1$ regardless of the value of $\zeta_{i}$.
	
	Redefining next the local fields $\mu(\sigma_{i})$, such that
	\begin{equation}
	\mu(\sigma_{i}) =
	\begin{cases}
	-\mu_{A}, &\text{if $\sigma_{i}=1$,}\\
	+\mu_{B}, &\text{if $\sigma_{i}=-1$,}\\
	0,    &\text{if $\sigma_{i}=0$.}
	\end{cases}
	\end{equation}
	we cast the grand-canonical  partition function into a form
	\begin{eqnarray}
	Z_{N}^{(BEG)}=\sum_{\{\sigma_{i}\}} \exp \left[\sum_{i=1}^{N}(-\beta J_{i,i+1}\sigma_{i}\sigma_{i+1} - \beta\mu(\sigma_{i})\sigma_{i}) \right],
	\end{eqnarray}
	which can now be conveniently written  as 
	the trace of a product of transfer matrices, 
	\begin{eqnarray}
	\label{ZBEG}
	Z_{N}^{(BEG)}={\rm Tr} \prod_{i=1}^{N} V_{i,i+1},
	\end{eqnarray}
	with $V_{i, i+1}$  given explicitly by
	\begin{eqnarray}
	V_{i,i+1} = \exp \left[-\beta J_{i,i+1}\sigma_{i}\sigma_{i+1}-\beta\left(\mu(\sigma_{i})\sigma_{i} + \mu(\sigma_{i+1})\sigma_{i+1} \right)/2\right].
	\end{eqnarray}
	In the thermodynamic limit,  the expressions for the pressure given by the grand-canonical 
	partition functions of Model I and by (\ref{ZBEG}) become identical, if we set $E_{1} = E_{2} = 0$, and $E_{3} \to \infty$. 
	For such values of the parameters, the transfer matrix $V_{i, i+1}$ attains the following form
	\begin{eqnarray}
	\label{transmat1}
	V_{i,i+1} = \begin{pmatrix}
	z_{A}    & \sqrt{z_{A}} & (1-\zeta_{i})\sqrt{z_{A}z_{B}} \\
	\sqrt{z_{A}} & 1            & \sqrt{z_{B}} \\
	(1-\zeta_{i})\sqrt{z_{A}z_{B}} & \sqrt{z_{B}} & z_{B}
	\end{pmatrix}.
	\end{eqnarray}
	We introduce next the following shortenings: $\sqrt{z_{A}} = x$, $\sqrt{z_{B}} = y$ and $1 - \zeta_{i} = \epsilon_{i}$. Then, the transfer matrix $V_{i, i+1}$ (\ref{transmat1}) can be simply written as
	\begin{eqnarray}
	\label{transmat2}
	V_{i,i+1} \equiv V_{\epsilon_{i}} = \begin{pmatrix}
	x^{2}    &     x      &  x y \epsilon_{i} \\
	x        &     1      & y \\
	x y \epsilon_{i}&     y      & y^{2}
	\end{pmatrix},
	\end{eqnarray}
	where $x$ and $y$ are real and positive definite, and random variable $\epsilon_{i}$ obeys
	\begin{equation}
	\epsilon_{i} =
	\begin{cases}
	0, &\text{with probability $p $,}\\
	1, &\text{with probability $1-p\equiv q$.}
	\end{cases}
	\end{equation}
	As a consequence, each $V_{\epsilon_{i}}$ (\ref{transmat2}) equals either 
	\begin{eqnarray}
	\label{transmatv0}
	V_{0} = \begin{pmatrix}
	x^{2}    &     x      &  0 \\
	x        &     1      & y \\
	0&     y      & y^{2}
	\end{pmatrix},
	\end{eqnarray}
	with probability $p$ or to
	\begin{eqnarray}
	\label{transmatv1}
	V_{1} = \begin{pmatrix}
	x^{2}    &     x      &  x y  \\
	x        &     1      & y \\
	x y &     y      & y^{2}
	\end{pmatrix},
	\end{eqnarray}
	with probability $q = 1-p$, respectively. The matrices $V_{\epsilon_{i}}$ are real and symmetric, and have non-negative entries.
	
	Calculation of the disorder-averaged pressure in Model I thus amounts to finding the Lyapunov exponent $\gamma$,
	\begin{eqnarray}
	\label{gamma0}
	\gamma = \lim_{N\to\infty} \frac{1}{N} \ln{ {\rm Tr} \prod_{i=1}^{N} V_{\epsilon_{i}}}.
	\end{eqnarray}
	of a product of random, uncorrelated $3 \times 3$ matrices of the form (\ref{transmat2}). 
	As was pointed to us by J.-M. Luck \cite{luck}, in the case at hand 
	a very singular feature of the model is that the matrix $V_{1}$ has rank $1$. As a matter of fact,  this very circumstance 
	allows for an exact calculation of the Lyapunov exponent. 
	
	The matrix $V_{1}$ has only one nonzero eigenvalue,  $= 1+x^2+y^2$, while other two are equal to 0, and the eigenvector corresponding to the nonzero eigenvalue is
	\begin{eqnarray}
	\vec{u} = \begin{pmatrix}
	x \\
	1 \\
	y
	\end{pmatrix}.
	\end{eqnarray}
	In other words, $V_{1}$ is a multiple of the orthogonal projector onto the direction of the vector $\vec{u}$. In addition, the kernel of the matrix $V_{1}$ is a subspace orthogonal to ${\vec{u}}$. It can be defined, for example, by the following two vectors $\vec{v}$ and $\vec{w}$:
	\begin{eqnarray}
	\vec{v} = \begin{pmatrix}
	1 \\
	-x \\
	0
	\end{pmatrix}
	\qquad \text{and} \qquad
	\vec{w} = \begin{pmatrix}
	0 \\
	-y \\
	1
	\end{pmatrix}.
	\end{eqnarray}
	Introduce next a matrix $P$ such that
	\begin{eqnarray}
	P  = (\vec{u} \, \, \vec{v} \, \, \vec{w}) =
	\begin{pmatrix}
	x  &  1 &  0\\
	1  & -x & -y\\
	y  &  0 &  1
	\end{pmatrix},
	\end{eqnarray}
	with its inverse matrix being
	\begin{eqnarray}
	P^{-1}  = \frac{1}{\lambda}
	\begin{pmatrix}
	x  &  1 &  y    \\
	1 + y^{2}  & -x & -x y  \\
	-x y  & -y &  1 + x^{2}
	\end{pmatrix},
	\end{eqnarray}
	where
	\begin{eqnarray}
	\lambda = 1 + x^{2} + y^{2}.
	\end{eqnarray}
	In the basis $\{\vec{u}, \vec{v}, \vec{w} \}$, the matrices $V_{1}$ and $V_{0}$  become, respectively,
	\begin{eqnarray}
	W_{1} = P^{-1} V_{1} P = \lambda
	\begin{pmatrix}
	1  &  0 &  0    \\
	0  &  0 &  0  \\
	0  &  0 &  0
	\end{pmatrix},
	\end{eqnarray}
	and
	\begin{eqnarray}
	W_{0} &=& P^{-1} V_{0} P \nonumber \\ &=& \frac{1}{\lambda}
	\begin{pmatrix}
	1 + 2 (x^{2} + y^{2}) + x^{4} + y^{4}  &  - x y^{2}       &  - x^{2} y    \\
	- x y^{2} (1 - x^{2} + y^{2})  &  x^{2} y^{2}     &  - x y (1+y^{2})  \\
	- x^{2} y (1 + x^{2} - y^{2})  &  - x y (1+x^{2}) &  x^{2} y^{2}
	\end{pmatrix}.
	\end{eqnarray}
	Let us define next a sequence of vectors
	\begin{eqnarray}
	{A}_{i} = \begin{pmatrix}
	{a}_{i} \\
	{b}_{i}  \\
	{c}_{i}
	\end{pmatrix},
	\end{eqnarray}
	such that
	\begin{eqnarray} \label{recmat}
	{A}_{i} = W_{\epsilon_{i}} {A}_{i-1},
	\end{eqnarray}
	with
	\begin{eqnarray}
	A_{0} = \begin{pmatrix}
	1 \\
	0  \\
	0
	\end{pmatrix}.
	\end{eqnarray}
	The entries $a_{i}$ are evidently positive. As a consequence, the Lyapunov exponent $\gamma$ in (\ref{gamma0}) 
	takes the form
	\begin{eqnarray}
	\label{gamma}
	\gamma = \lim_{N\to\infty} \frac{1}{N} \ln{{a}_{N}} = \lim_{N\to\infty} \frac{1}{N} \sum_{i=1}^{N}\ln \frac{{a}_{i}}{{a}_{i-1} }.
	\end{eqnarray}
	We notice that once $\epsilon_{i} = 1$, we have $A_{i} = \lambda {a}_{i-1} {A}_{0}$, and therefore $a_{i} = \lambda {a}_{i-1}$, such that 
	the contribution of each site with  $\epsilon_{i} = 1$ to the sum 
	in Eq. (\ref{gamma}) is $\ln{\lambda}$, $\lambda = 1+ x^2 + y^2$. Importantly, the vector ${A}_{i}$ is proportional to ${A}_{0}$, irrespective of ${A}_{i-1}$. This resetting to a fixed direction is, in fact, the key feature allowing for an exact calculation of the Lyapunov exponent. One example of such a  situation was discovered long ago in Ref. \cite{Domb59}, which analyzed  the frequency spectrum of a chain of light and heavy beads connected by identical springs, in the limit when the masses of the heavy beads are infinitely large. 
	
	To proceed, 
	it is convenient to renumber the sites along the chain according to the last occurrence of $\epsilon_{i} = 1$. In this procedure, any site in the chain gets a label $j$ with probability $q p^{j}$ (where $j \geq 0$). In doing so, we have 
	\begin{eqnarray}
	\label{gamma1}
	\gamma = q \left(\ln{\lambda} +\sum_{j \geq 1} p^{j}\ln{\frac{a_{j}}{a_{j-1}}}\right).
	\end{eqnarray}
	Setting for further convenience
	\begin{eqnarray}
	a_{j} = \frac{t_{j}}{\lambda},
	\end{eqnarray}
	the expression (\ref{gamma1}) can be simplified to give
	\begin{eqnarray}
	\label{gamma2}
	\gamma = q^{2} \left(\ln{\lambda} +\sum_{j \geq 1} p^{j}\ln{t_{j}}\right).
	\end{eqnarray}
	Next, it follows from  (\ref{recmat}) that $t_{j}$ obeys the four-site recursion
	\begin{eqnarray}
	\label{recct}
	t_{j+3} - \lambda t_{j+2} + x^2 y^2 t_{j+1} + x^2 y^2 t_{j} = 0, 
	\end{eqnarray}
	with
	the initial conditions
	\begin{eqnarray}
	\label{initcondt}
	t_{-2} = t_{-1} =1 \qquad \text{and} \qquad t_{0} = \lambda.
	\end{eqnarray}
	It is rather straightforward to find first few terms in this recursion by just iterating the initial conditions, which gives, e.g.,  
	\begin{eqnarray}
	\label{2sol}
	t_{1} = \lambda^{2} - 2 x^2 y^2, \qquad t_{2} = \lambda^{3} - (3 \lambda  - 1) x^2 y^2,
	\end{eqnarray}
	and so on. The general solution for an arbitrary $j$ can be found by standard means, e.g. 
	from the characteristic polynomial $\mathcal{Q}(\eta)$ for $W_0$ (or alternatively for $V_0$). In this representation, 
	\begin{eqnarray}
	\label{sert}
	t_{j} = \sum_{k=1,2,3} \alpha_{k} \eta_{k}^{j+2},
	\end{eqnarray}
	where $\eta_k$ are the solutions of characteristic equations $\mathcal{Q}(\eta)=0$:
	\begin{eqnarray}
	\eta^{3} - \lambda \eta^{2} + (\eta + 1) x^2 y^2 = (\eta - \eta_{1})(\eta - \eta_{2})(\eta - \eta_{3}) . 
	\end{eqnarray}
	Note that the exponent $(j+2)$ in (\ref{sert}) is chosen for a mere convenience.
	Further on, since $V_{0}$ is a symmetric matrix, the eigenvalues $\eta_{k}$ ($k=1,2,3$) are real, and we order them according to
	\begin{eqnarray}
	\eta_{3} < 0 < \eta_{2} < \eta_{1}, 
	\end{eqnarray}
	such that the Perron-Frobenius eigenvalue $\eta_{1}$ is the largest in absolute value. All three solutions  $\eta_{k}$ are defined as
	\begin{eqnarray}
	\label{etas}
	\eta_{1} &=& 2 \sqrt{r} \cos{\left(\frac{1}{3} \arccos{(X)}\right)} + \frac{1}{3} \left(1+x^2+y^2\right), \nonumber \\
	\eta_{2} &=& 2 \sqrt{r} \cos{\left(\frac{1}{3} \arccos{(X)} -\frac{2 \pi}{3} \right)} + \frac{1}{3} \left(1+x^2+y^2\right), \nonumber \\
	\eta_{3} &=& 2 \sqrt{r} \cos{\left(\frac{1}{3} \arccos{(X)} -\frac{4 \pi}{3} \right)} + \frac{1}{3} \left(1+x^2+y^2\right),
	\end{eqnarray}
	with
	\begin{eqnarray}
	r &=& \frac{1}{9} \left(x^2 \left(x^2 - \left(y^2-2\right)\right) +\left(y^2+1\right)^2\right), \nonumber \\
	q &=& \frac{1}{54} \left( x^2 \left( x^2 \left(2 x^2 - 3\left(y^2-2\right)\right)
	-3 \left(y^2 \left(y^2 + 8\right)-2\right) \right)+2 \left(y^2+1\right)^3\right), \nonumber \\
	X &=& \frac{q}{r^{3/2}} . \nonumber
	\end{eqnarray}

	The amplitudes $\alpha_{k}$ can be determined from the initial conditions (\ref{initcondt}). Therefore, after some algebra, 
	we find
	\begin{eqnarray}
	\label{alpha1}
	\alpha_{k} = \frac{\eta_{k}^{2} - x^2 y^2}{\lambda \eta_{k}^{2} - (2 \eta_{k} + 3) x^2 y^2} \,,
	\end{eqnarray}
	or, equivalently,
	\begin{eqnarray}
	\label{alpha2}
	\alpha_{k} \eta_{k} =1 + \frac{2 x^2 y^2}{\lambda \eta_{k}^{2} - (2 \eta_{k} + 3) x^2 y^2}.
	\end{eqnarray}
	The expression in (\ref{gamma2}), together with (\ref{recct}) and (\ref{initcondt}), (or with (\ref{sert}) and (\ref{alpha1})), respectively, provides an exact value of the Lyapunov exponent.
	The Lyapunov exponent is evidently a symmetric function of $x$ and $y$. It is a monotonically decreasing  function of $p$, 
	which interpolates between
	the value $\gamma = \ln{\lambda}  = \ln(1 + x^2 + y^2)$ (since $\lambda$ is the largest eigenvalue of $V_{1}$), which value is attained for $p=0$, and the value 
	$\gamma = \ln{\eta_{1}}$, (recall that $\eta_{1}$ is the largest eigenvalue of $V_{0}$), for $p=1$.
	Consider last its behavior in some limiting situations:

	\begin{enumerate}[(a)]
		\item For $p \to 0$ (i.e., $q \to 1$), keeping only the term $j=1$ in (\ref{gamma2}), we obtain the following expansion:
		\begin{eqnarray}
		\label{gammaq}
		\gamma = \ln{\lambda} + p\ln\left(1 - \frac{2 x^2 y^2}{\lambda^{2}}\right) + \ldots \qquad (p \to 0).
		\end{eqnarray}
		The term linear in $p$ is negative. Higher-order corrections are of order $O(p^{2})$.
		
		\item For $p \to 1$ (i.e., $q \to 0$), a large number of terms, [of order $O(1/(1-p))$], contributes to the sum in (\ref{gamma2}). For large $j$, it is legitimate to approximate $t_{j} \approx \alpha_{1} \eta_{1}^{j+2}$. Hence,  the following expansion holds:
		\begin{eqnarray}
		\label{gammap}
		\gamma = \ln{\eta_{1}} + (1-p) \ln{\alpha_{1}\eta_{1}} + \ldots \qquad (p \to 1).
		\end{eqnarray}
		The term linear in $(1-p)$ is positive, as a consequence of (\ref{alpha2}), in which the denominator is positive for $k=1$.
		Higher-order corrections are of order $O((1-p)^{2})$.
		\item When $x$ and $y$ are small, the Lyapunov exponent exhibits a weak linear dependence on $p$. We have that here $\eta_{1} \approx \lambda - 2 x^2 y^2$, (where $x^2 y^2$ is very small), and, more generally,
		\begin{eqnarray}
		\gamma \approx \ln{\lambda} - 2 x^2 y^2 p \approx x^{2} + y^{2} - 2 p x^{2} y^{2},
		\end{eqnarray}
		which is corroborated by the expansions (\ref{gammaq}) and (\ref{gammap}).
		\item When $x$ and $y$ are both large, the dependence of the Lyapunov exponent on $1-p$ is also linear. Assume, for simplicity, that the ratio
		\begin{eqnarray}
		g = \frac{y}{x} < 1
		\end{eqnarray}
		is fixed. We have $\eta_{1} \approx x^{2}$, $\eta_{2} \approx y^{2}$, $\alpha_{1} \eta_{1} \approx 1$, $\alpha_{2} \eta_{2} \approx 1$, while $\eta_{3}$ is negligibly small, such that
		\begin{eqnarray}
		t_{j} \approx x^{2 (j+1)} + y^{2 (j+1)}.
		\end{eqnarray}
		Inserting the latter estimate into (\ref{gamma2}), we obtain after some algebra
		\begin{eqnarray}
		\label{gammafin}
		\gamma \approx \ln{x^{2}} + (1-p)^{2} \sum_{j \geq 0} g^{j} \ln{(1 + g^{2(j+1)})}.
		\end{eqnarray}
		The leading logarithmically divergent contribution is independent of $p$. The expansion (\ref{gammaq}) becomes
		\begin{eqnarray}
		\gamma \approx \ln{x^{2} + y^{2}} + p \ln{\frac{x^{4} + y^{4}}{(x^{2} + y^{2})^{2}}} \qquad (p \to 0),
		\end{eqnarray}
		in agreement with (\ref{gammafin}). Then, the expansion (\ref{gammap}) becomes
		\begin{eqnarray}
		\gamma \approx \ln{x^{2}} + \frac{2 (1-p) y^{2}}{x^{2}(x^{2} + y^{2})} \qquad (p \to 1),
		\end{eqnarray}
		whereas the correction term in (\ref{gammafin}) has a factor $(1-p)^{2}$, showing that the limits $x, y \to \infty$ and $p \to 1$ do not commute. 
	\end{enumerate}
	
	\begin{figure*}
		\includegraphics[width=0.4\textwidth,height=0.3\textwidth]{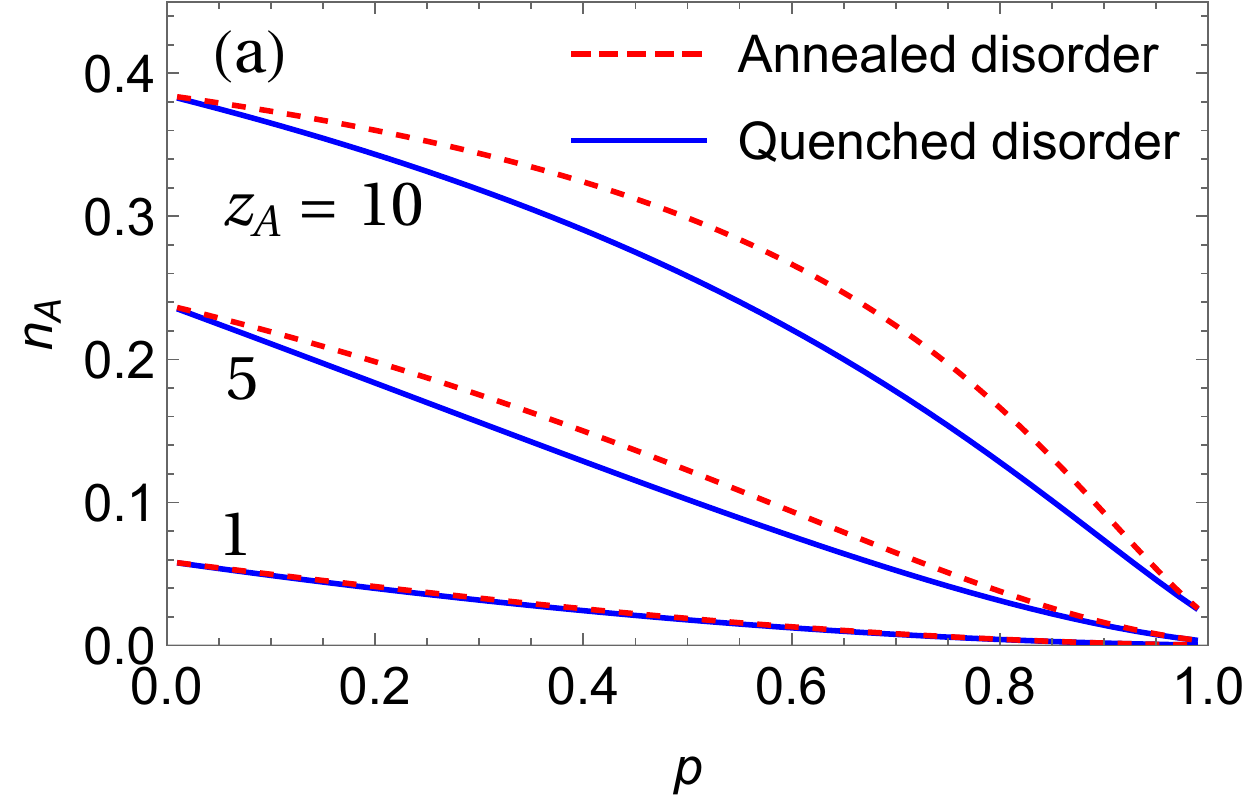}
		\includegraphics[width=0.4\textwidth,height=0.3\textwidth]{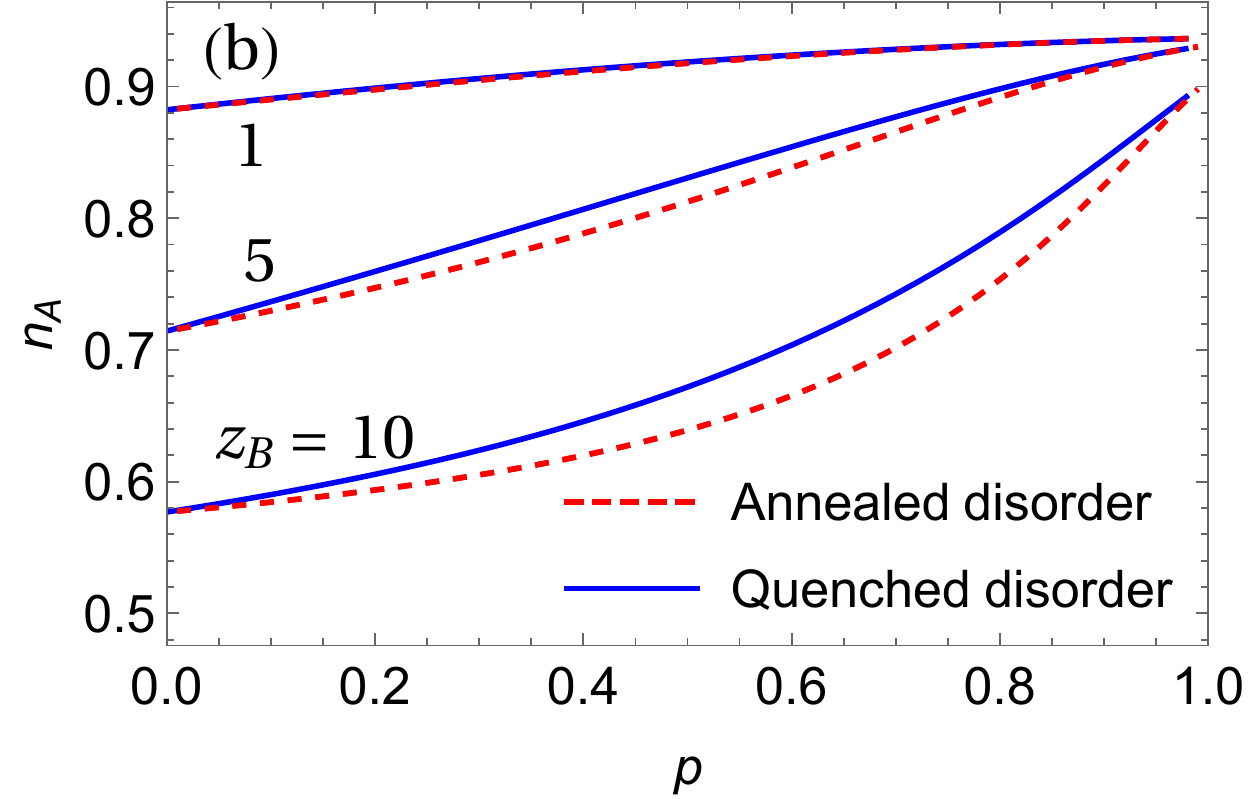}
		\includegraphics[width=0.4\textwidth,height=0.3\textwidth]{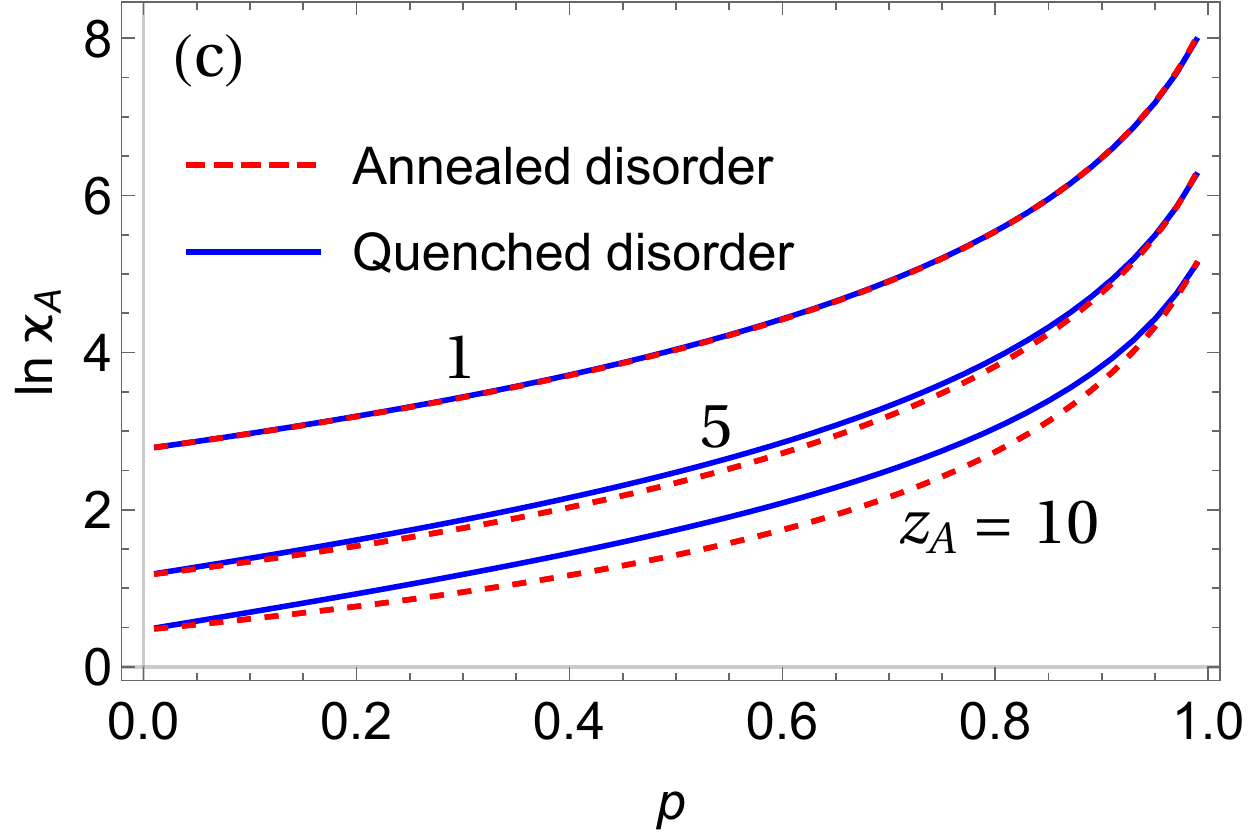}
		\includegraphics[width=0.4\textwidth,height=0.3\textwidth]{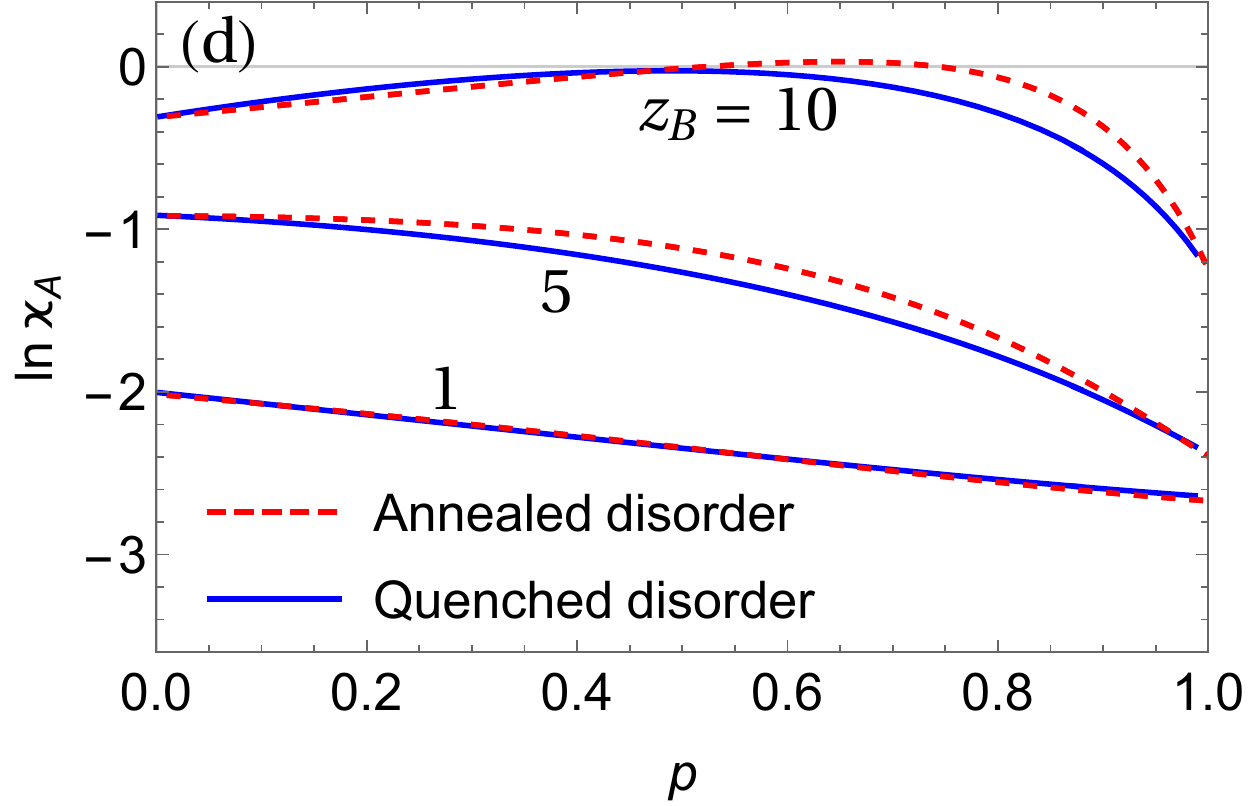}
		\caption{{\em Model I. Annealed (red dashed curves) versus quenched (blue solid curves) disorder.}
			(a), (b) Disorder-averaged density $n_A$ as a function of the mean concentration $p$ of catalytic bonds for three values of $z_A$ and $z_B = 15$ (a)
			and for three values of $z_B$ and $z_A = 15$ (b).
			(c), (d) Logarithm of the compressibility $\varkappa_{A}$ as a function of the mean concentration $p$ of catalytic bonds for three values of $z_A$ and $z_B = 15$ (c)
			and for three values of $z_B$ and $z_A = 15$ (d).
			\label{figsm1}
		}
	\end{figure*}
	
	Therefore we obtain the following expression for the disorder-averaged pressure per site in the case of quenched disorder :
	\begin{eqnarray}
	\label{LyapPress}
	\beta P^{(quen)} = (1-p)^{2} \left(\ln{\lambda} + \sum_{j = 1}^{\infty} p^{j} \ln{\left(\sum_{k=1,2,3} \frac{(\eta_{k}^{2} - x^2 y^2) \eta_{k}^{j+2}}{\lambda \eta_{k}^{2} - (2 \eta_{k} + 3) x^2 y^2}\right)} \right),
	\end{eqnarray}
	which can be rewritten in terms of the original variables as
	\begin{align}
	\label{LyapPress2}
	\beta P^{(quen)} &= (1-p)^{2} \ln{(1 + z_{A} + z_{B})}+\nonumber\\
	&+ (1-p)^{2} \sum_{j = 1}^{\infty} p^{j} \ln{\left(\sum_{k=1,2,3} \frac{(\eta_{k}^{2} - z_{A} z_{B}) \eta_{k}^{j+2}}{(1 + z_{A} + z_{B}) \eta_{k}^{2} - 2 (z_{A} z_{B}) \eta_{k} - 3 z_{A} z_{B}}\right)},
	\end{align}
	where $\eta_{k}$ ($k=1,2,3$) are defined in (\ref{etas}).

	\subsection{Model I. Annealed versus quenched disorder}\label{SMIc}
	
	Here we present an additional figure,  complementary to Figs. \ref{fig3} and \ref{fig4}. 
	In Fig.~\ref{figsm1} we provide a  comparison of the behavior of the thermodynamic properties in the case of  annealed (red dashed curves) and of quenched disorder (blue curves). We depict in Figs.~\ref{figsm1} (a) and \ref{figsm1} (b) the disorder-averaged density $n_A$ as a function of the mean concentration $p$ of catalytic bonds  
	for different values of activities $z_{A}$ and fixed $z_{B}$. In Figs. \ref{figsm1} (c) and \ref{figsm1} (d) we plot a logarithm 
	of 
	the compressibility $\varkappa_{A}$ as a function of $p$ for three values of $z_A$ [and fixed $z_B = 15$; Fig.~\ref{figsm1} (c)], 
	and three values of $z_B$ [and fixed $z_A = 15$; Fig.\ref{figsm1} (d)]. As we have already remarked, the behavior appears to be surprisingly similar, and only a noticeable difference emerges at intermediate $p$ and large values of the activity.

	\section{Model II}\label{SMII}
	
	\subsection{Annealed disorder}\label{SMIIa}
	The grand-canonical partition function $\langle Z^{(II)}_{N}[\eta_{i}]\rangle_{\eta}$ of Model II, averaged directly
	over the spatial distribution of the sites with catalytic properties, obeys
	\begin{eqnarray}
	\langle Z^{(II)}_{N}[\eta_i]\rangle_{\eta} &=& \sum_{\{n_{i},m_{i}\}} \exp\left(\beta \mu_{A}\sum_{i}n_{i}\right) \exp\left(\beta \mu_{B}\sum_{i}m_{i}\right) \prod_{i}^{N} (1 - n_{i} m_{i})  \nonumber \\ &\times& \prod_{i}^{N} \Big(p \ (1-n_{i}m_{i-1})(1-n_{i}m_{i+1})(1-m_{i}n_{i-1})(1-m_{i}n_{i+1}) + 1 -p \Big).
	\end{eqnarray}
	For Boolean variables $n_i$ and $m_i$, which assume only values $0$ and $1$, the term in the second line can be formally rewritten as
	\begin{eqnarray}
	\Big(p \ (1-n_{i}m_{i-1})(1-n_{i}m_{i+1})(1-m_{i}n_{i-1})(1-m_{i}n_{i+1}) + 1 -p \Big) \equiv  
	(1-p)^{\Psi_i}, \nonumber
	\end{eqnarray}
	where $\Psi_i $ is a Boolean function of the form
	\begin{align}
	\Psi_i = 1-(1-n_{i}m_{i-1})(1-n_{i}m_{i+1})(1-m_{i}n_{i-1})(1-m_{i}n_{i+1}) \,.
	\end{align}
	This function can be equal to only $0$ or $1$, depending on the values of the occupation variables. As a consequence,
	the disorder-averaged grand-canonical partition function of Model II reads
	\begin{eqnarray}
	Z^{(II)}_{N} = \langle Z^{(II)}_{N}[\eta_i]\rangle_{\eta} = \sum_{\{n_{i},m_{i}\}} z_{A}^{\sum_{i=1}^{N}n_{i}} z_{B}^{\sum_{i=1}^{N}m_{i}} \left(\prod_{i}^{N} (1 - n_{i} m_{i})\right) (1-p)^{\sum_{i=1}^{N}\Psi_i}.
	\end{eqnarray}
	In order to calculate $Z^{(II)}_{N}$, we pursue the same strategy as we employed in the case of Model I, i.e., we seek an appropriate recursion scheme obeyed by this property. To this end, we introduce auxiliary grand-canonical partition functions, i.e., 
	grand-canonical partition functions with a fixed occupation of the last site $i=N$. Let $Z_{N}^{(A)}$ correspond to the situation when this last site is occupied by an $A$ particle, and $Z_{N}^{(B)}$ to the situation when this site is occupied by a $B$ particle. Then, we have for $Z^{(II)}_{N}$ that
	\begin{eqnarray}
	\label{recc1}
	Z^{(II)}_{N}=Z_{N}^{(0)} + Z_{N}^{(A)} + Z_{N}^{(B)} = Z^{(II)}_{N-1} + Z_{N}^{(A)} + Z_{N}^{(B)}, \ \text{for $N\geq2$.}
	\end{eqnarray}
	Recurrence relations obeyed by the auxiliary partition functions 
	can be pursued further if we take into account that 
	a particle which resides on a catalytic site, can interact with its both neighbors. As a consequence, if the site $i=N$ is occupied by  an $A$ particle (the same for a $B$ particle), then
	\begin{eqnarray}
	Z_{N}^{(A)} &=& Z_{N}^{(A,\ 0)} + Z_{N}^{(A,\ A)} + Z_{N}^{(A,\ B)} \nonumber \\ &=& Z_{N}^{(A,\ 0)} + Z_{N}^{(A,\ A)} + z_{A}(1-p)^{2} Z_{N-1}^{(B,\ 0)} + z_{A}(1-p)^{2} Z_{N-1}^{(B,\ B)} + z_{A}(1-p) Z_{N-1}^{(B,\ A)}, \nonumber
	\end{eqnarray}
	where 
	\begin{eqnarray}
	Z_{N}^{(A,\ 0)}=z_{A} Z_{N-1}, \qquad  Z_{N}^{(A,\ A)}=z_{A} Z_{N-1}^{(A)}, \nonumber
	\end{eqnarray}
	\begin{eqnarray}
	Z_{N}^{(B,\ 0)}=z_{B} Z_{N-1}, \qquad Z_{N}^{(B,\ B)}=z_{B} Z_{N-1}^{(B)}, \nonumber
	\end{eqnarray}
	\begin{eqnarray}
	Z_{N}^{(B,\ A)} = Z_{N}^{(B)} - z_{B} Z_{N-2} - z_{B} Z_{N-1}^{(B)} \nonumber.
	\end{eqnarray}
	Gathering these terms, we find that the auxiliary grand-canonical partition functions satisfy for $N\geq4$ the following recursions :
	\begin{eqnarray}
	\label{reccA}
	Z_{N}^{(A)} = z_{A} Z_{N-2} + z_{A} Z_{N-1}^{(A)} + z_{A}(1-p) Z_{N-1}^{(B)} - z_{A}z_{B}\ p \ (1-p) \left(Z_{N-3} + Z_{N-2}^{(B)} \right),
	\end{eqnarray}
	\begin{eqnarray}
	\label{reccB}
	Z_{N}^{(B)} = z_{B} Z_{N-2} + z_{B} Z_{N-1}^{(B)} + z_{B}(1-p) Z_{N-1}^{(A)} - z_{A}z_{B}\ p \ (1-p) \left(Z_{N-3} + Z_{N-2}^{(A)} \right), 
	\end{eqnarray}
	which are to be complemented by the  initial conditions
	\begin{eqnarray}
	Z_{1} &=& 1+z_{A}+z_{B}, \nonumber \\
	Z_{1}^{(A)} &=& z_{A}, \qquad Z_{2}^{(A)}=z_{A}\left(1 + z_{A} + z_{B}(1-p)^{2}\right), \nonumber \\
	Z_{3}^{(A)} &=& z_{A} (1 + z_{A} + z_{B} + z_{A} (1 + z_{A} + z_{B} (1-p)^{2}))  \\ \nonumber &+& z_{A} z_{B} (1-p)^{2} (1 + z_{B} + z_{A} (1-p)),
	\label{initcond}
	\end{eqnarray}
	and similar conditions for $Z_{N}^{(B)}$ with $N = 1, 2, 3$.
	
	To solve the recurrence relations (\ref{recc1}) -- (\ref{reccB}), we resort to a standard  technique 
	of  generating functions. In doing so, we find that
	$\mathcal{Z}_{l} = \sum_{N=1}^{\infty}Z_{N} l^{N}$ obeys
	\begin{eqnarray}
	\label{relateq}
	\mathcal{Z}_{l} = \frac{l \mathcal{L}_{1}(l)}{\mathcal{L}_{2}(l)},
	\end{eqnarray}
	where
	\begin{eqnarray}
	\mathcal{L}_{1}(l) &=& \frac{1 + z_{A} + z_{B}}{z_{A}z_{B}} + p(1-p)(z_{A} + z_{B}) - p \left(2 (2-p) + (1-p) p (z_{A}^{2} + z_{B}^{2}) \right) l  \nonumber \\ &-&p \left((2-p) \left(1 + (1-p)(z_{A} + z_{B})\right) + (1-p)^{2} p z_{A} z_{B} (z_{A} + z_{B})\right) l^{2}  \nonumber \\ & +& (1-p) p^{2}\left(z_{A}^{2} + z_{B}^{2} + 2 (1-p)z_{A} z_{B}\right) l^{3} + (1-p) p^{2} z_{A}z_{B} \left(1 + z_{A} + z_{B}\right) l^{4}, \nonumber\\
	\label{L2sites}
	\mathcal{L}_{2}(l) &=& \frac{1}{z_{A} z_{B}} - \frac{1 + z_{A} + z_{B}}{z_{A}z_{B}} l + (2-p) p l^{2} + p \left(2-p + (1-p)^{2} (z_{A} + z_{B})\right) l^{3} \nonumber \\ &-& (1-p)^{2} p^{2} z_{A} z_{B} l^{4} - (1-p)^{2} p^{2} z_{A} z_{B} l^{5}.
	\end{eqnarray}
	Note that in this case the denominator is a quintic equation of $l$ which has five roots $l_{i}$, $i=1,2,\ldots,5$. Therefore,  
	expression (\ref{relateq}) can be cast into the form
	\begin{flalign}
	\label{Zttsites}
	\mathcal{Z}_{l} = \sum_{N=1}^{\infty} \left[ \gamma_{1}\left(\frac{l}{l_{1}}\right)^{N} + \gamma_{2}\left(\frac{l}{l_{2}}\right)^{N} + \gamma_{3}\left(\frac{l}{l_{3}}\right)^{N} + \gamma_{4}\left(\frac{l}{l_{4}}\right)^{N} + \gamma_{5}\left(\frac{l}{l_{5}}\right)^{N}\right].
	\end{flalign}
	and the grand-canonical partition function, in principle, can be formally written as
	\begin{eqnarray}
	\label{Znsites}
	Z^{(II)}_{N} = \frac{\gamma_{1}}{l_{1}^{N}} +  \frac{\gamma_{2}}{l_{2}^{N}} +  \frac{\gamma_{3}}{l_{3}^{N}} +  \frac{\gamma_{4}}{l_{4}^{N}} +  \frac{\gamma_{5}}{l_{5}^{N}}.
	\end{eqnarray}
	Here, however,
	the coefficients $\gamma_{i}$, $i=1,\ldots, 5$ will evidently have a more complicated structure as compared to (\ref{alphas}) 
	and 
	the roots $l_{i}$ can be found analytically only in some very special case; indeed,
	only certain classes of quintic equations can be solved algebraically in terms of the root extractions. In general, we will have to resort to a numerical analysis.
	
	\subsubsection{Symmetric case}
	
	Luckily, Eq. (\ref{L2sites})  can be solved analytically in the important symmetric case $z_{A} = z_{B} = z$.
	In this case the quintic equation factorises into a product of quadratic and cubic equations 
	\begin{eqnarray}
	\label{factorL2}
	\mathcal{L}_{2}(l) = \frac{1}{z^{2}}\Big(1 - p z l - p (1 - p) z^{2} l^{2}\Big) \Big(1 - (1 + (2 - p) z) l - p z (1 - (1 - p)z) l^{2} + p (1 - p) z^{2} l^{3}\Big) \,,
	\end{eqnarray}
	whose solutions can be written in an explicit form
	As in the previously considered cases, we are interested in the smallest positive solution of Eq. (\ref{factorL2}). It can be shown that $l_{1} > l_{4} > l_{2}  > 0 >  l_{3} > l_{5}$, where $l_{4}$ and $l_{5}$ are the solutions of the quadratic equation in (\ref{factorL2}), while $l_{1}$, $l_{2}$, and $l_{3}$ are the solutions of the cubic equation. We note that $|l_{5}| > |l_{3}| > l_{2}$ and thus $l_{2}$ is the smallest, by absolute value, solution of Eq. (\ref{factorL2}). 
	
	We introduce the following shortenings :
	\begin{eqnarray}
	\label{varrqx}
	r_{2} &=& \frac{3 (1 + 2 z) - \ p (2 + z (11 - 5 p - (1 - p)^{2} z))}{27 \, p \, (1-p)^{2} z^{2}},\nonumber \\
	q_{2} &=& \frac{-9 + 7 p + 3 (1 - p)(6 - 7p) z + 3 (1 - p)^{2} (6 - 5 p) z^{2} + 2 (1 - p)^{3} \, p \, z^{3}}{54 (1 - p)^{3} \, p \, z^{3}}, \nonumber \\
	X_{2} &=& \frac{q_{2}}{r_{2}^{3/2}}.
	\end{eqnarray}
	Since $q_{2}^{2} - r_{2}^{3} < 0$ 
	for all $z > 0$,  all three roots of the cubic polynomial in (\ref{factorL2}) are real and can be conveniently written as 
	\begin{eqnarray}
	l_{1,3} &=& \pm 2\sqrt{r_{2}}\cos\left(\pm\frac{\pi}{6} + \frac{1}{3}\arcsin(X_{2})\right) - \frac{1}{3} \left(1 - \frac{1}{(1-p) z}\right), \nonumber\\
	\label{t2s}
	l_{2} &=& 2\sqrt{r_{2}}\sin\left(\frac{1}{3}\arcsin(X_{2})\right) - \frac{1}{3} \left(1 - \frac{1}{(1-p) z}\right).
	\end{eqnarray}
	Finally, the annealed disorder-averaged  grand partition function pressure in the thermodynamic limit $N \to \infty$ is determined by $l_2$ and is given as follows:
	\begin{equation}
	\label{smpartannealedII}
	Z^{(II)}_{N} = \exp\Bigg(- N \Bigg[2 \sqrt{r_2} \sin\left(\frac{1}{3} \arcsin\left(\frac{q_2}{r_2^{3/2}}\right)\right)  - \frac{1}{3} \left(1 - \frac{1}{(1 - p) z}\right)\Bigg]\Bigg) \,.
	\end{equation}
	Then, the disorder-averaged pressure in this case is given by
	\begin{eqnarray}
	\label{Pannsites}
	P^{(ann)}=-\frac{1}{\beta}\ln\left[2\sqrt{r_{2}}\sin\left(\frac{1}{3}\arcsin(X_{2})\right) - \frac{1}{3} \left(1 - \frac{1}{(1-p) z}\right) \right].
	\end{eqnarray}
	Consider the limits $p\ll1$ and $p\sim1$, for which we find the following expressions:
	\begin{eqnarray}
	\label{np0sites}
	n^{(ann)}(p) = \frac{2z}{1 + 2z} - 2 z^{2} \frac{4 + 5 z}{(1 + 2 z)^{4}} \, p + \mathcal{O}(p^2),
	\end{eqnarray}
	for $p\ll1$, and
	\begin{eqnarray}
	\label{np1sites}
	n^{(ann)}(p) {=} \frac{1}{2}\left(1 - \frac{1 - z}{\sqrt{1 + z(6 + z)}}\right) {+} \frac{4 z^{2}}{(1 {+} z (6 {+} z))^{3/2}}(1{-}p)^2 {+} \mathcal{O}((1{-}p)^3)
	\end{eqnarray}
	in the limit $p\sim1$, respectively. Note that the leading term in (\ref{np1sites}) coincides with the leading term
	in (\ref{nanb1}), which describes the behavior of the total density in the case of annealed disorder in Model I.
	
	\subsection{Quenched disorder}\label{SMIIb}
	
	We fix positions of the catalytic sites and introduce a set of $N_{nc}+1$ intervals $\{l_{n}\}$ connecting noncatalytic sites. Each interval $l_{n}$ is defined as $l_{n}=X_{n} - X_{n-1}$ (with $X_{0}=0$) and $l_{N_{nc}+1}= N + 1 - X_{N_{nc}}$, where $\{X_{n}\}$, $n=1, 2,\ldots, N_{nc}$ are the positions of the noncatalytic sites. 
	A logarithm of the grand-canonical partition function, averaged over all possible 
	placements of noncatalytic sites, is given by
	\begin{eqnarray}
	\label{quenpart1}
	\langle \ln Z^{(II)}_{N}[\eta_{i}] \rangle_{\eta} = \sum_{N_{nc}=0}^{N} p^{N-N_{nc}} (1-p)^{N_{nc}} \sum_{\{l_{n}\}} \ln Z_{N}(\{l_{n}\}),
	\end{eqnarray}
	where the sums are to be performed subject to a "conservation" law-type constraint
	\begin{eqnarray}
	l_{1} + l_{2} + l_{3} +\cdots+ l_{N_{nc}+1} = N+1,\ \ \ \text{where \  $l_{i}\geq1$}.
	\end{eqnarray}
	Further on, a logarithm of the grand-partition function of the entire chain containing $N$ sites, splits naturally into a sum of logarithms of completely catalytic $K$-clusters 
	\begin{equation}
	\langle\ln Z^{(II)}_{N}[\eta_{i}] \rangle_{\eta} = \sum_{N_{nc}=0}^{N} p^{N-N_{nc}}(1-p)^{N_nc}\sum_{K=1}^{N} N_{K}(N_{nc}|N) \ln Z_{K},		
	\end{equation}
	where $N_{K}(N_{nc}|N)$ is the total number of $K$-clusters, 
	\begin{eqnarray}
	N_{K}(N_{nc}|N)=\sum_{\{l_{n}\}}\mathcal{N}_{K}(\{l_{n}\}|N).
	\end{eqnarray}
	Then, the disorder-averaged pressure follows
	\begin{equation}
	\label{defpressites}
	\beta P^{(quen)}(p)=\lim_{N\rightarrow \infty} \frac{1}{N}\sum_{K=1}^{N} {\omega}_{K,N}(p)\ln Z_{K},
	\end{equation}
	where $\omega_{K,N}(p)$ is the statistical weight of $K$-clusters in a chain comprising $N$ sites, 
	which is given by
	\begin{eqnarray}
	\omega_{K,N}(p) = \sum_{N_{nc}=0}^{N} p^{N-N_{nc}}(1-p)^{N_{nc}}N_{K}(N_{nc}|N).
	\end{eqnarray}
	As shown in Ref. \cite{Osh2}, in the case of a chain with catalytic sites there are two types of intervals, which are formed on this chain, and the combinations of these intervals will form all possible clusters:
	\begin{enumerate}[(1)]
		\item The number of $K$ clusters starting from any boundary site (``surface''):
		\begin{eqnarray}
		J^{(S)}_{(n)}(\{l_{n}\}|K|N) = 2 \left( \prod_{i=1}^{N} \delta(l_{i}\geq2) \right) \delta(l_{N+1},1) \delta(l_{1}+l_{2}+\cdots+l_{n},K),
		\end{eqnarray}
		\item The number of $K$ clusters that are entirely within the chain and do not include the boundary sites (``bulk''):
		\begin{eqnarray}
		J^{(B)}_{(n)}(\{l_{n}\}|K|N) = \sum_{r=1}^{N_{nc}-n} \delta(l_{r},1) \left( \prod_{i=r+1}^{n+r} \delta(l_{i}\geq2) \right)  \delta(l_{r+n+1},1) \delta(l_{r+1}+l_{r+2}+\cdots+l_{r+n}+1,K).
		\end{eqnarray}
	\end{enumerate}
	Therefore, the total number of all $K$ clusters consisting of $n$ intervals in a given realization of a random chain containing $N_{nc}$ noncatalytic sites is given by
	\begin{eqnarray}
	\mathcal{N}_{K}^{(n)}(\{l_{n}\}|N) = J^{(S)}_{(n)}(\{l_{n}\}|K|N) + J^{(B)}_{(n)}(\{l_{n}\}|K|N),
	\end{eqnarray}
	where
	\begin{eqnarray}
	J^{(S)}_{(n)}(\{l_{n}\}|K|N) = 2 \begin{pmatrix} K-1-n \\ n-1 \end{pmatrix}  \begin{pmatrix} N-K-1 \\ N_{nc}-n-1 \end{pmatrix} ,
	\end{eqnarray}
	\begin{eqnarray}
	J^{(B)}_{(n)}(\{l_{n}\}|K|N) = (N_{nc}-n) \begin{pmatrix} K-2-n \\ n-1 \end{pmatrix}  \begin{pmatrix} N-K-1 \\ N_{nc}-n-2 \end{pmatrix}.
	\end{eqnarray}
	Subsequent summation of $\mathcal{N}_{K}^{(n)}(\{l_{n}\}|N)$ over all intervals $ \{l_{n} \} $, according to the ``conservation'' law, and thereafter summation over all possible numbers of intervals $n$ in a $K$ cluster, lead to the statistical weights of the $K$ clusters (\ref{weightkn}) and (\ref{weightnn}).

	We focus on the symmetric case $z_{A}=z_{B}=z$ and use the previously obtained expression for the logarithm of the grand partition function (\ref{LogZN}). After some algebra, we find that the disorder-averaged pressure for Model II is given by
	\begin{eqnarray}
	\label{Psitesquen}
	\frac{1}{N}  \left\langle \ln Z^{(II)}_{N}[\eta_i] \right\rangle_{\eta}&=&\beta P^{(quen)}(p) = (1-p)^{3}\ln{(1 + 2z)} + p (1 - p)^{2} \ln{\left(\frac{1 + 3 z + \sqrt{1 + z(6 + z)}}{2\sqrt{1 + z(6 + z)}}\right)} \nonumber \\
	&-& p (p^{2} - 3 p + 3) \ln{\left(\frac{\sqrt{1 + z(6 + z)} - (1 + z)}{2z}\right)} - \frac{p (1-p)^4}{\sqrt{p (4-3p)}}  \nonumber \\
	&\times& \sum_{m=0}^{\infty} \left(\frac{1}{X_{+}^m} - \frac{1}{X_{-}^m} \right) \ln{\left(1-(-1)^{m+1} \frac{1 + 3 z - \sqrt{1 + z(6 + z)}}{1 + 3 z + \sqrt{1 + z(6 + z)}}\left( \frac{t_{2}}{t_{1}} \right)^{m+3}\right)},
	\end{eqnarray}
	where
	\begin{eqnarray}
	X_{\pm}^m = \frac{1}{2 p (1-p)} (- p \pm \sqrt{p (4- 3p)}).
	\end{eqnarray}
	From Eq.~(\ref{Psitesquen}) it is possible to find the average particle density, obtained by differentiation with respect to the chemical potential $\mu$ [$z=\exp{(\beta \mu)}$]:
	\begin{eqnarray}
	\label{nsitesquen}
	n^{(quen)}(p) &=& \frac{2 z (1-p)^{3}}{1 + 2 z} - 4 p \, z \frac{(1+z)(1-p (2-p)) - (4 -5 p + 2 p^{2})\sqrt{1 + z(6 + z)}}{( 1 + z (6 + z)) (1 - z + \sqrt{1 + z(6 + z)})} - \frac{z \, p (1-p)^4}{\sqrt{p (4-3p)}}  \nonumber \\
	&\times& \sum_{m=0}^{\infty} \left(\frac{1}{X_{+}^m} - \frac{1}{X_{-}^m} \right) \left(1 - (-1)^{m+1} \frac{1 + 3 z - \sqrt{1 + z(6 + z)}}{1 + 3 z + \sqrt{1 + z(6 + z)}} \left(\frac{t_{2}}{t_{1}}\right)^{m + 3}\right)^{-1} \frac{4(-1)^{m+2} (t_{2}/t_{1})^{m + 2}}{\sqrt{1 + z(6 + z)}} \nonumber \\ &\times& \left(\frac{4 z (t_{2}/t_{1})}{(1 + 3 z + \sqrt{1 + z(6 + z)})^{2}} \frac{1 + 3 z - \sqrt{1 + z(6 + z)}}{1 + 3 z + \sqrt{1 + z(6 + z)}} \frac{(3 + m)(1 - z)}{(1 + z + \sqrt{1 + z(6 + z)})^{2}}\right).
	\end{eqnarray}
	
	The asymptotic behavior of the disorder-averaged particles density $n^{(quen)}(p)$ (\ref{nsitesquen}) for the small concentration of the catalytic sites $p\ll1$ obeys
	\begin{eqnarray}
	\label{nqs0}
	n^{(quen)}(p) = \frac{2z}{1 + 2z} + \frac{1}{2} \left(1 + \frac{6}{1 + 2z} + \frac{1-z}{\sqrt{1 + z (6 + z)}} - 8\frac{1 + 3z (1 + z) (2 + z)}{1 + 2 z (4 + z) (1 + z)^{2}}\right) \, p + \mathcal{O}(p^2),
	\end{eqnarray}
	and in the limit $p\sim1$ one has:
	\begin{eqnarray}
	\label{nqs1}
	n^{(quen)}(p) = \frac{1}{2}\left(1 - \frac{1 - z}{\sqrt{1 + z (6 + z)}}\right) - \frac{1}{\sqrt{1 + z (6 + z)}}\left(1 - \frac{1 + 3z}{\sqrt{1 + z (6 + z)}}\right) (1-p)^{2} + \mathcal{O}((1-p)^3).
	\end{eqnarray}
	
	\subsubsection{Mapping of Model II with quenched disorder onto spin-1 model}
	
	Similarly to the approach described in Sec.~\ref{SMIb}, we seek here an appropriate representation of our Model II with quenched disorder in terms of 
	the classical spin $S=1$ model. We assign to each site of a chain
	a three-state variable $\sigma_{i}$, ($i = 1, \ldots, N$), such that
	{\setlength{\belowdisplayskip}{5pt} \setlength{\belowdisplayshortskip}{5pt}
		\setlength{\abovedisplayskip}{5pt} \setlength{\abovedisplayshortskip}{5pt}
	\begin{equation}
	\sigma_{i} =
	\begin{cases}
	+1, &\text{if site $i$ is occupied by an $A$ particle,}\\
	-1, &\text{if site $i$ is occupied by a $B$ particle ,} \\
	0, &\text{if site $i$ is vacant.}
	\end{cases}
	\end{equation}}
	Standard occupation numbers $n_{i}$ and $m_{i}$ are expressed through 
	$\sigma_{i}$ via
	\begin{eqnarray}
	\label{occupnumbers_s}
	n_{i} = \frac{\sigma_{i} + \sigma_{i}^{2}}{2}, \qquad m_{i} = \frac{-\sigma_{i} + \sigma_{i}^{2}}{2}.
	\end{eqnarray}
	The boundary conditions for this model are $\sigma_{N+1}=\sigma_{1}$.
	
	The coupling constant between the occupation variables at the nearest-neighboring sites
	is modified, as compared to the model with catalytic bonds, to take into account 
	a circumstance 
	that here the reaction occurs once either of a dissimilar species resides on a catalytic site. In this case, we have
	\begin{equation}
	J_{i,j} =
	\begin{cases}
	-E_{1} \ (\to 0), &\text{for $A-A$ neighbors,}\\
	-E_{2} \ (\to 0), &\text{for $B-B$ neighbors,}\\
	+\eta_{i} \eta_{j} E_{3} \ (E_{3}\to \infty), &\text{for $A-B$ or $B-A$ neighbors,}\\
	0,  &\text{otherwise.}
	\end{cases}
	\end{equation}
	In parentheses we indicate the limiting values of these coupling constants. 
	The Hamiltonian of such a system is defined as
	\begin{eqnarray}
	\label{hamilts1_s}
	{\cal H} = \sum_{<ij>}^{N}[-n_{i} n_{j} E_{1} - m_{i} m_{j} E_{2} + \eta_{i} \eta_{j} (n_{i} m_{j} + n_{j} m_{i}) E_{3}] - \sum_{i=1}^{N}(\mu_{A}n_{i} + \mu_{B}m_{i}),
	\end{eqnarray}
	where summation in the first term is held again by over all pairs of the nearest-neighboring sites. We  rewrite next the Hamiltonian (\ref{hamilts1_s}) replacing the occupation numbers $n_{i}$ and $m_{i}$ by the expressions (\ref{occupnumbers_s}). In doing so, we have
	\begin{eqnarray}
	\label{hamilt2_s}
	{\cal H}=&{-}& \frac{E_{1}{+}E_{2}{+}2\eta_{i{-}1} \eta_{i}E_{3}}{8}\sum_{i=1}^{N}\sigma_{i{-}1}\sigma_{i} {-} \frac{E_{1}{+}E_{2}{+}2\eta_{i} \eta_{i{+}1}E_{3}}{8}\sum_{i=1}^{N}\sigma_{i}\sigma_{i{+}1} {-}\frac{E_{1}{+}E_{2}{-}2\eta_{i{-}1} \eta_{i} E_{3}}{8}\sum_{i=1}^{N}\sigma_{i{-}1}^{2}\sigma_{i}^{2} \nonumber \\ &{-}& \frac{E_{1}{+}E_{2}{-}2\eta_{i} \eta_{i{+}1} E_{3}}{8}\sum_{i=1}^{N}\sigma_{i}^{2}\sigma_{i{+}1}^{2} {-}\frac{E_{1}{-}E_{2}}{4}\sum_{i=1}^{N}(\sigma_{i}\sigma_{i{+}1}^{2}{+}\sigma_{i{+}1}\sigma_{i}^{2}) {-}\frac{\mu_{A}{-}\mu_{B}}{2}\sum_{i=1}^{N}\sigma_{i} {-}\frac{\mu_{A}{+}\mu_{B}}{2}\sum_{i=1}^{N}\sigma_{i}^{2},
	\end{eqnarray}
	where the  parameters entering the Hamiltonian are identified as
	\begin{eqnarray}
	\label{interparam_s}
	J_{1}&=&\frac{E_{1}+E_{2}+2\eta_{i-1} \eta_{i} E_{3}}{8}, J_{2} = \frac{E_{1}+E_{2}+2\eta_{i} \eta_{i+1}E_{3}}{8}, \, K_{1} = \frac{E_{1}+E_{2}-2\eta_{i-1} \eta_{i}E_{3}}{8}, \nonumber \\ K_{2} &=& \frac{E_{1}+E_{2}-2\eta_{i} \eta_{i+1}E_{3}}{8}, \,  C=\frac{E_{1}-E_{2}}{4}, \,
	H=\frac{\mu_{A}-\mu_{B}}{2}, \, \text{and} \, \, \Delta=-\frac{\mu_{A}+\mu_{B}}{2} ,
	\end{eqnarray}
	to match the standard definition of the general spin-$1$ model \cite{Blum71,Bax82}. 
	Again, in order to resort to the transfer matrix representation, 
	we introduce the local fields $\mu(\sigma_{i})$ as
	\begin{equation}
	\label{localfields}
	\mu(\sigma_{i}) =
	\begin{cases}
	-\mu_{A}, &\text{if $\sigma_{i}=1$,}\\
	+\mu_{B}, &\text{if $\sigma_{i}=-1$,}\\
	0,    &\text{if $\sigma_{i}=0$.}
	\end{cases}
	\end{equation}
	Therefore, the grand-canonical 
	partition function writes
	\begin{eqnarray}
	Z_{N}^{(BEG)}=\sum_{\{\sigma_{i}\}} \exp \left[\sum_{i=1}^{N}(-\beta J_{i-1,i}\sigma_{i-1}\sigma_{i}-\beta J_{i,i+1}\sigma_{i}\sigma_{i+1} - \beta\mu(\sigma_{i})\sigma_{i}) \right],
	\end{eqnarray}
	or, equivalently,
	{\setlength{\belowdisplayskip}{5pt} \setlength{\belowdisplayshortskip}{5pt}
		\setlength{\abovedisplayskip}{5pt} \setlength{\abovedisplayshortskip}{5pt}
	\begin{eqnarray}
	\label{ZBEG_s}
	Z_{N}^{(BEG)}={\rm Tr} \prod_{i=1}^{N} V_{i-1,i} V_{i,i+1},
	\end{eqnarray}}
	where the transfer matrices $V_{i, j}$ are given explicitly by
	\begin{eqnarray}
	\label{transmat1_s}
	V_{i,j} = \begin{pmatrix}
	z_{A}^{1/2}    & z_{A}^{1/4}  & (1{-}\eta_{i})(1{-}\eta_{j})(z_{A}z_{B})^{1/4} \\
	z_{A}^{1/4} & 1            & z_{B}^{1/4} \\
	(1{-}\eta_{i})(1{-}\eta_{j})(z_{A}z_{B})^{1/4} & z_{B}^{1/4} & z_{A}^{1/2}
	\end{pmatrix},
	\end{eqnarray}
	once we set $E_{1} = E_{2} = 0$, and $E_{3} \to \infty$. Note also that
	here the subscripts $i, j$ indicate the pairs of nearest-neighboring sites. Since the transfer matrices $V_{i, j}$ have the similar structure the grand partition function (\ref{ZBEG_s}) can be rewritten in the following form:
	\begin{eqnarray}
	\label{ZBEG_ss}
	Z_{N}^{(BEG)}={\rm Tr} \prod_{i=1}^{N} (V_{i,i+1})^{2}.
	\end{eqnarray}
	Note that here the transfer matrices are not statistically independent and have sequential pairwise correlations.
	
	\subsection{Model II. Annealed versus quenched disorder}\label{SMIIc}
	
\begin{figure*}
		\includegraphics[width=0.4\textwidth,height=0.3\textwidth]{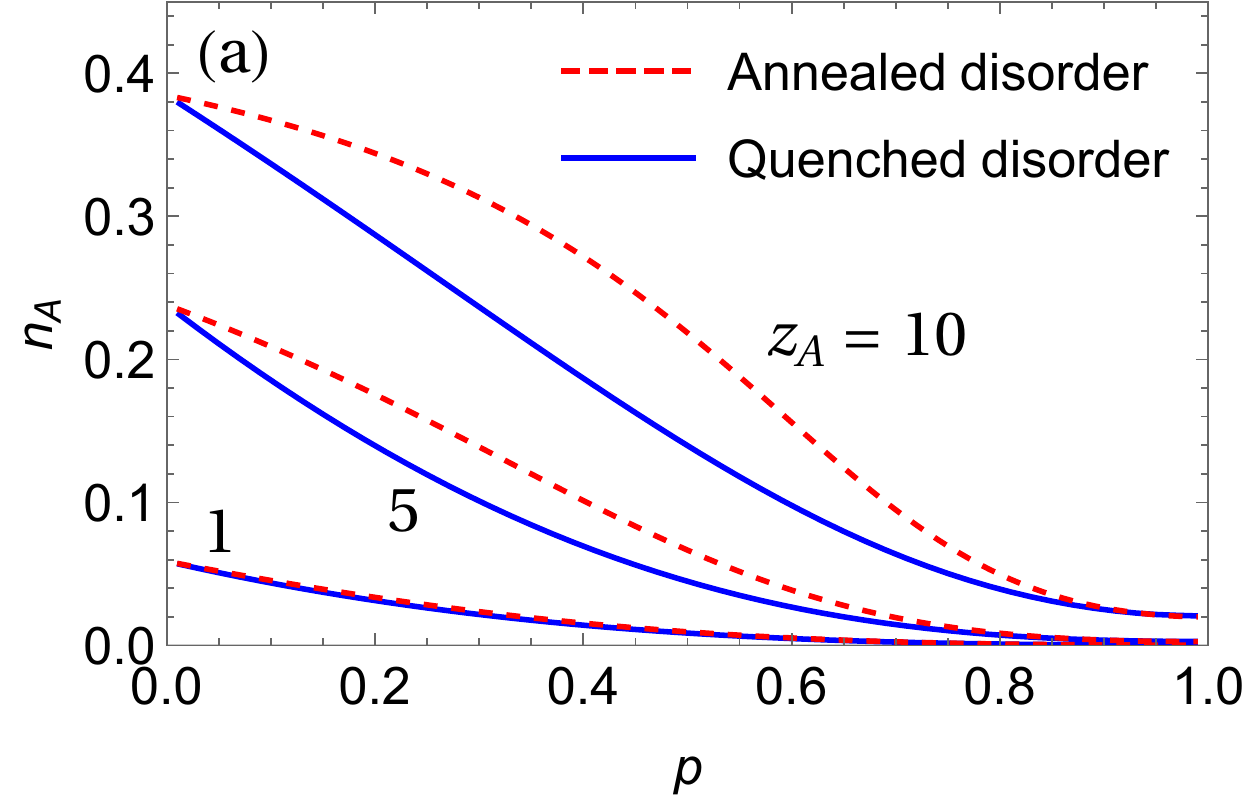}
		\includegraphics[width=0.4\textwidth,height=0.3\textwidth]{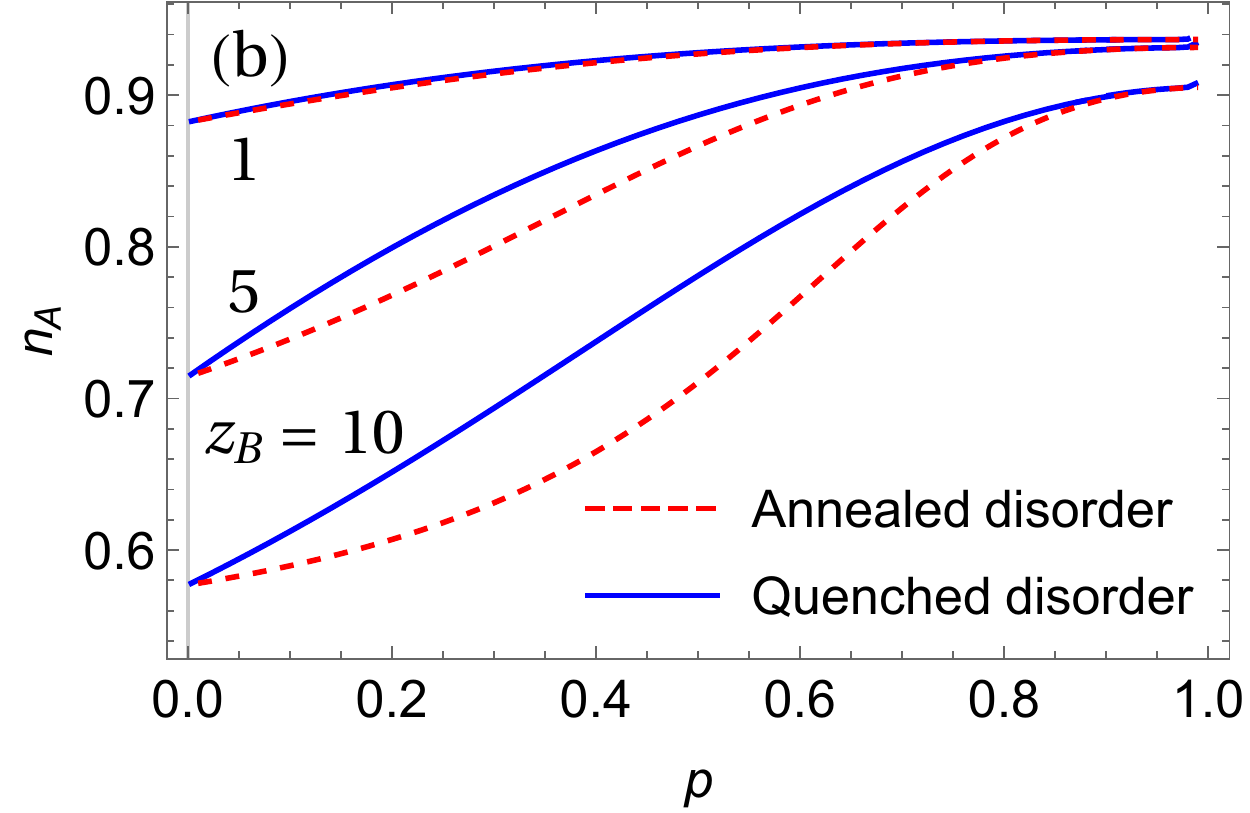}
		\includegraphics[width=0.4\textwidth,height=0.3\textwidth]{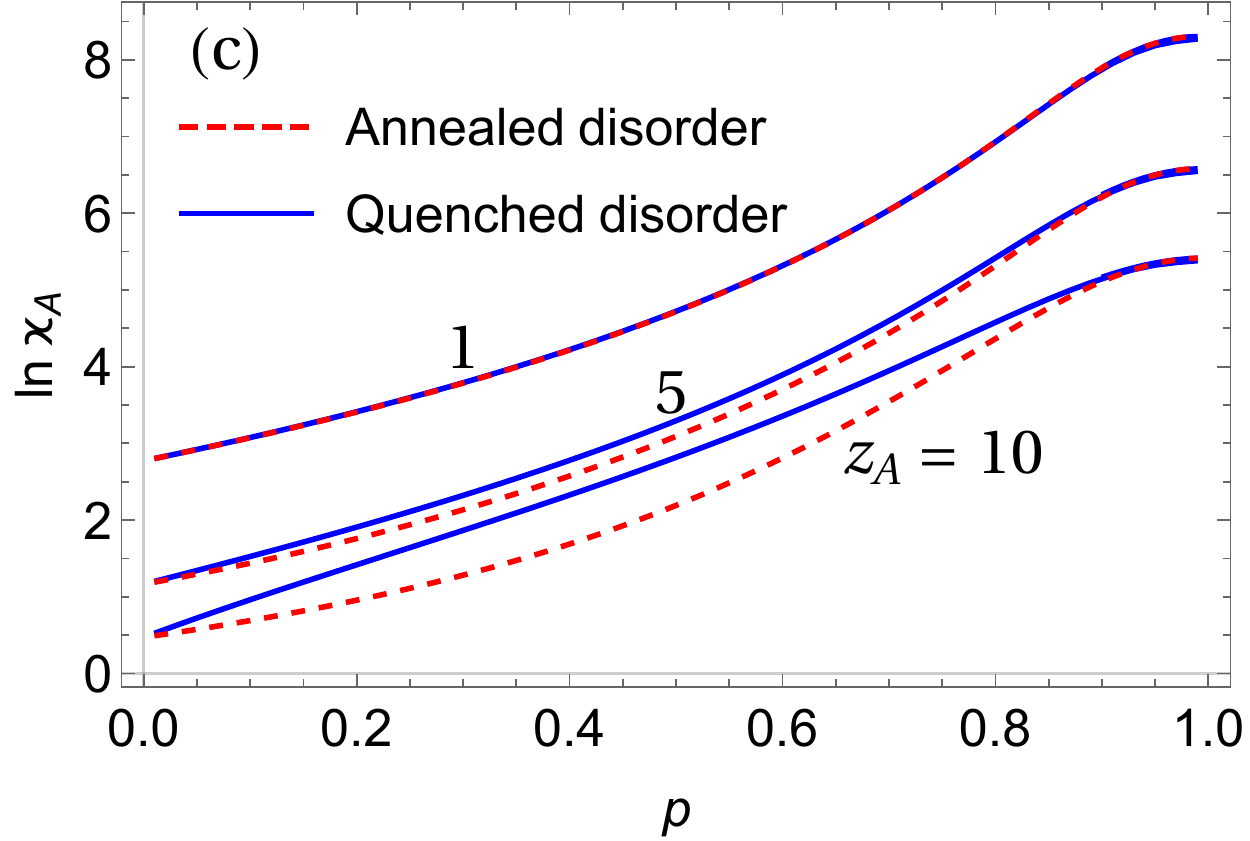}
		\includegraphics[width=0.4\textwidth,height=0.3\textwidth]{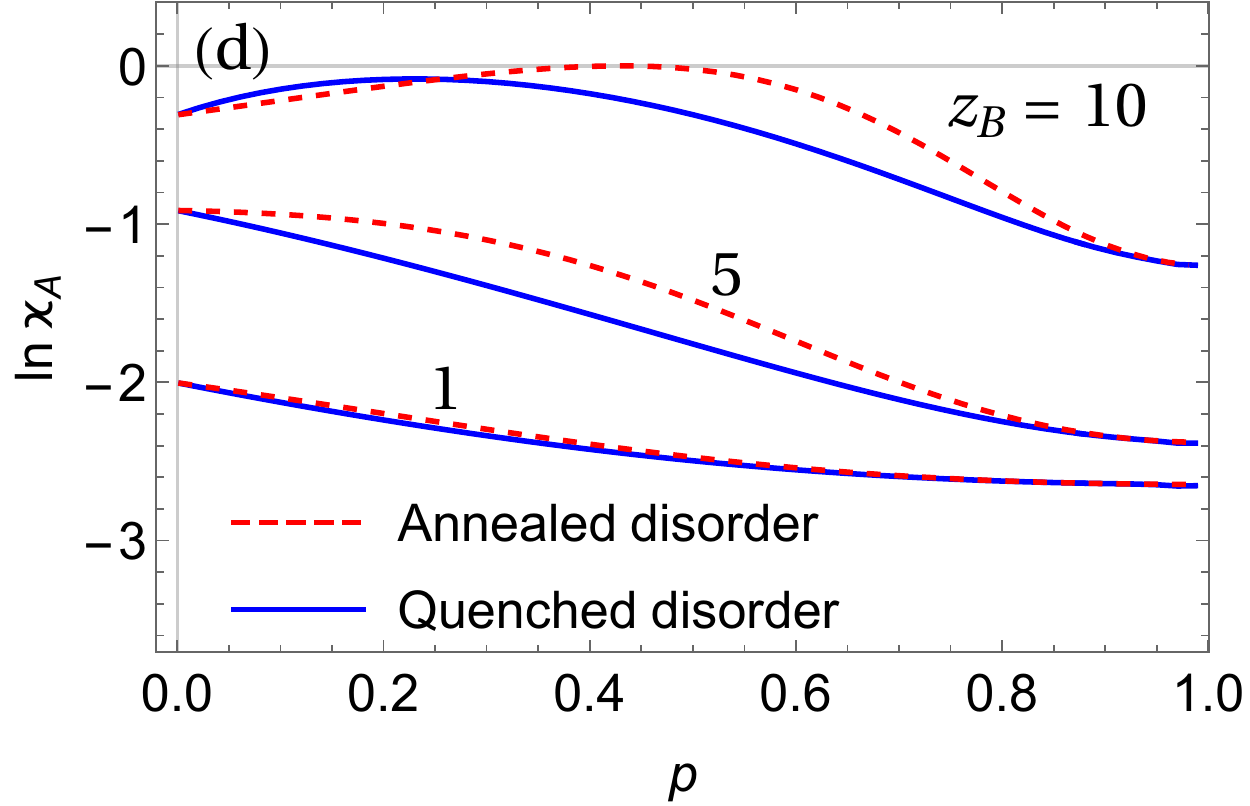}
		\caption{{\em Model II. Annealed (red dashed curves) versus quenched (blue solid curves) disorder.}
			(a), (b) Disorder-averaged density $n_A$ as a function of the mean concentration $p$ of catalytic sites for three values of $z_A$ and $z_B = 15$ (a)
			and for three values of $z_B$ and $z_A = 15$ (b).
			(c), (d) Logarithm of the compressibility $\varkappa_{A}$ as a function of the mean concentration $p$ of catalytic sites for three values of $z_A$ and $z_B = 15$ (c)
			and for three values of $z_B$ and $z_A = 15$ (d).
			\label{figsm2}
		}
	\end{figure*}

	In Fig.~\ref{figsm2} we compare the behavior of the thermodynamic properties for Model II with annealed (red dashed curves) and quenched disorder (blue solid curves). In Figs.~\ref{figsm2} (a) and \ref{figsm2} (b) we depict 
	the disorder-averaged density $n_A$  as a function of the mean concentration $p$ of the catalytic sites for several values of $z_A$ and fixed $z_B = 15$ [Fig.~\ref{figsm2} (a)] and for several values of $z_B$ and fixed $z_A = 15$ [Fig.~\ref{figsm2} (b)]. Figures \ref{figsm2} (c) and \ref{figsm2} (d) present a logarithm of the 
	the compressibility $\varkappa_{A}$ again a function of the mean concentration $p$ of the catalytic sites. We observe that, qualitatively, the behavior is very similar to the one found in Model I. However, quantitatively, in Model II
	the difference between the cases of annealed and quenched disorder is more pronounced than in Model I, especially at intermediate concentrations $p$ of the catalytic sites and high values of activities.

\twocolumngrid


\end{document}